\documentclass[twocolumn]{svjour3}         
\smartqed  
\usepackage{graphicx}
\usepackage{amssymb}
\usepackage{amsmath}
\usepackage{subfigure}
\usepackage{relsize}
\usepackage{booktabs}
\begin{document}

\title{Effect of size of nanoparticles on the Wormlike micelle-nanoparticle system
}


\author{Sk. Mubeena         
}


\institute{Sk. Mubeena \at
              IISER-Pune, 900 NCL Innovation Park, Dr. Homi Bhaba Road,  Pune-411008, India. \\
              Tel.: +91 9130731891\\
              \email{mubeena@students.iiserpune.ac.in}           
}
\date{\today}

\maketitle

\begin{abstract}

            We investigate the effect of the size of nanoparticles on the behaviour of equilibrium polymers (Wormlike micelles) and nanoparticle system. The self-organised structures of nanoparticles in the system show a morphological change from percolating networks to non-percolating clusters with an increase in the minimum approaching distance (EVP-excluded volume parameter) between nanoparticles and the equilibrium polymers. The shape of the nanoparticle clusters (nanorods, nanosheets, etc.) depends on the density of the polymer matrix, irrespective of the size of nanoparticles. We show that with an increase in the nanoparticle size, the value of EVP at which the nanoparticle structure undergoes the morphological change shifts to lower values. We also report that with the increase in nanoparticle size, the packing of nanoparticles decreases. Hence, they do not form a well-defined structure with nanoparticles of bigger size. This decrease in the packing is due to the decrease in the surface to volume ratio which in turn decreases the surface interactions.

\keywords{sel-assembly, computational modelling, Wormlike micelles, polymer-nanocomposites, equilibrium polymers, polymer templating }
\end{abstract}

\section{Introduction}

       In nature, there exist many of the complex fluids that show size dependent behaviour. The interaction between these particles not only depend on their chemistry and surface charge but also on their size and shape \cite{hassanabadi2014effect}. It is the suitable choice of the size and shape of the particles that engineer the whole rheological behaviour of toothpaste, paints, food texture, etc. \cite{anvari2014rheology,nelson2017design,beckett2009chocolate}. The classic effects of the size of the particles can be seen in the transition from micro-composites to nanocomposites leading to super-enhanced material properties that has galvanised the whole economy by adding nanofillers in various polymeric matrices \cite{revilla2007comparative,gao2004clay,tjong2006structural}. The topic of nanocomposites of polymers has transformed into an active research area. The literature suggests that the dramatic change in the mechanical and viscoelastic properties of a polymer nanocomposite is due to the smaller size of nanoparticles (NPs) having a high surface to volume ratio with properties having a quantum physics origins \cite{vollath2004synthesis,wang2006photoluminescence,vollath2004oxide}. There are reports of an increase in the tensile strength of beads/polyester composite with the increase in the size of glass beads \cite{hassanabadi2014effect}. Ng. et. al observed a higher failure strain and scratch resistance of $TiO_2$ (32 nm)/epoxy composite in comparison to the micron size $TiO_2$ composite ~\cite{ng1999synthesis}.

         Apart from the improved mechanical and viscoelastic properties, polymer nanocomposites are also explored for nanofabrication and investigation having fundamental interests in equilibrium and non-equilibrium physics ~\cite{muthukumar1997competing,yang1998hierarchically,liu2003nanofabrication}. Most of these studies are concerned with polymer matrices. There also exists some reports on the matrices involving thermotropic nematic liquid crystals ~\cite{hegmann2007nanoparticles,bisoyi2011liquid}. In a nematic matrix, particle dispersion gives rise to a long-range interparticle interaction due to the elasticity of the matrix ~\cite{poulin1994dispersion,ramaswamy1996power,voltz2006director}. There are reports of formation of chain-like arrays of colloidal silicone droplets having dipolar interaction and formation of 2-dimensional crystal with quadrupolar interaction ~\cite{loudet2000colloidal,loudet2004colloidal,muvsevivc2006two}. An increase in the solid moduli of the elastic matrix on the inclusion of colloidal particles in nematic, lamellar and smectic phases are also observed ~\cite{basappa1999structure,shouche1994effect,zapotocky1999particle}. An experimental investigation involving Wormlike micellar matrix showed that larger sized NPs get excluded by the Wormlike micellar domain while smaller particles get inside the matrix making it swell~\cite{sharma2009self}. 

         There exist very few studies concerning Wormlike micelles (or equilibrium polymers) and NP composites and the literature is devoid of a systematic study on the effect of the size of NPs.  In this paper, we will study the system with respect to the change in the size of NPs and report the morphological transformations with system parameters. The percolating network-like structures of nanoparticles inside a wormlike micellar matrix is already reported experimentally \cite{kamenda2011self}. The system also shows usniformly mixed state and non-percolating clusters of nanoparticles for lowest and high values of EVP \cite{arxive}. In this paper we show that with increase in the size of nanoparticles, the value of EVP at which the morphological changes occurs gets shifted to the lower values. Moreover, we also report that the larger size particles are unable to form a well-packed or a well-defined structure. This is attributed to the decrease in surface to volume ratio of nanoparticles with increase in size.

\section{Model and method}


         Our model is a modified version of the model presented by \cite{rahul,chatterji2003statistical}. In our model, Wormlike micelles (WLM)are represented by a chain of beads which evolves under the given potentials to form wormlike micellar chains. In the process of coarse graining from amphiphilic molecules to a bead as a group of amphiphilic molecules, we ignore all chemical details and only use the details which are relevant at the mesoscopic scale. Thus every single spherical particle which we call monomer represents a group of amphiphilic molecules that form a micellar chain. A precisely same model has also been used in \cite{mubeena2015hierarchical}. We have kept the size of each monomer $\sigma$ as the system unit and all the distances are measured from centre-to-centre unless, otherwise mentioned.
\subsection{The potentials:}
\begin{figure}
\centering
\includegraphics[scale=0.15]{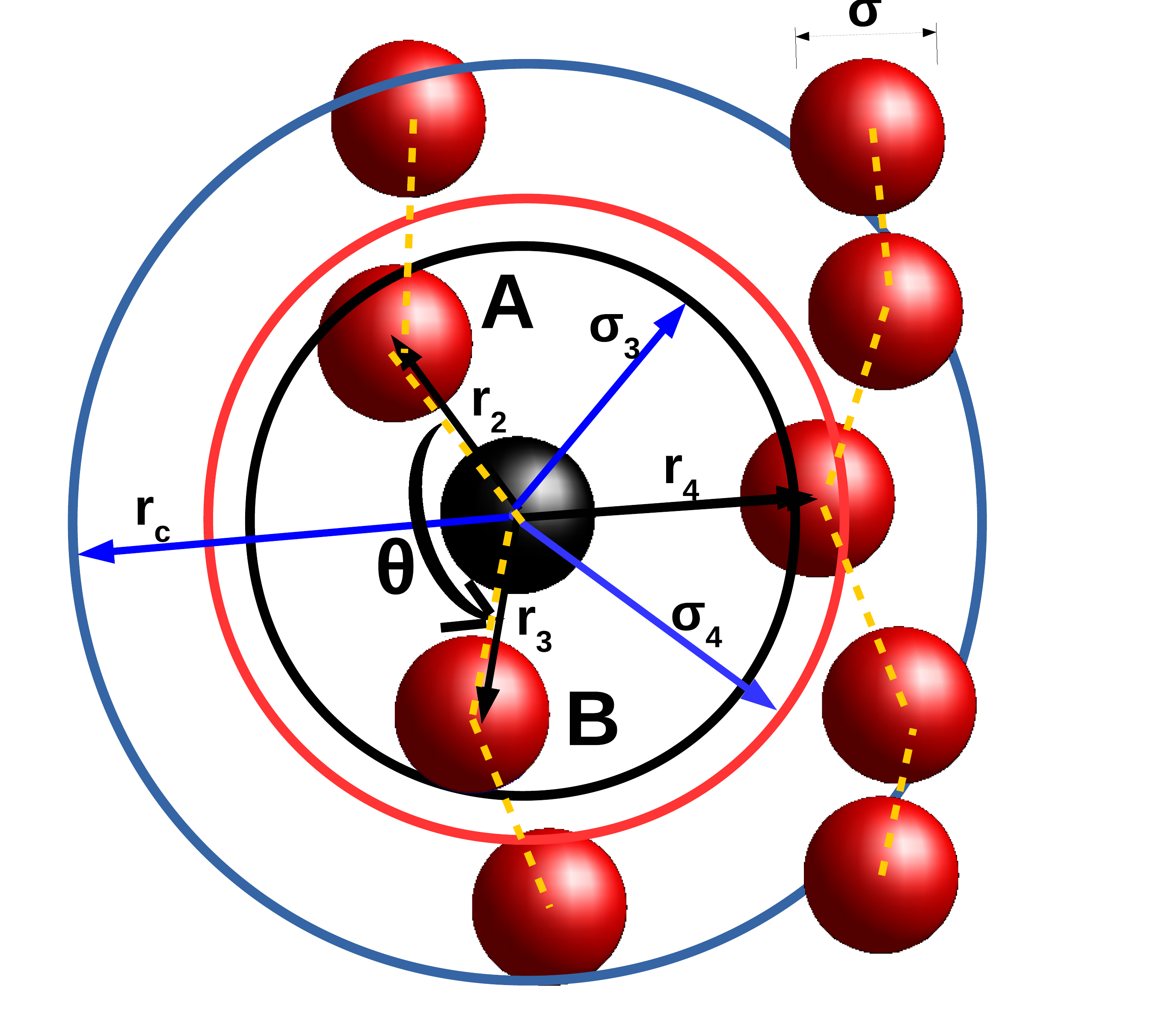}
\includegraphics[scale=0.2]{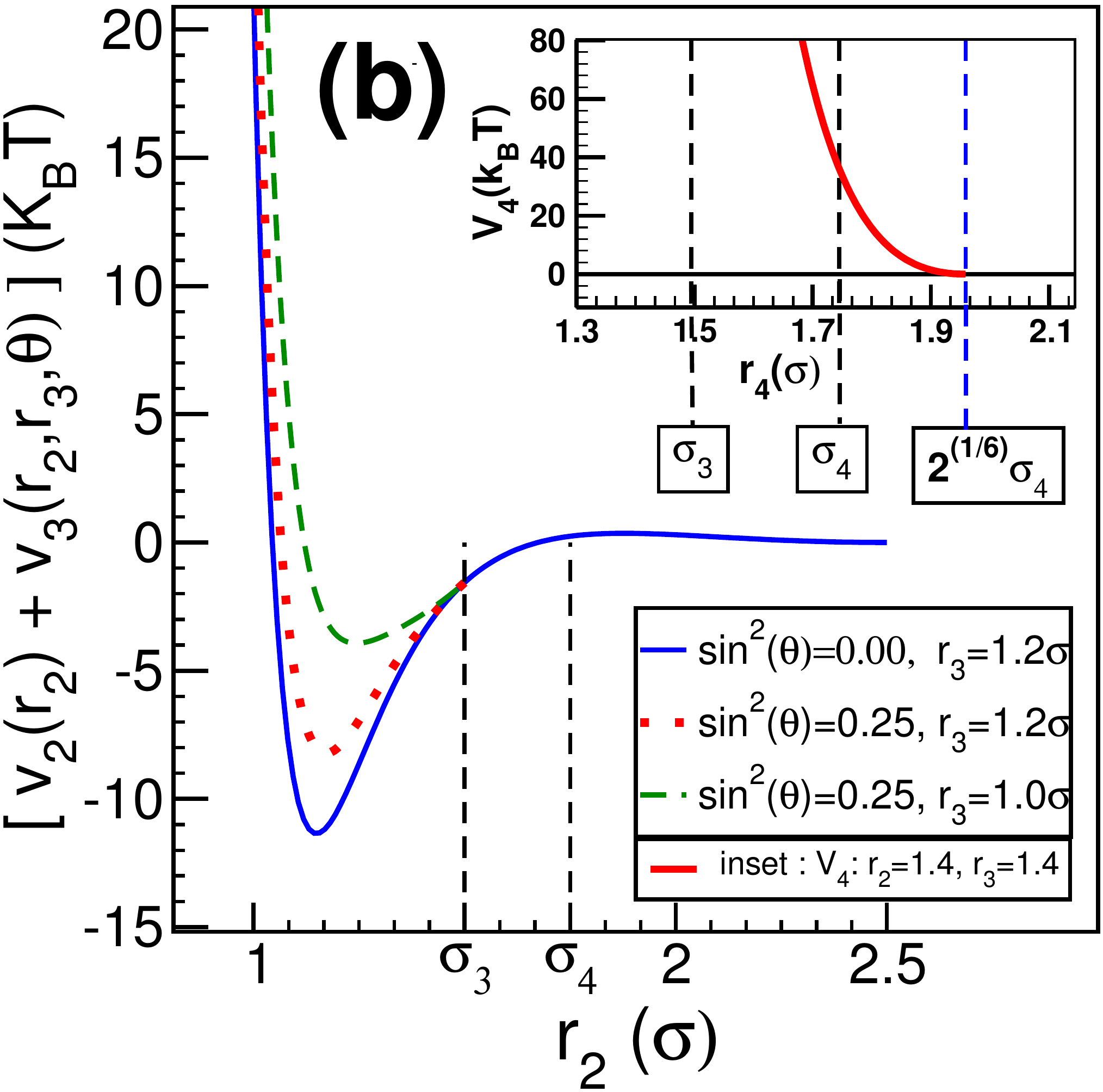}
\caption{(a) The figure shows the coarse-grained model of Wormlike micelles where the spherical monomers with diameter $\sigma$ self-assemble to form a chain-like structure. All the distances are shown with respect to the central monomer shown in black. These monomers are acted upon by two body potential $V_2$ having a cutoff range of $r_c$. A three-body potential $V_3$ acts on a triplet with a central monomer bonded by two other monomers A and B at distances $r_1$ and $r_2$, respectively, that forms an angle $\theta$ at the central monomer and has a cutoff range of $\sigma_3$. To avoid branching in chains, there is a four-body potential $V_4$ which acts when a monomer approaches another monomer that is already part of a chain (shown at a distance $r_4$ from the central monomer) and thus tries to form a branch. The potential $V_4$ is a shifted Lennard-Jones potential having a cutoff distance $2^(1/6)\sigma_4$. (b) The plot of $V_2+V_3$ shown for different values of $sin^2\theta$. The inset shows a shifted Lennard-Jones potential $V_4$.}
\label{fig:model}
\end{figure}

\begin{itemize}
\item {
A two body attractive Lennard-Jones Potential superimposed with an exponential term:\\
   \begin{equation}
V_2 = \epsilon [ (\frac{\sigma}{r_2})^{12} - (\frac{\sigma}{r_2})^6 + \epsilon_1 e^{-a r_2/\sigma}]; 
\, \forall  r < r_c.
\label{eq1}
\end{equation}
    Where $r_2$ is the distance between two interacting monomers, with $\epsilon=110k_BT$, and the cutoff distance $r_c=2.5\sigma$ . The exponential term in the above potential creates a maximum at $r_2=1.75\sigma$ which acts as a potential barrier to joining or breaking of monomers from chains. We set $\epsilon_{1}=1.34\epsilon$ and $a= 1.72$.
}
\item{
Now a three-body potential $V_3$ is added to provide semiflexibility to chains.
\begin{equation}
V_3 = \epsilon_3 (1 - \frac{r_{2}}{\sigma_3})^2(1 - \frac{r_{3}}{\sigma_3})^2 \sin^2(\theta); 
\, \forall r_{2},r_{3} < \sigma_3. 
\label{eq2}
\end{equation}

      Here $r_{2}$ and $r_{3}$ are the distances of the two monomers from the central monomer which forms a triplet with an angle $\theta$ subtended by $\vec{r_{1}}$ and $\vec{r_{2}}$. Here $\epsilon_{3}=6075k_{B}T$ and the cutoff distance for this potential defined to be $\sigma_3=1.5\sigma$. Thus monomers at a distance less than or equal to $\sigma_3$ are considered to be bonded/part of a chain. The terms leading to $\sin^2(\theta)$ are necessary to ensure that the potential and force goes smoothly to zero at the cutoff of $\sigma_3$. This potential acts only if a monomer has two bonded neighbours at a distance below $\sigma_3$.
}

\item{ Using potentials $V_2$ and $V_3$, monomers do self-assemble to form chains, but these chains can also join to give rise to branches. To overcome this problem, it was necessary to repel any particle which tries to form a branch. Thus we introduce a four-body term which is a shifted Lennard-jones repulsive potential $V_{4}$\\
\begin{equation}
V_4 = \epsilon_4 (1 - \frac{r_{2}}{\sigma_3})^2(1 - \frac{r_{3}}{\sigma_3})^2 \times V_{LJ}(\sigma_4,r_4) 
\label{eq3}
\end{equation}
 Here $r_{2}$ and $r_{3}$ are the distances of the two monomers from central monomer which forms a triplet, while $r_{4}$ is the distance of the fourth monomer which tries to approach the central monomer which already has two bonds, as shown in Fig.~\ref{fig:model}(a). We set the cutoff distance for this potential to be $2^{1/6}\sigma_4$ where, $\sigma_4=1.75\sigma$ (note $\sigma_{4}>\sigma_{3}$). The large value of  $\epsilon_4=2.53 \times 10^5 k_BT$ is necessary to ensure enough repulsion in case $(1 - \frac{r_{2}}{\sigma_3})^2 << 1$ and/or $(1 - \frac{r_{3}}{\sigma_3})^2 << 1$. These terms ensure that the potential and the force goes smoothly to zero at the cutoff of $2^{1/6}\sigma_4$.
}
\end{itemize}

      Figure~\ref{fig:model}(a) explains the above model. In this figure The red spheres denote the micellar monomers having diameter $\sigma$ and all the distances are shown with respect to the central monomer in black. These monomers are acted upon by a two-body potential $V_2$ having a cutoff range of $r_c$. A three-body potential $V_3$ is acting on a triplet with the central monomer (shown in black) bonded by two other monomers forming an angle $\theta$ at the central monomer and having a cutoff range of $\sigma_3$. In addition to these potentials, there is a four body-potential $V_4$ which is a shifted Lennard-Jones potential introduced to prevent branching and having a cutoff distance $\sigma_4$.

The potential $V_2$+$V_3$ is shown in Fig.~\ref{fig:model}(b) for different values of $\sin^2\theta$ with the inset showing the four body shifted Lennard-Jones potential $V_4$.

   Using Metropolis Monte Carlo technique, the model WLM system is evolved. The system produces an exponential distribution of its chain length and shows an isotropic-nematic transition, thereby, confirming the characteristic properties of the WLM system (or equilibrium polymers) ~ \cite{mubeena2015hierarchical}.

 To explore the self-assembly of NPs inside the WLM matrix, we add model NPs in the model WLM system. The NPs are modelled by a Lennard-jones potential given by,
\begin{equation}
V_{2n} = \epsilon_n[(\frac{\sigma_{n}}{r_n})^{12} - (\frac{\sigma_{n}}{r_n})^6],      \forall     r_n   <=   r_{cn}
\end{equation}
 Where, $\sigma_n$ and ${r_n}$ are the size of NPs and the distance between the two interacting particles, respectively. The cutoff distance for the interaction is kept to be $r_{cn}=2\sigma_n$ and the strength of the attraction is $\epsilon_n=11k_BT$.
The interaction between NPs and micellar monomers is modelled by Lennard-Jones potential shifted in such a way that only the repulsive part is within the cutoff range. This is given by,
\begin{equation}
V_{4n} = \epsilon_{4n}[(\frac{\sigma_{4n}}{r_n})^{12} - (\frac{\sigma_{4n}}{r_n})^6],  \forall r_n <= 2^{1/6}\sigma_{4n},
\end{equation}
The strength of repulsion for the above potential is kept to be $\epsilon_{4n}=30k_BT$. The parameter $\sigma_{4n}$ sets the minimum approaching distance between monomers and NPs, therefore it is used as an excluded volume parameter (EVP) between NPs and monomers.

       In order to study the system behaviour of the model NPs in the model WLM matrix, the monomers and NPs are randomly initialized in a $30\times 30\times 60\sigma^3$ simulation box. The system is evolved using Metropolis Monte Carlo technique. As the system under consideration is quite dense and Monte Carlo moves were inefficient to move the particles, a semi-grand canonical scheme is used for the NPs. According to this scheme, the model polymeric system is first evolved with a few hundred (200-300) of NPs in it for $10^5$ Monte Carlo Steps (MCSs). Then the Monte Carlo steps are coupled with semi-grand canonical steps only on irandomly chosen NPs such that the frequency of semi-grand canonical steps on NPs is 300 for every 50 Monte Carlo steps. Each successful attempt of addition or removal of a NP is penalised by an energy gain or loss of $\mu_n$, which sets the chemical potential of the system. The same method has laso been used in \cite{mubeena2015hierarchical}. The value of $\mu_n$ is kept to be $-8k_BT$.

\section{Results ::}

        We study in detail the self-assembly of the WLM-NP system and report the morphological transformations of the self-assembled NP structures with a change in $\sigma_{4n}$ for the size of particles $\sigma_n>1.5\sigma$. The same has been reported earlier for $\sigma_n=1.5\sigma$. The study confirms the results with 10 different independent runs initializing the mixture in a randomly mixed state (similar to the in-situ preparation of NPs inside a polymeric matrix), all converging to the statistically similar morphological states. These structures are claimed to be kinetically arrested states as there is a very slow variation in the number of nanoparticles in the system. This is because, as the system evolves the nanoparticles clusters get more and more packed accommodating more particles inside the box. However, as the system volume is constant the number of nanoparticles eventually comes to a constant value but after a very long run. This has been tested with all the values of parameters and with MCSs as long as $(4-8)\times 10^6$ iteration and it is observed that after around $2\time 10^6$ iterations there is effectively no change in the system architecture. Moreover, Convergence of all the independent runs starting with a randomly mixed state of polymeric chains and NPs confirm the existence of the metastable state. The percolating networks of nanoparticles inside Wormlike micelles is also reported experimentally\cite{sharma2009self}.

\begin{figure}
\includegraphics[scale=0.2]{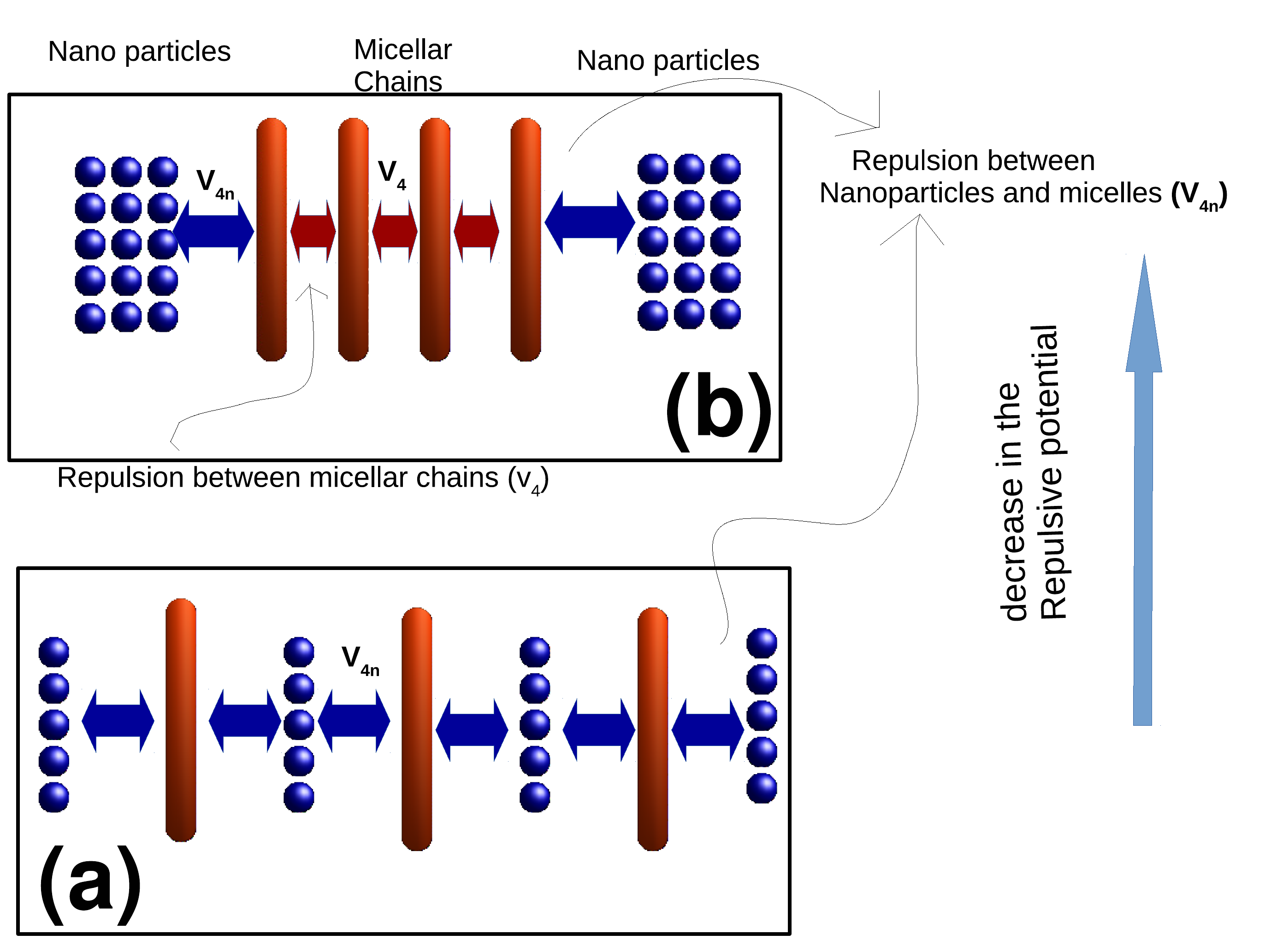}
\caption{(a) The figure shows a schematic representation of a uniformly mixed state of micellar chains and NPs where there exists a large number of contacts between NPs and micellar chains. The repulsion between NPs and micellar chains are shown by the blue arrows. (b) With the increase in $\sigma_{4n}$, the system reorganises to form clusters of chains and NPs that reduces the repulsive potential $V_{4n}$ at the cost of a small increase in $V_4$. The repulsion between chains due to $V_4$ is shown by the red arrows. Thus, the rearrangement of the system architecture with the change in $\sigma_{4n}$ is a result of the competition between $V_4$ and $V_{4n}$.}
\label{trans}
\end{figure}

       For small particles $\sigma_n=1.5\sigma$, it is already shown that the system forms a uniformly mixed state for the minimum possible value of $\sigma_{4n}=1.25\sigma$ ($(\sigma_{4n})_{min}=(\sigma+\sigma_n)/2$) \cite{arxive}. Further increase in $\sigma_{4n}$ leads to the formation of percolating network of NPs and polymeric chains. This transformation from a uniformly mixed state to a system spanning network is due to the competition between the repulsive potentials $V_4$  and $V_{4n}$. A uniformly mixed state of polymeric chains and NPs reduces the chain-chain number of contacts and hence reducing $V_4$ as shown in Fig.\ref{trans}(a). In case of very lowe value of $\sigma_{4n}$, there is enough volume to accomodate nanoparticles inside the simulation box such that the nanoparticles and micellar chains can maintain a distance $> \sigma_{4n}$ to avoid repulsion (due to $V_{4n}$). However, with the increase in $\sigma_{4n}$, this leads to a very high repulsive potential $V_{4n}$ due to increased contacts between NPs and chains. Therefore, the chains and NPs reorganise to form clusters as shown in Fig.\ref{trans}(b) similar to microphase separation. These clusters join to form an interpenetrating network-like structure of micellar chains and NPs. For an intermediate value of $\sigma_{4n}$, the volume fraction of NPs is higher than the volume fraction of monomers. However, the volume fraction of NPs decreases with increase in $\sigma_{4n}$. Therefore, for a higher value of $\sigma_{4n}$, the energy of NPs is higher than monomer chains because of low NP density. The micellar chains get aligned with increasing the distance between chains so that the repulsive potential $V_4$ is reduced. Hence, the NP network breaks into non-percolating clusters of NPs with the shape anisotropy of the NPs governed by the density of the matrix of polymers. With the increase in the polymeric matrix volume fraction, the shape anisotropy of the NP clusters increases giving rise to clusters from nanosheets to nanorods. These morphological transformations are studied for small particles $\sigma_n=1.5\sigma$. We study here the effect of increase in the size of particles.

       Thus, we see that the structural changes in the system architecture are mainly due to the competition between the repulsive interactions $V_4$ and $V_{4n}$.  Therefore, it is very convenient to analyse the system by including the total excluded volume fraction due to $V_4$ and $V_{4n}$ to the volume fraction of the matrix of polymers. We call it the effective volume fraction of the polymeric matrix. The current investigation is carried out on the same system, similarly varying $\sigma_{4n}$ and the monomer number density $\rho_m$ as earlier, but here we also examine the changes in system behaviour with a change in the NP size. Throughout this paper, the value of the chemical potential for NPs is kept fixed at $\mu_n=-8k_BT$. Thus, the current investigation have three parameters : NP size $\sigma_n$, EVP $\sigma_{4n}$ and micellar number density $\rho_m$. Before we proceed to the results of this investigation, lets first understand the procedure to calculate the effective volume of monomers.

\begin{figure}
\includegraphics[scale=0.2]{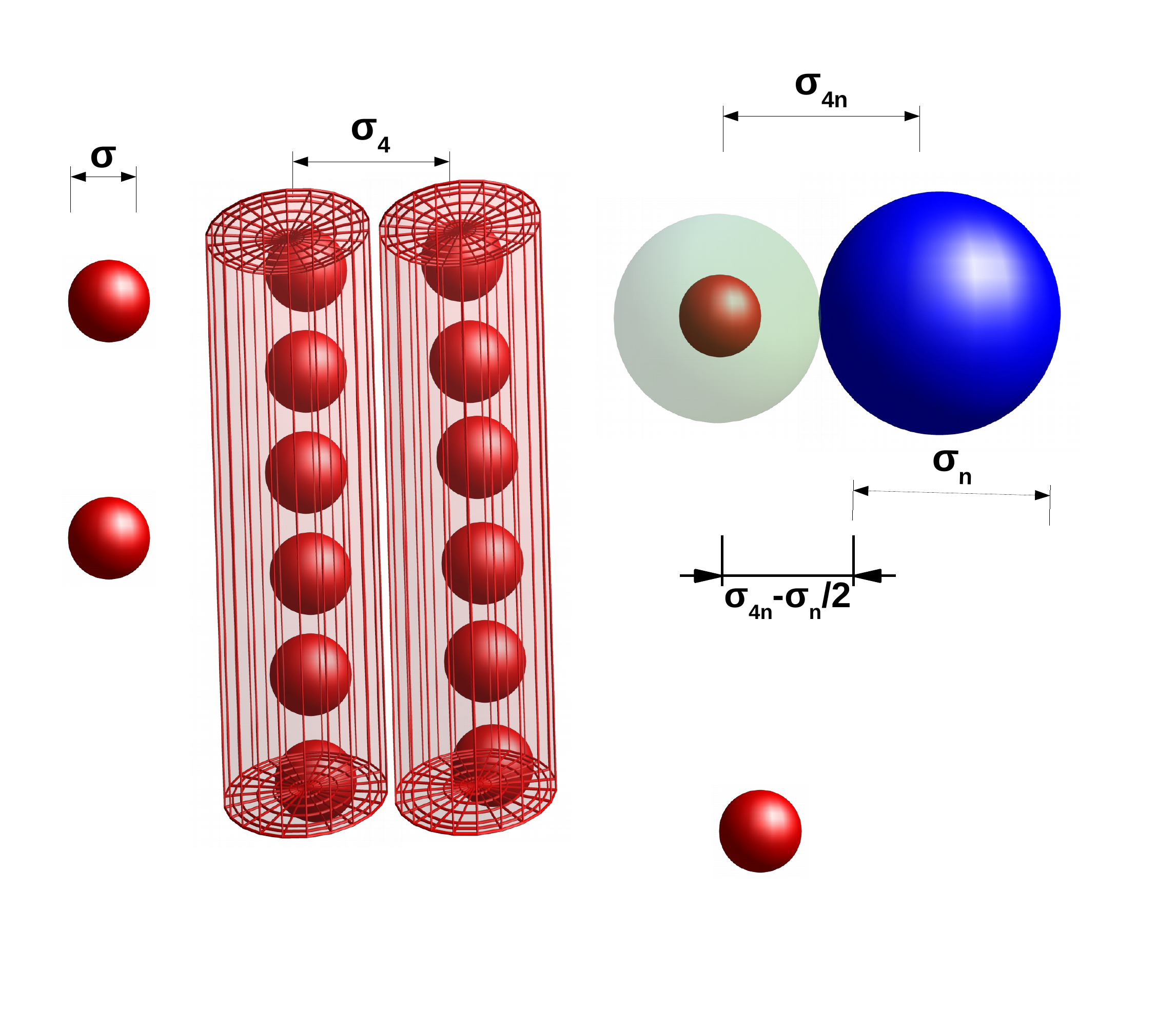}
\caption{The figure explains the calculation of effective volume of monomers. The red particles indicate monomers while the blue one shows a NP. Any two micellar chains under the repulsive interaction $V_4$ (within the cutoff distance of $2^{1/6}\sigma_4$ from each other) are considered as cylinders of radius $\sigma_4/2$ (shown as shaded cylindrical regions). If a monomer is within the cutoff distance $2^{1/6}\sigma_{4n}$ from a NP then, the monomer is considered as a sphere of radius $(\sigma_{4n}-\sigma_n)/2$ (shown by the shaded sphere).}
\label{eff_vol}
\end{figure}

       The parameter that defines the minimum approaching distance between micellar chains is $\sigma_4$ which is kept constant to be $1.75\sigma$. However, the parameter defining the minimum approaching distance between a monomer chain and a NP is defined as $\sigma_{4n}$, which we call as excluded volume parameter (EVP). Since the effective volume of monomers depends on the repulsive interactions in the system, it gets affected by the arrangement of the constituent particles in the system. Thus the change in the value of $\sigma_{4n}$ and the arrangement of the self-assembled micellar chains changes the effective volume of monomers. To calculate the effective volume of monomers, monomers which are part of a chain are grouped together and a detailed mapping of chains is done with respect to their distances from other chains and NPs. Any two micellar chains situated at a distance $r <= 1.75\sigma$ (under the range of repulsive potential $V_4$) from each other, are considered as cylinders of radius $\sigma_4/2$ and length equal to the length of the chain. Monomers situated at a distance $2^{1/6}\sigma_{4n}$ (cutoff distance for $\sigma_{4n}$) from NPs are considered as spheres of radius $\sigma_{4n}-\sigma_n/2$. If a monomer is found to be having a repulsive interaction with another monomer (from another chain) and NPs simultaneously, then whichever volume of the monomer is higher, is considered. This scheme of calculation of the effective volume of monomers is explained in Fig.\ref{eff_vol} and also has been used in \cite{arxive}.

\begin{figure}
\centering
\includegraphics[scale=0.2]{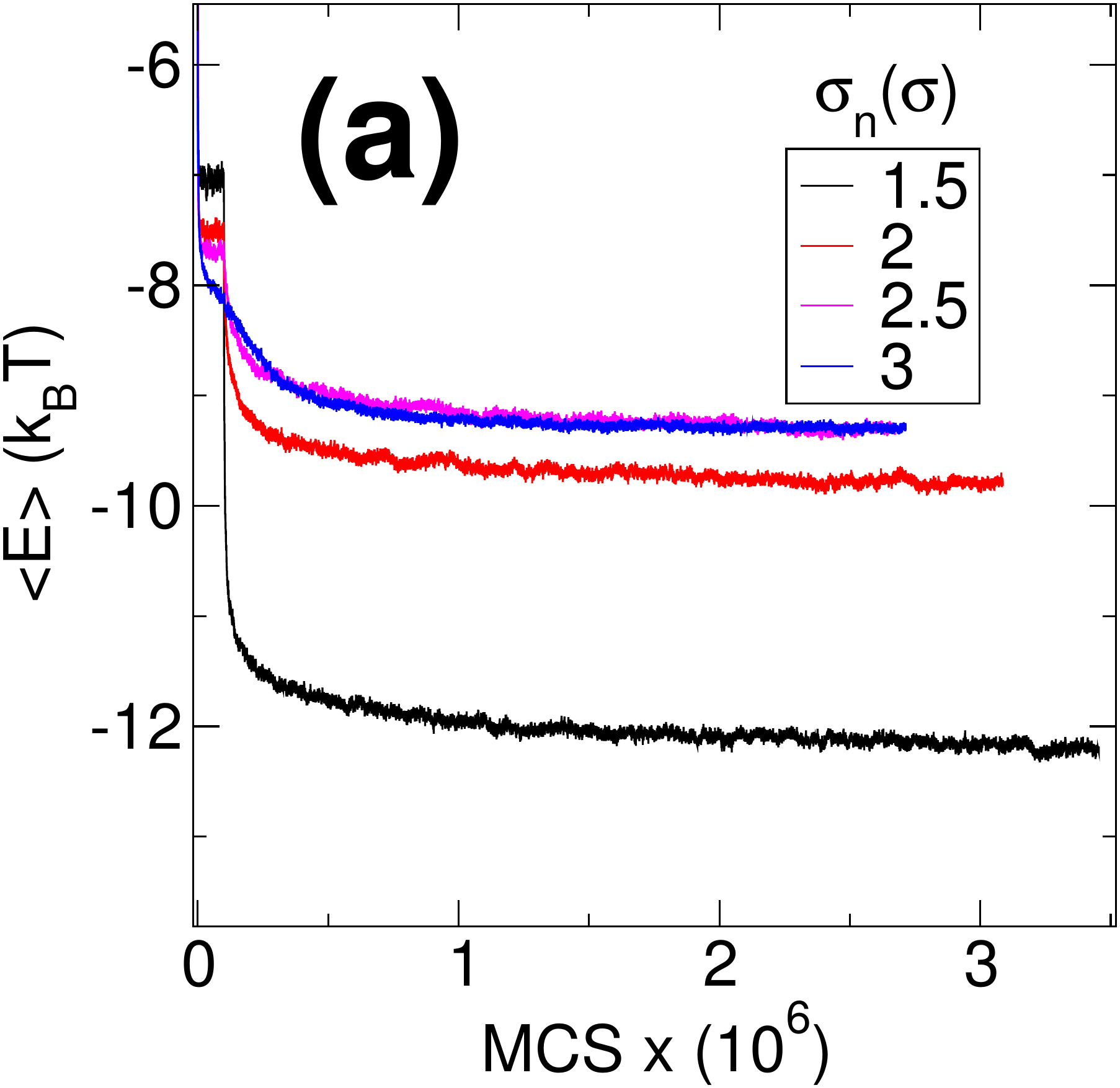}
\includegraphics[scale=0.25]{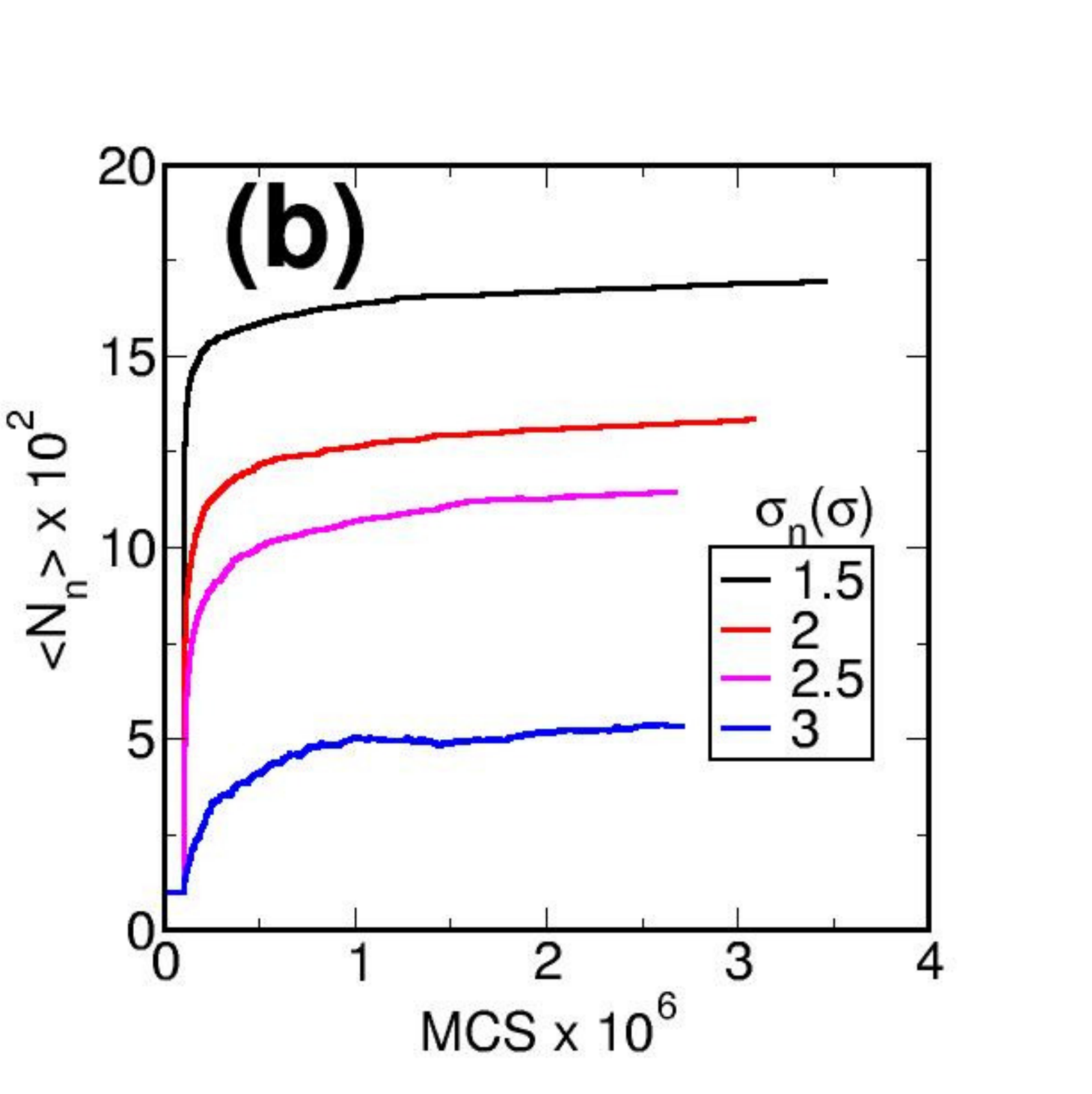}
\caption{The figure shows the evolution of (a) average energy of the system particles <E> and (b) the number of NPs $<N_n>$ for different values of NP size $\sigma_n$ as indicated by different symbols in each figure plotted against the no. of Monte Carlo Steps (MCSs). The system is subjected to Monte Carlo runs for first $10^5$ iterations and then a GCMC scheme is applied for the rest of the iterations. All the graphs show jumps in their values on switching on the GCMC scheme indicating a sudden increase in the number of NPs. After around $(2-3)\times 10^5$ of iterations, the value nearly becomes constant with a very slow variation in its value and the systems do not show any change in their morphological structures for observation time spanning to 1.5-4 millions of iterations. }
\label{energies}
\end{figure}

 To investigate the effect of the change in NP size on the model NP-micellar system, four different micellar number densities are considered, viz. $\rho_m=0.037\sigma^{-3}, 0.074\sigma^{-3}, 0.093\sigma^{-3}$ and $0.126\sigma^{-3}$. Now, for each number density of micelles considered, the EVP parameter $\sigma_{4n}$ is varied. Thus a set of runs were obtained with different values of $\rho_m$ and $\sigma_{4n}$ (for each value of $\rho_m$) for a fixed value of NP size $\sigma_n$. This set of runs is repeated for different size of NPs $\sigma_n$ varying from $1.5\sigma$ to $3.5\sigma$. For each run, the system was evolved with MCSs for first $10^5$ iterations and then the semi-GCMC scheme is switched on for the rest of the iterations. The systems are allowed to evolve long enough $(2-4)\times 10^6$ iterations) to ensure that the system morphology remains unchanged. Each run is further confirmed by using 10 independent runs all of which converge to statistically similar morphology and average thermodynamic quantities. The evolution of average energy per particle and the average number of NPs is shown in Fig.\ref{energies}(a) and (b), respectively, for $\rho_m=0.093\sigma^{-3}$ and $\sigma_{4n}=2\sigma$. Each figure shows graphs for different values of $\sigma_n$. In both the figures, at $10^5$ MCS, all the graphs show a jump in their values indicating an increase in the number of NPs as a result of switching on the GCMC scheme. It is observed that after around $(2-3)\times 10^5$ iterations the system does not show any morphological change in its self-assembled structure. After the systems show stable morphological structure, the representative snapshots are produced.

\begin{figure*}
\centering
\includegraphics[scale=0.2]{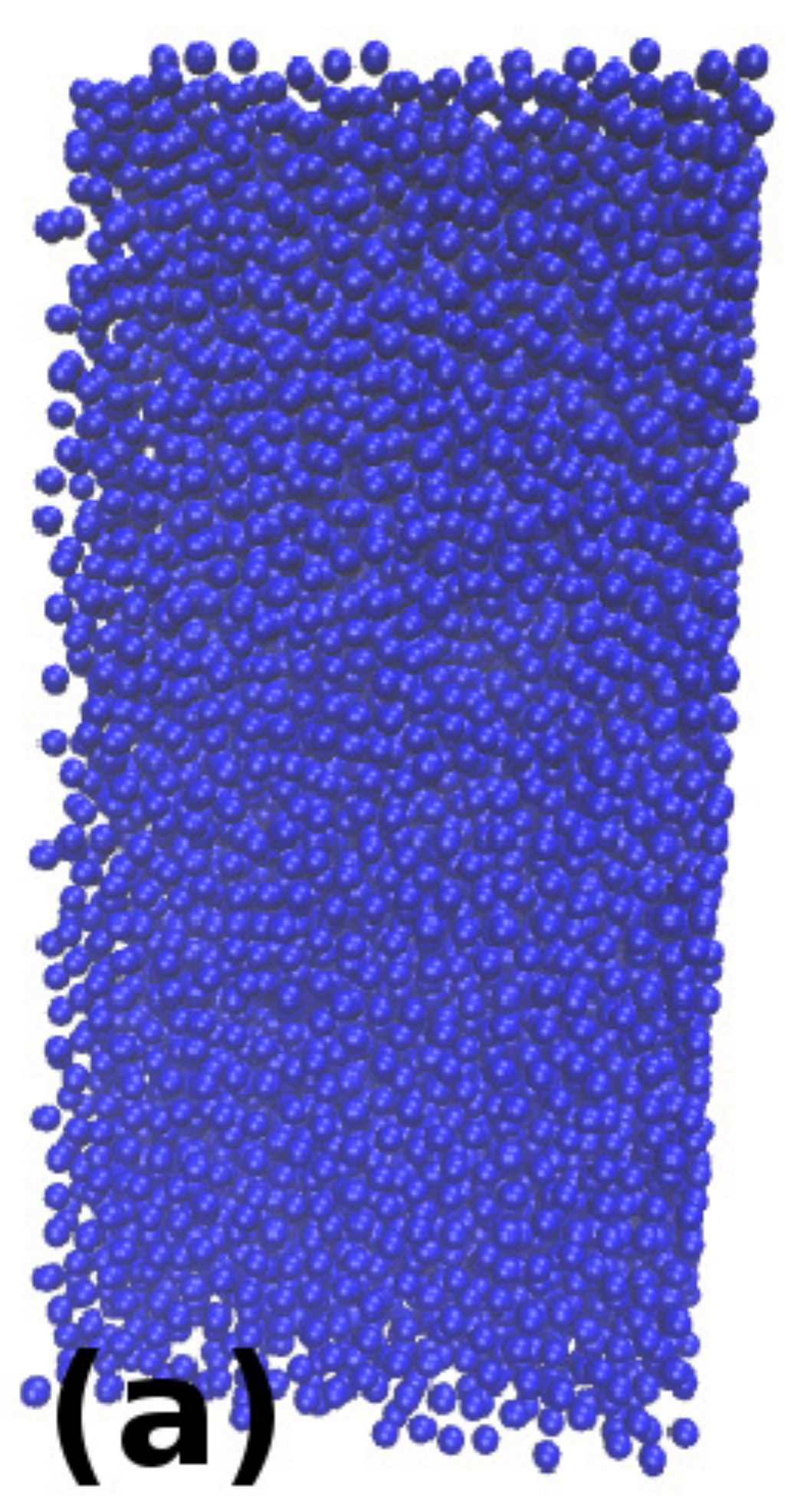}
\hspace{1cm}
\includegraphics[scale=0.2]{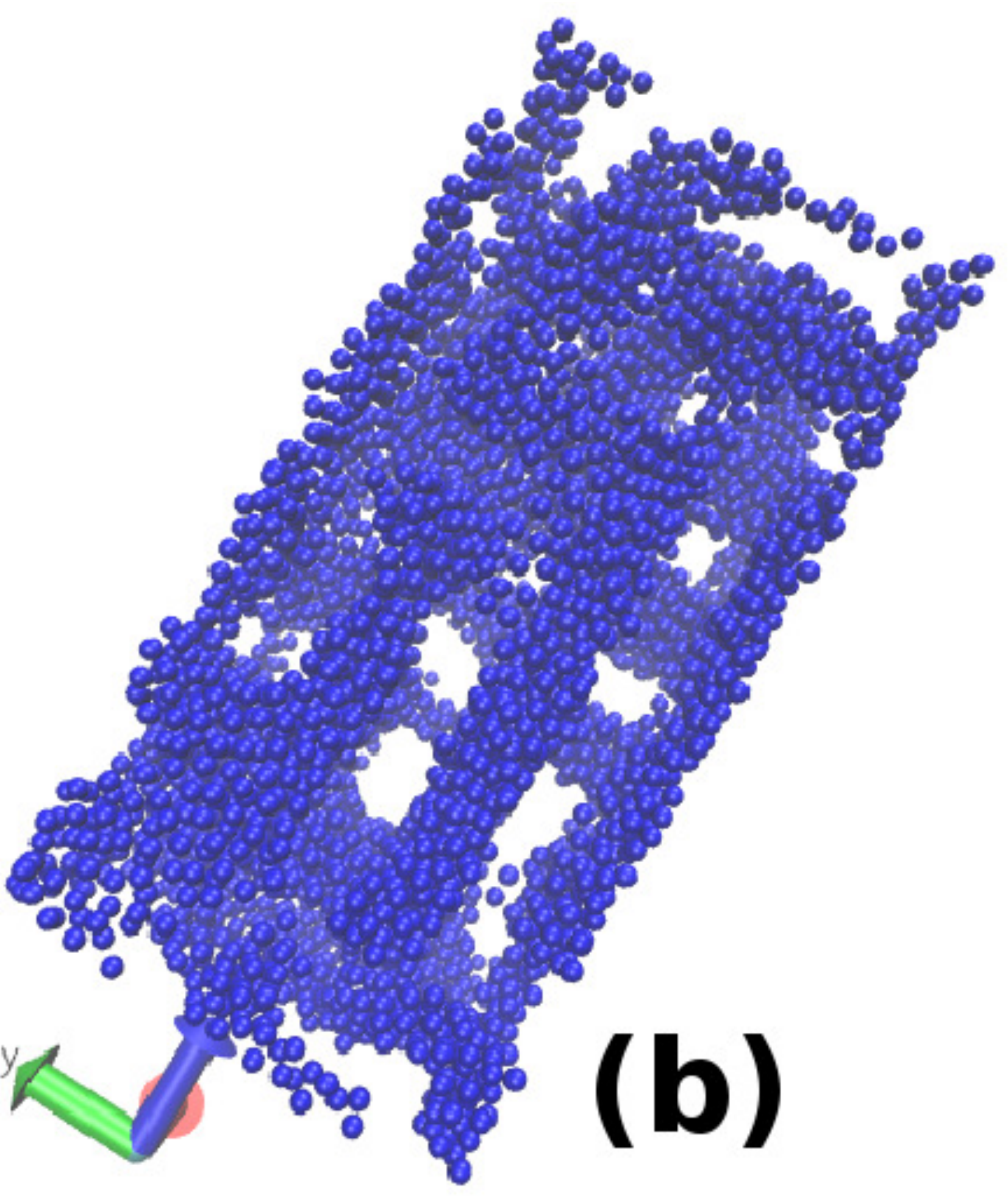}
\hspace{1cm}
\includegraphics[scale=0.2]{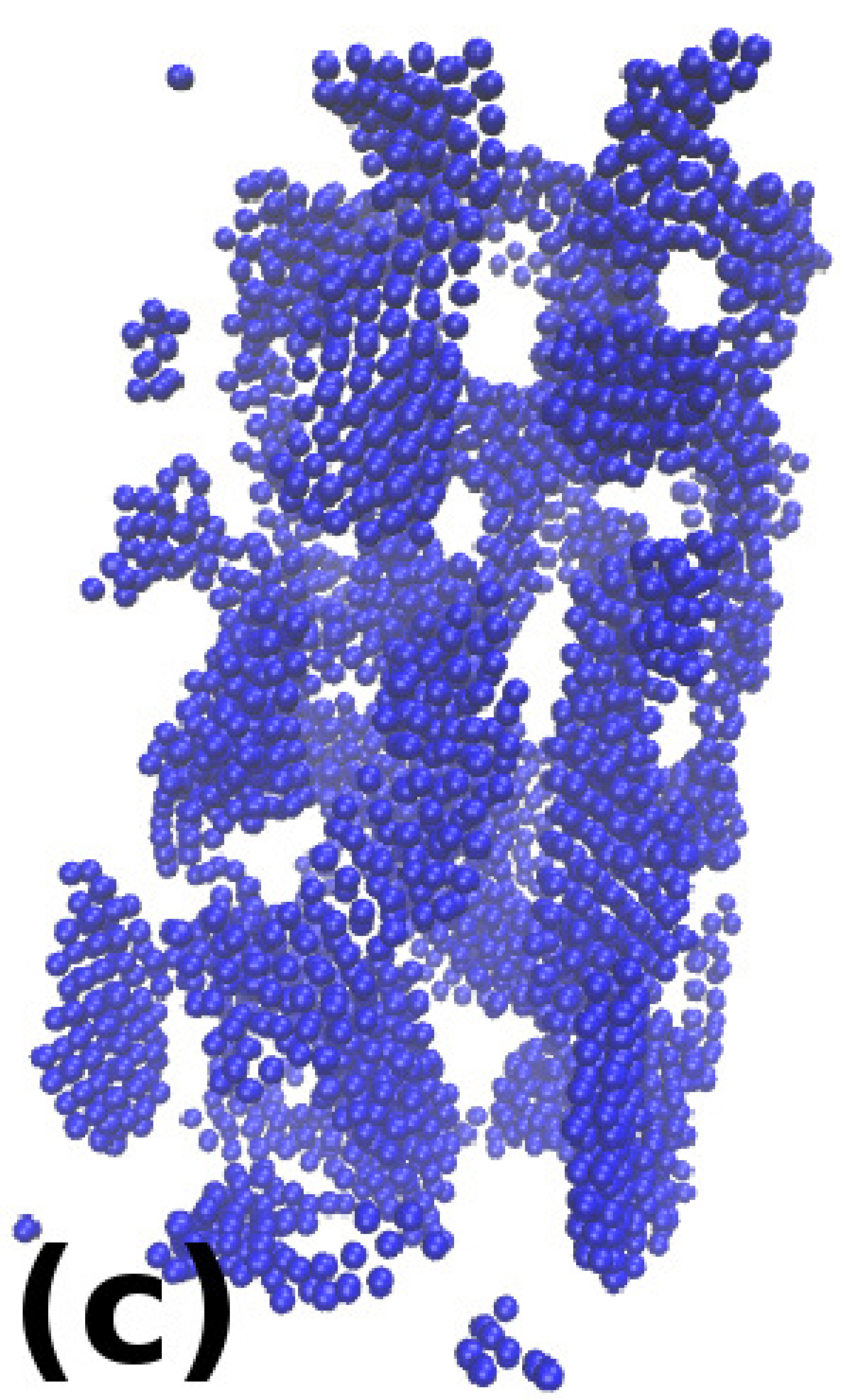}
\hspace{1cm}
\includegraphics[scale=0.2]{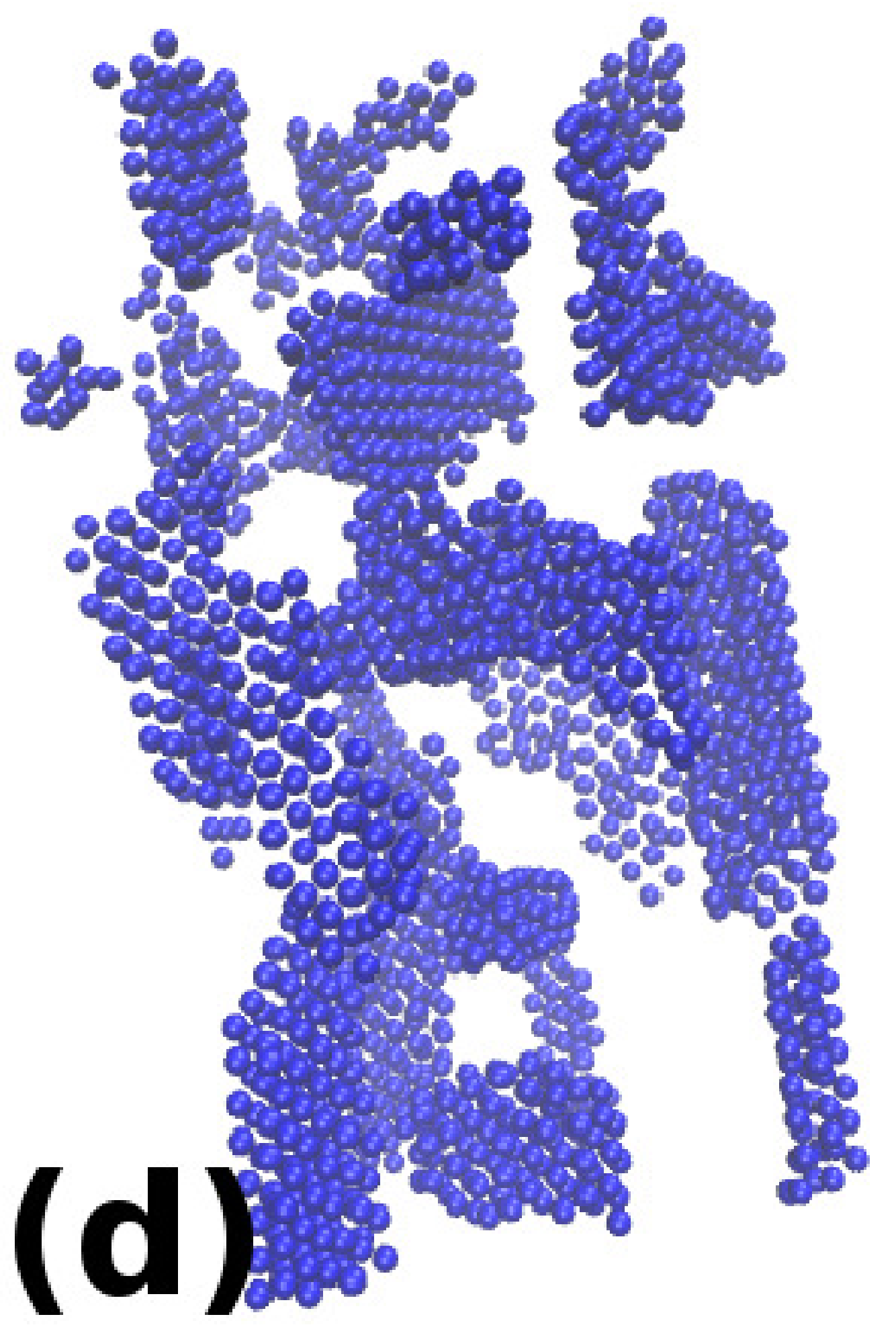} \\
\includegraphics[scale=0.3]{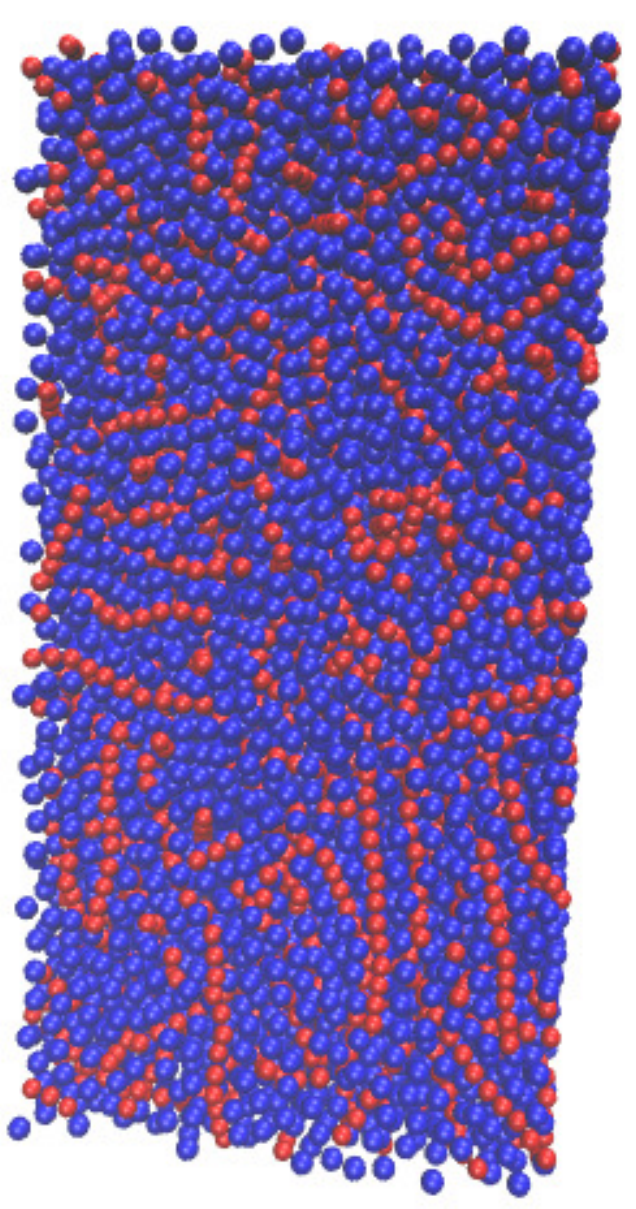}
\hspace{1cm}
\includegraphics[scale=0.3]{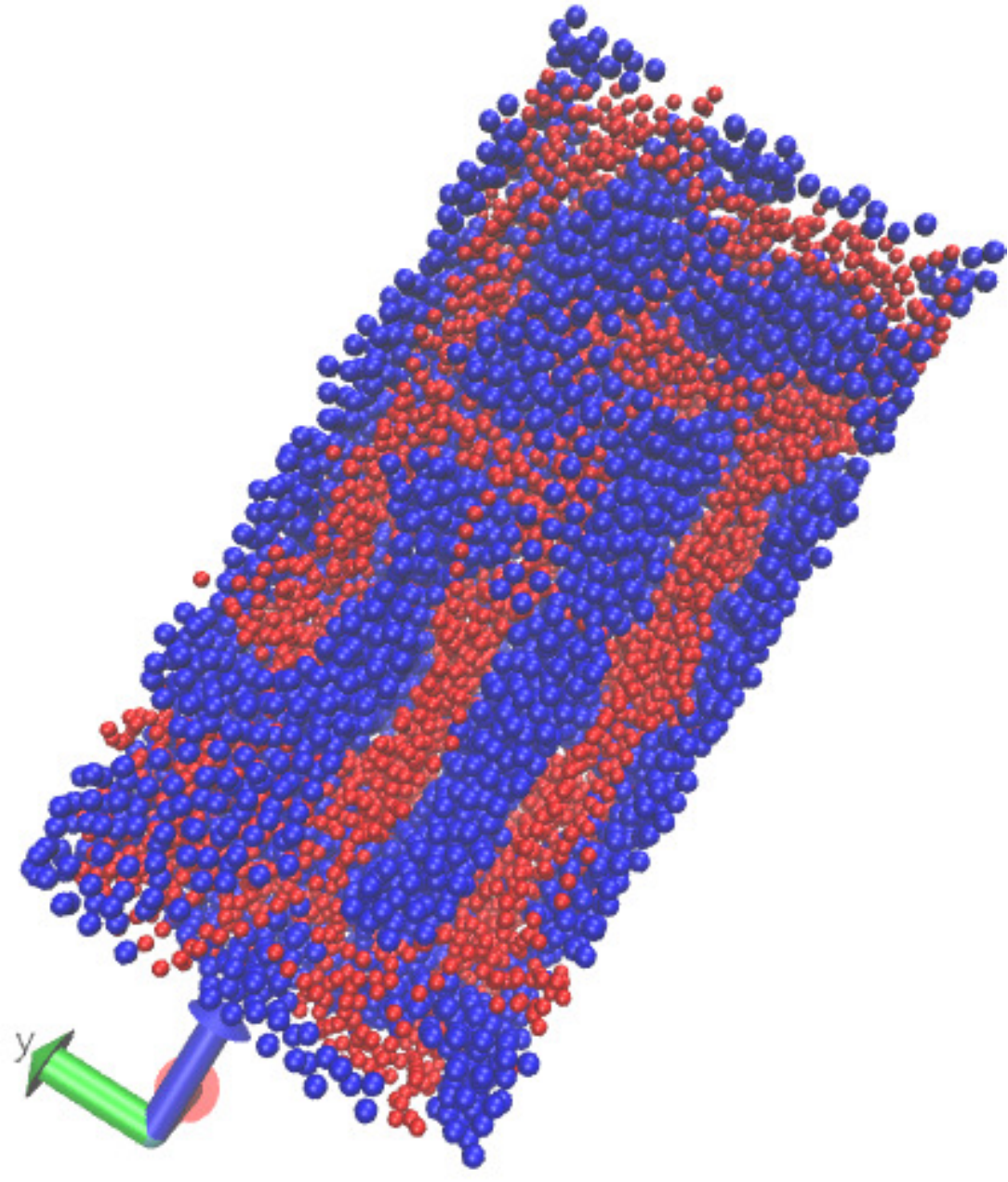}
\hspace{1cm}
\includegraphics[scale=0.3]{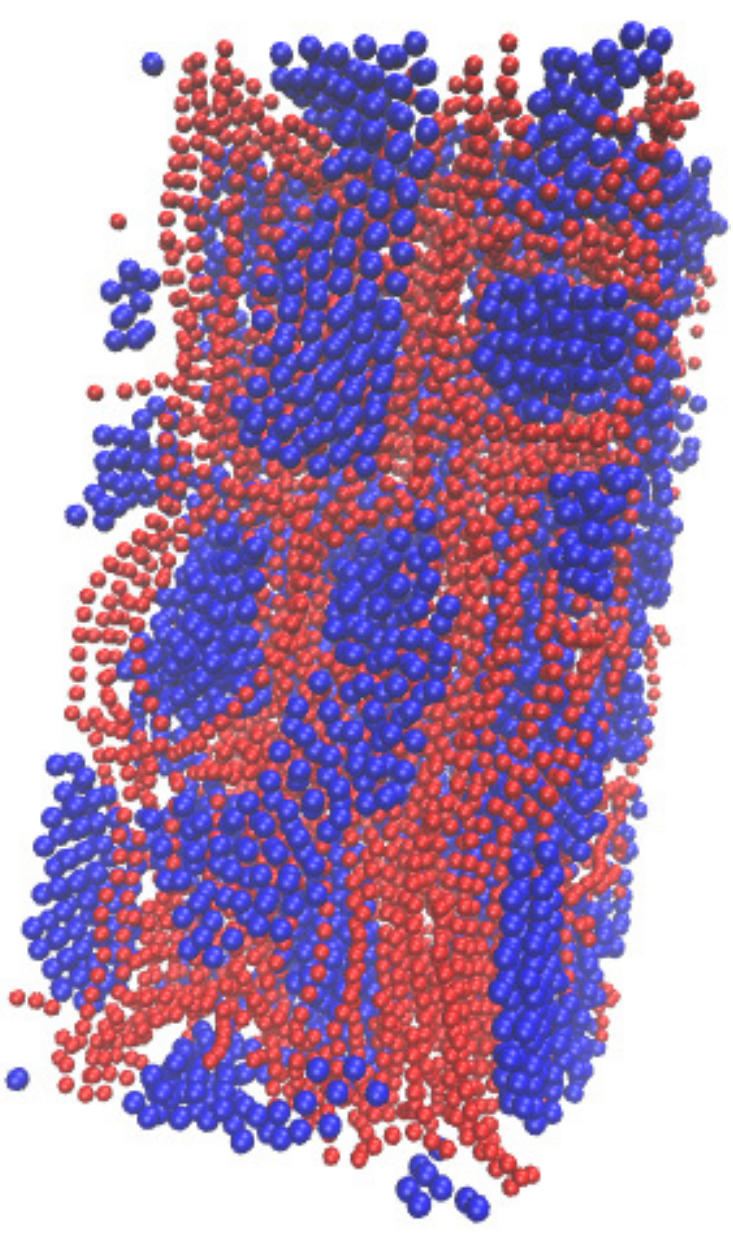}
\hspace{1cm}
\includegraphics[scale=0.3]{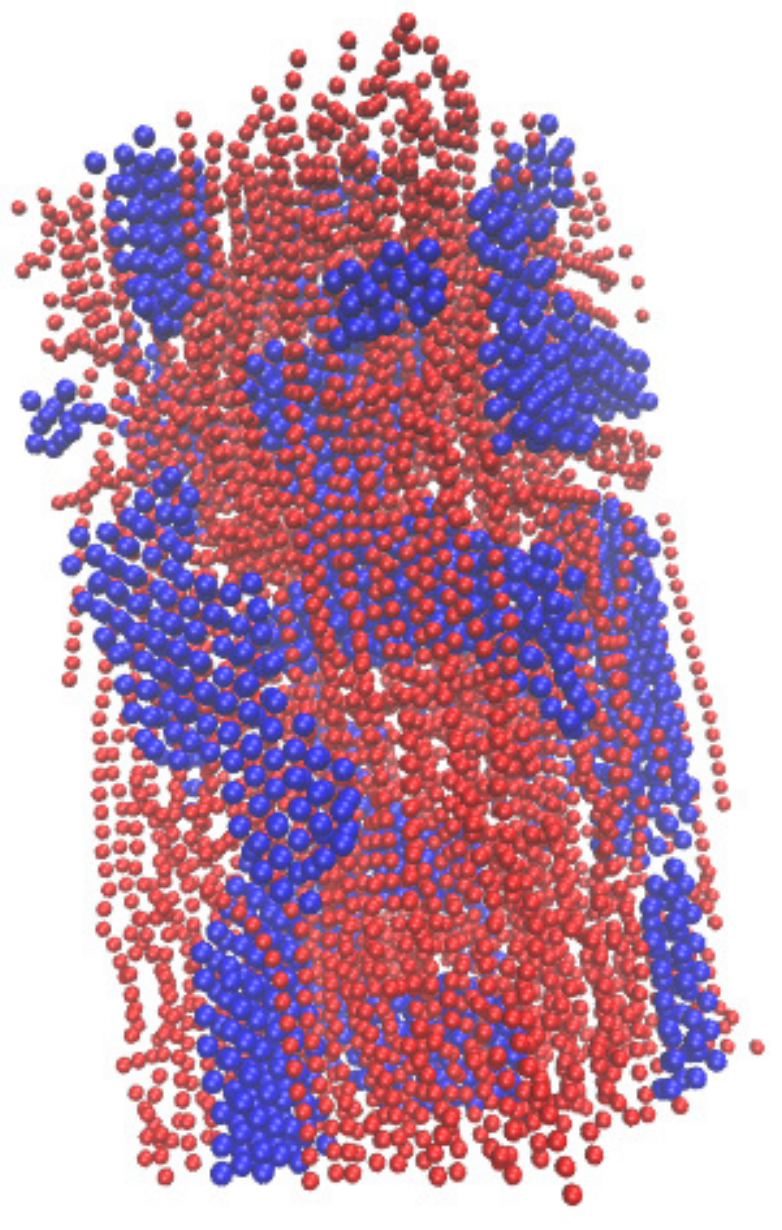}
\caption{ The figure shows four snapshots of NPs of size $\sigma_n=1.5\sigma$ for different values of $\sigma_{4n}$, (a) $1.25\sigma, (b) 2\sigma, (c) 2.5\sigma$ and (d) $3\sigma$. The corresponding systems of Wormlike micellar chains and NPs are shown in the same columns in the lower row. In the lower row, the monomers and NPs are indicated by red and blue particles, respectively. The snapshot for the lowest value of $\sigma_{4n}=1.25\sigma$ shows a uniformly mixed state of NPs. With the increase in $\sigma_{4n}$, the NPs form a percolating network with the network gradually breaking (as in (c)) and eventually forming non-percolating sheet-like structures of NPs as shown in the figure (d).}
\label{5000_150}
\end{figure*}

\begin{figure*}
\centering
\includegraphics[scale=0.2]{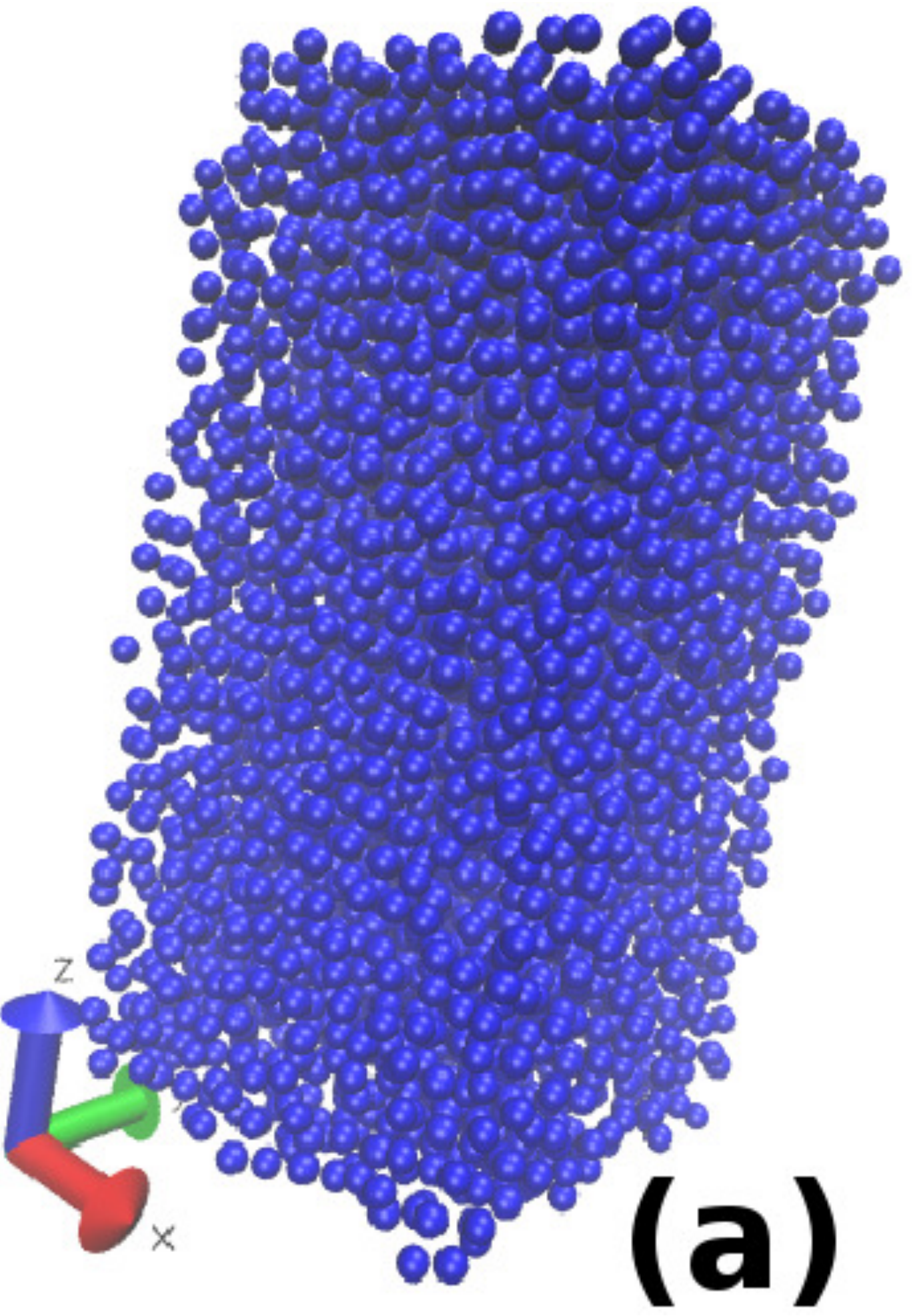}
\hspace{1cm}
\includegraphics[scale=0.2]{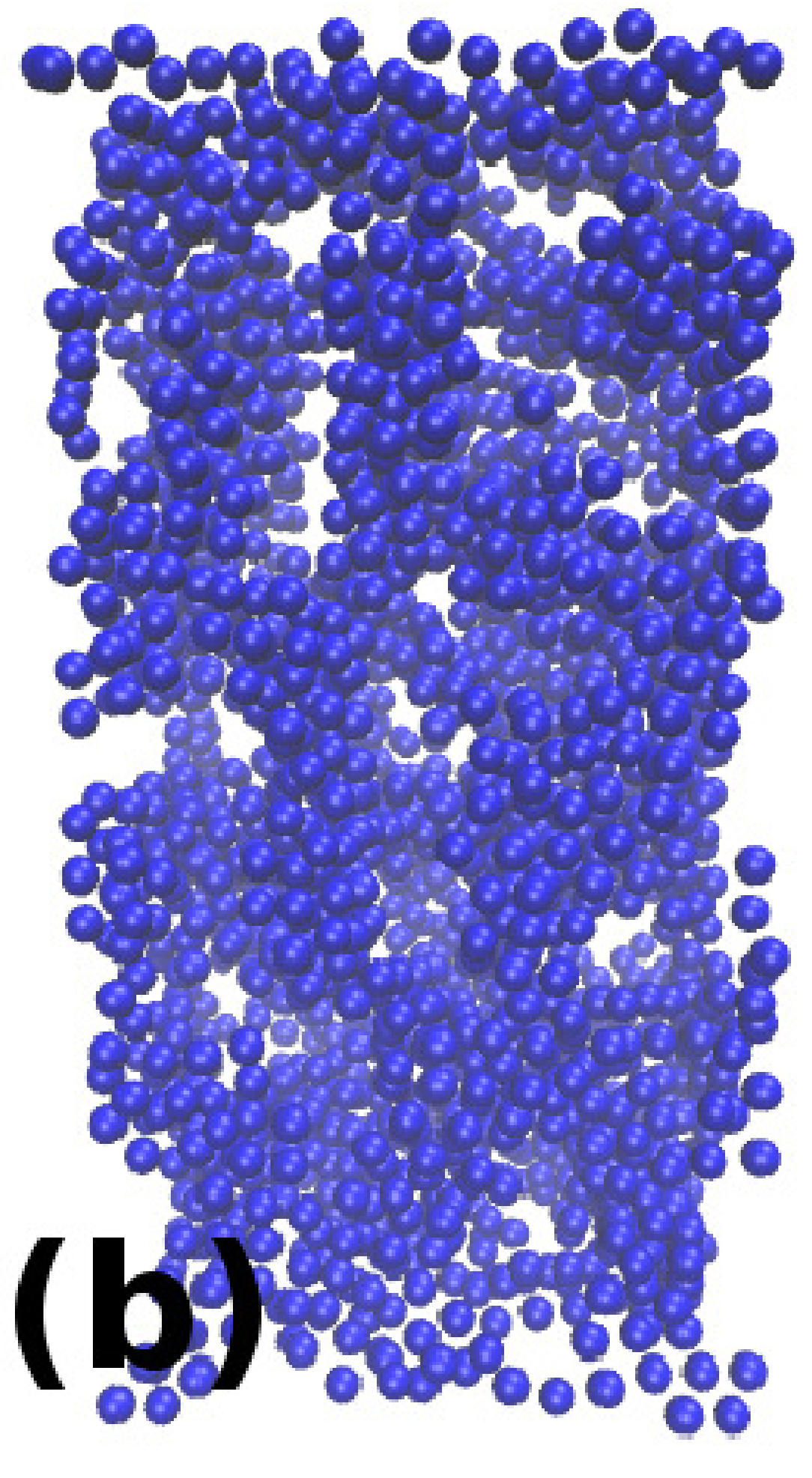}
\hspace{1cm}
\includegraphics[scale=0.2]{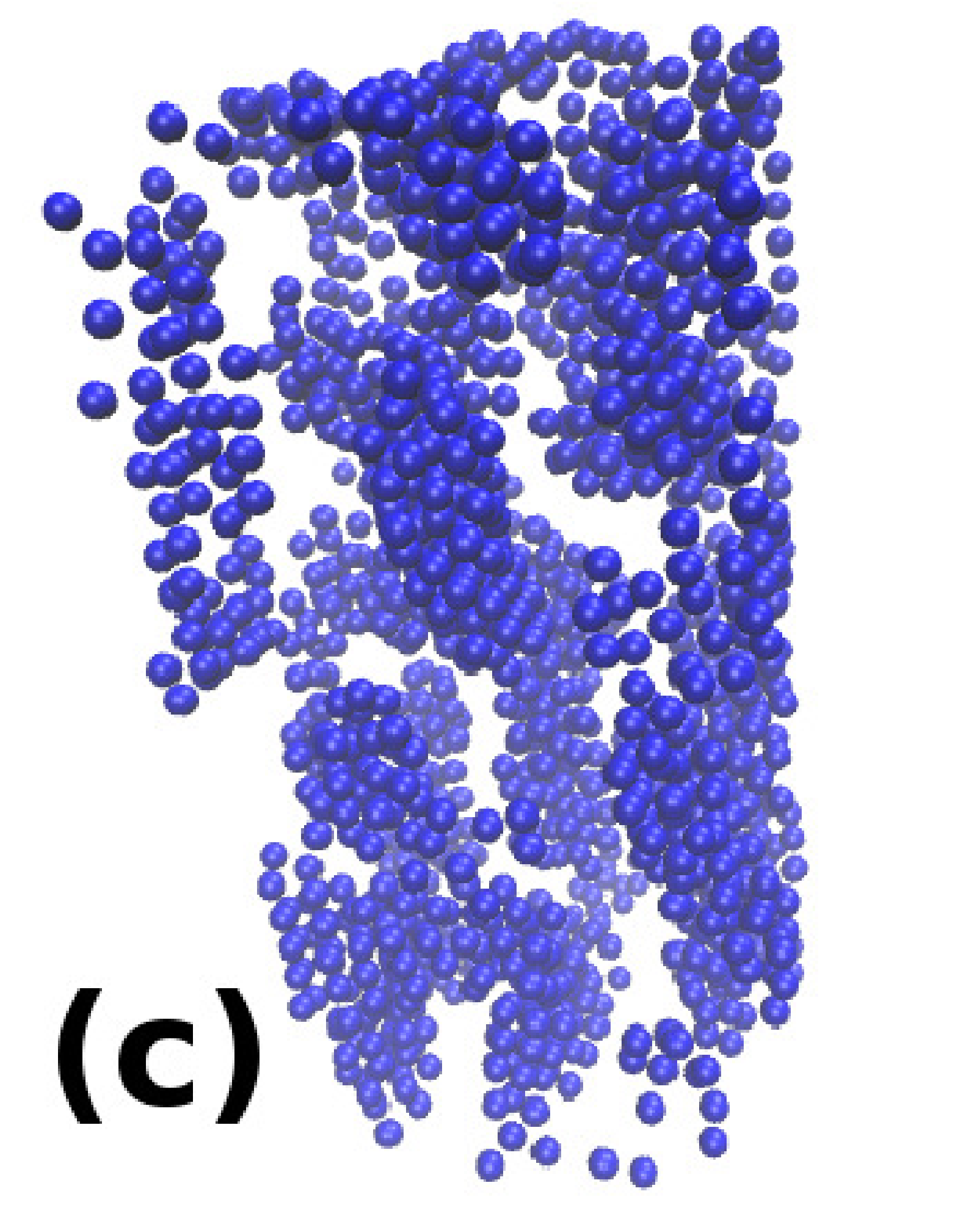}
\hspace{1cm}
\includegraphics[scale=0.2]{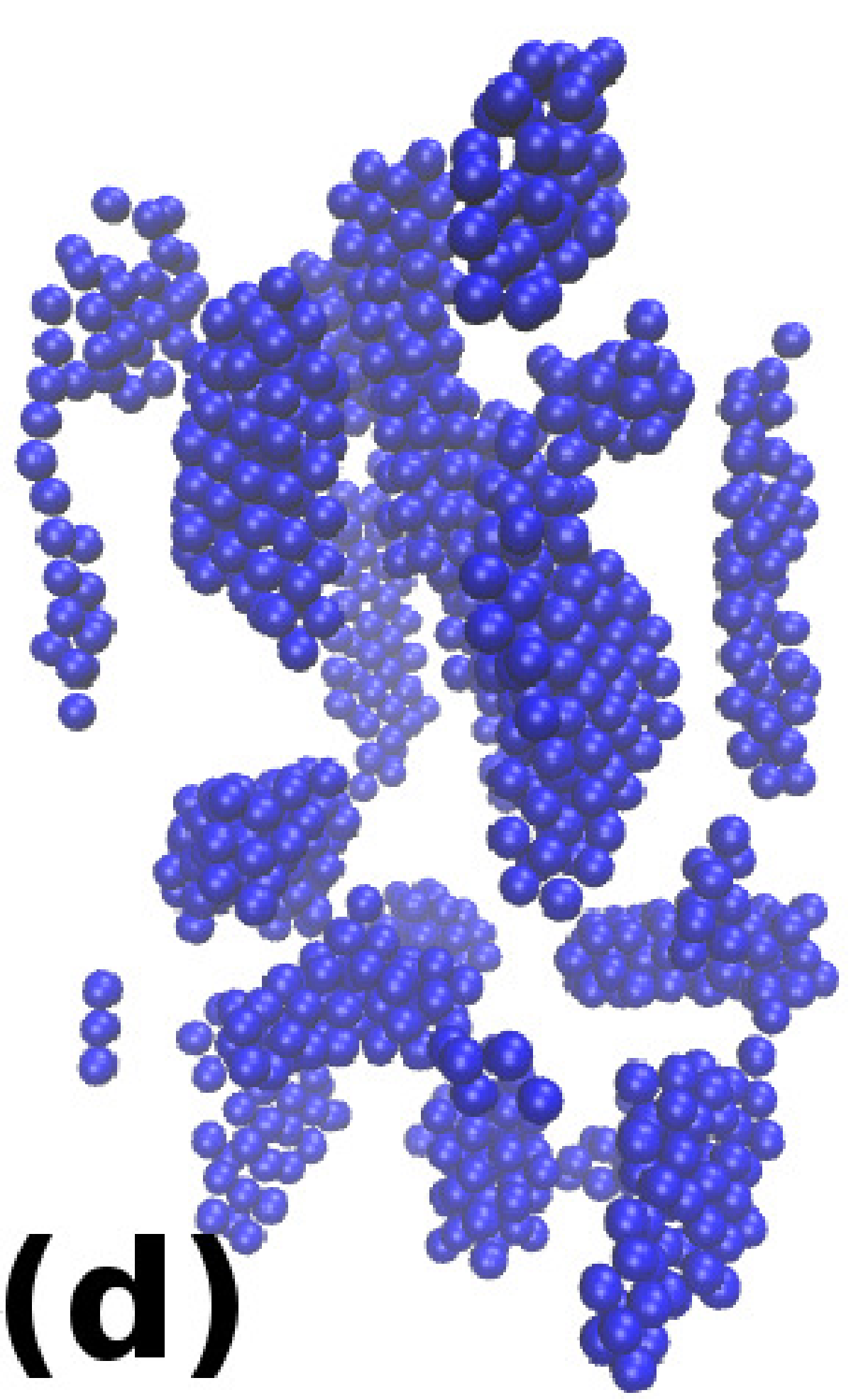} \\
\includegraphics[scale=0.3]{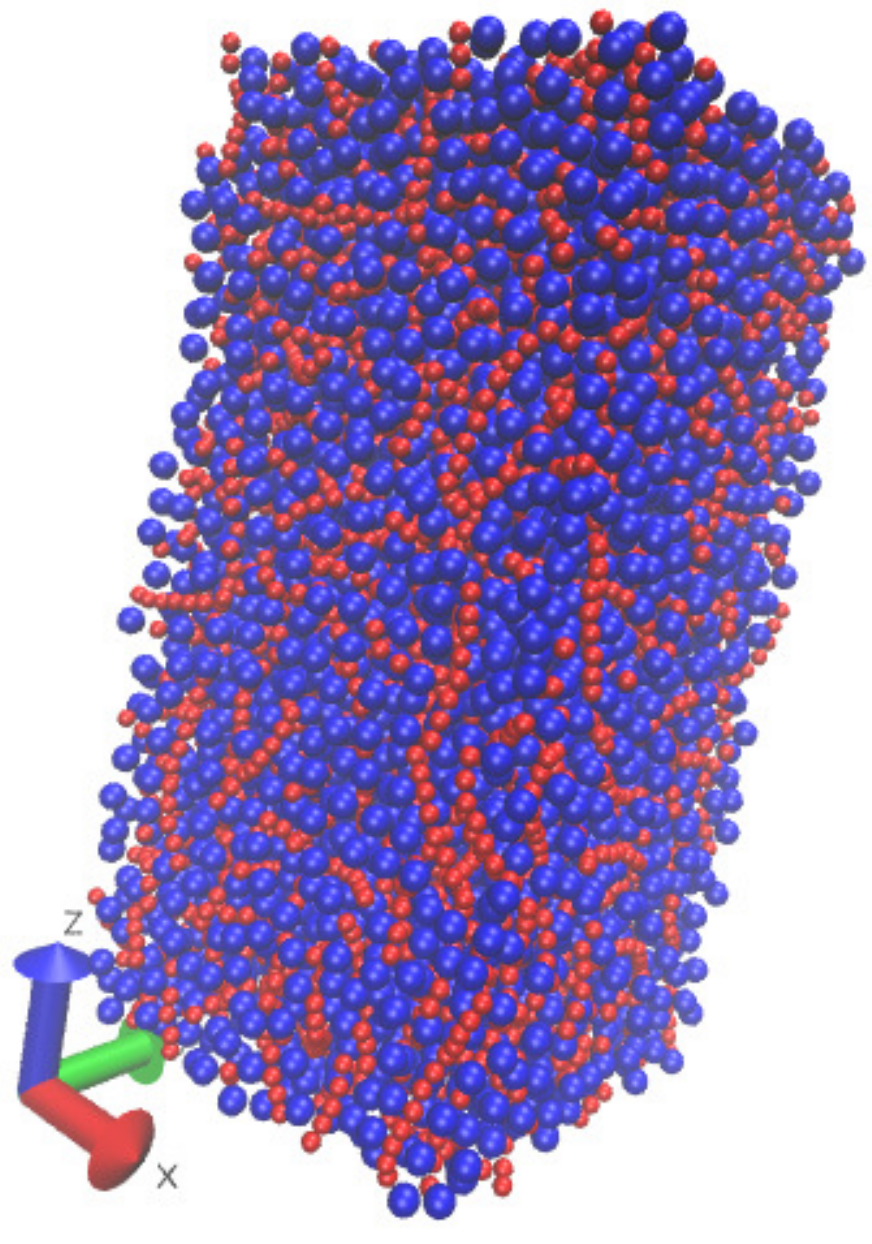}
\hspace{1cm}
\includegraphics[scale=0.3]{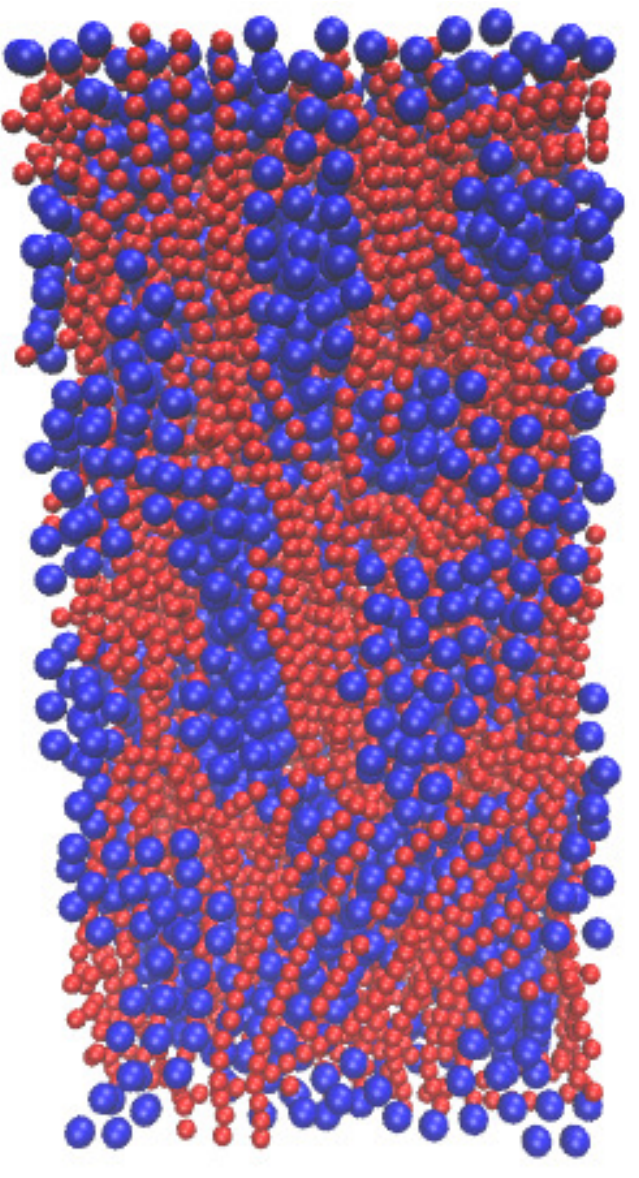}
\hspace{1cm}
\includegraphics[scale=0.3]{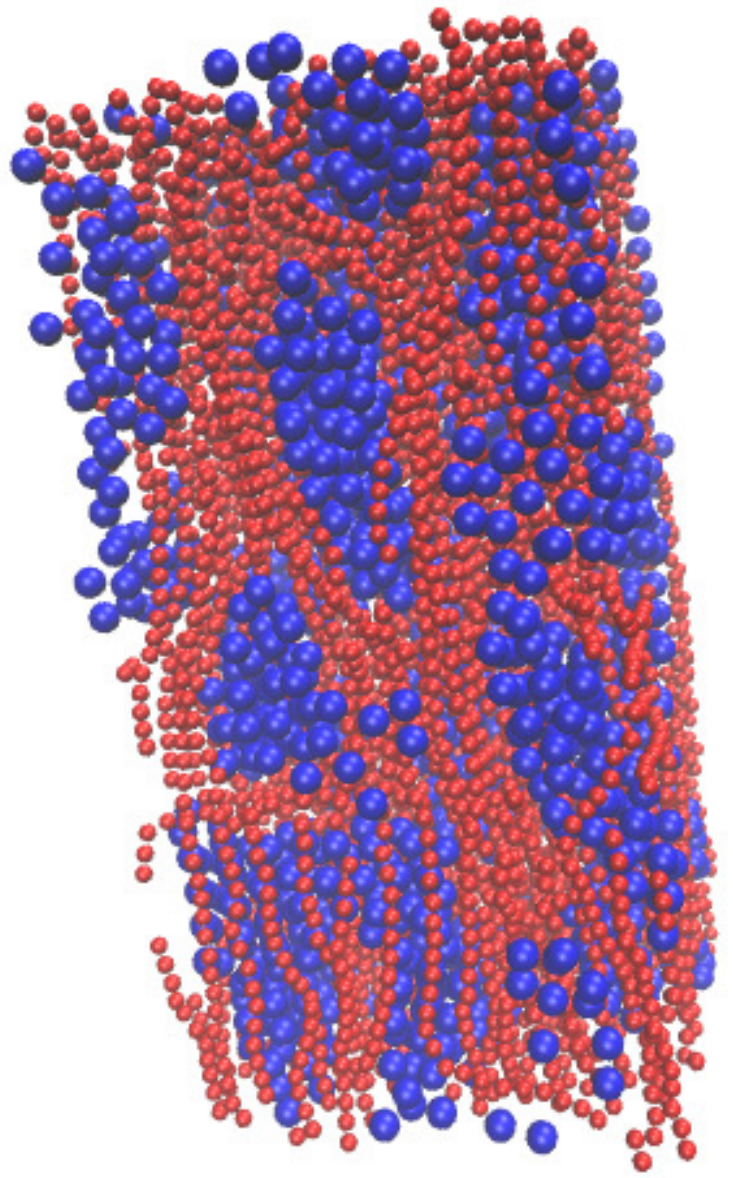}
\hspace{1cm}
\includegraphics[scale=0.3]{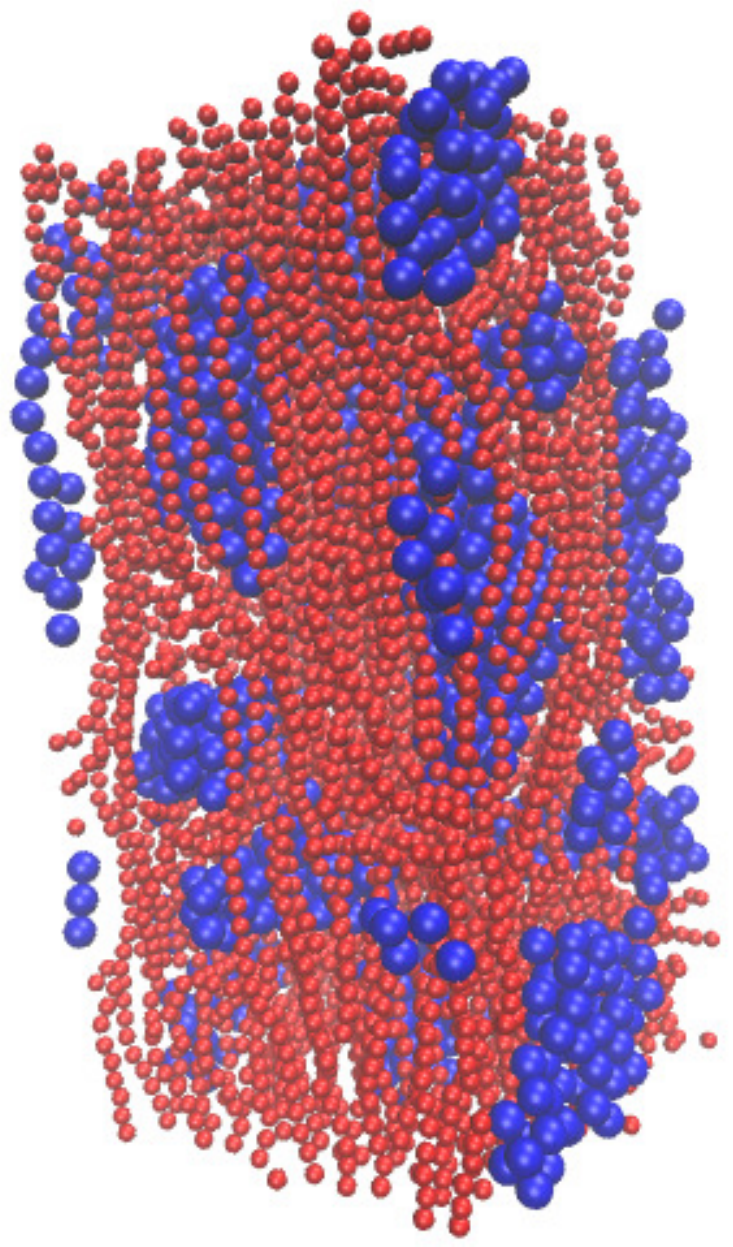}
\caption{The figure shows four snapshots of NPs of size $\sigma_n=2\sigma$ for different values of $\sigma_{4n}$, (a) $1.5\sigma, (b) 2\sigma, (c) 2.5\sigma$ and (d) $2.75\sigma$ respectively. The corresponding system for each NP snapshot is shown in the lower row, within the same column. The snapshots in the lower row show monomers and NPs indicated by red and blue colours, respectively. The snapshot for the lowest value of $\sigma_{4n}=1.5\sigma$ shows a uniformly mixed state of NPs and the polymeric chains. With an increase of $\sigma_{4n}$, the NPs form a percolating network as shown in (b). Further increase in $\sigma_{4n}$ form intermediate states where the NP networks gradually break as shown in (c). For $\sigma_{4n}=2.75\sigma$ the NP network transforms into non-percolating clusters as shown in (d). }
 
\label{5000_200}
\end{figure*}

\begin{figure*}
\centering
\includegraphics[scale=0.2]{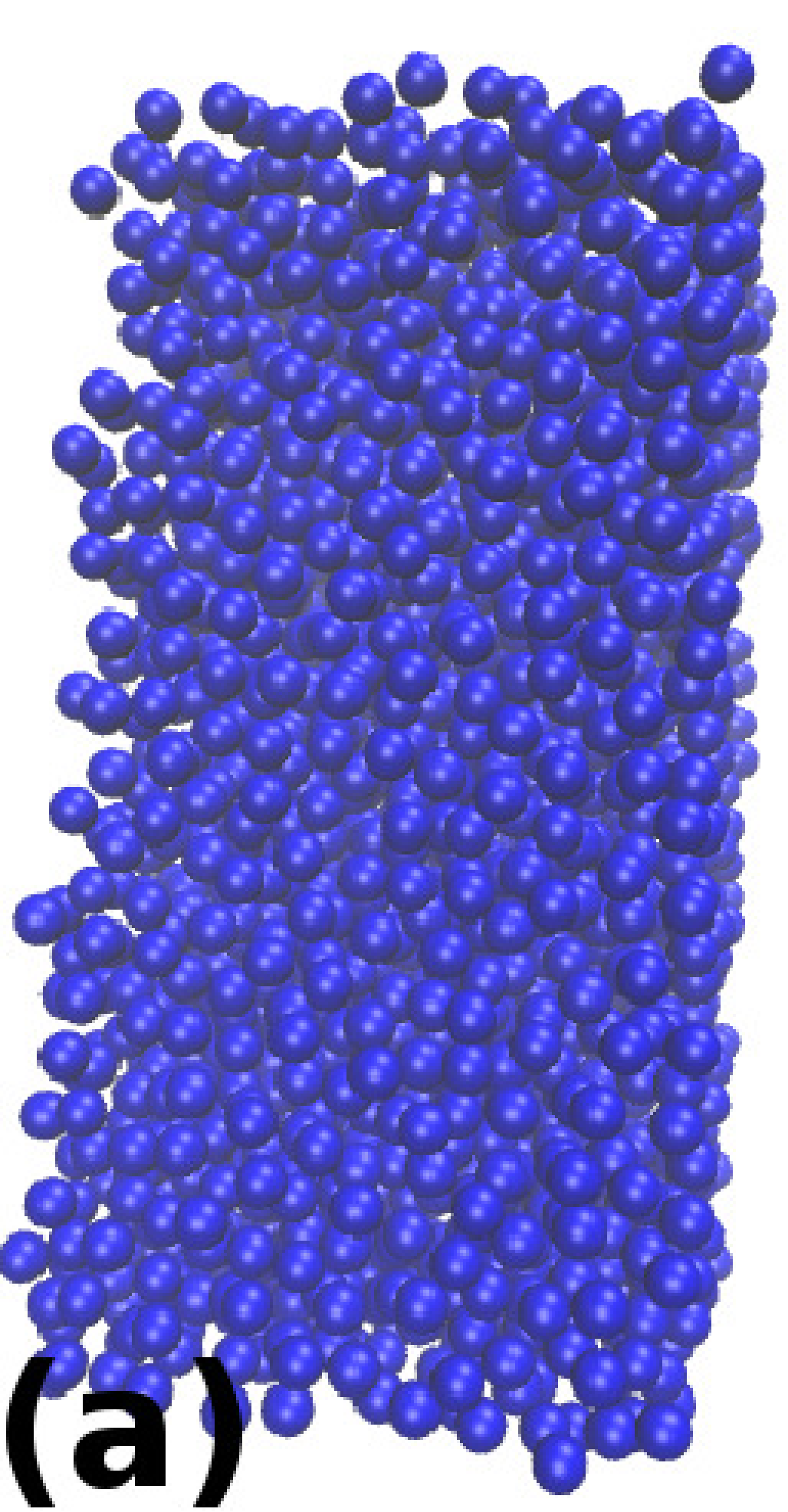}
\hspace{1cm}
\includegraphics[scale=0.2]{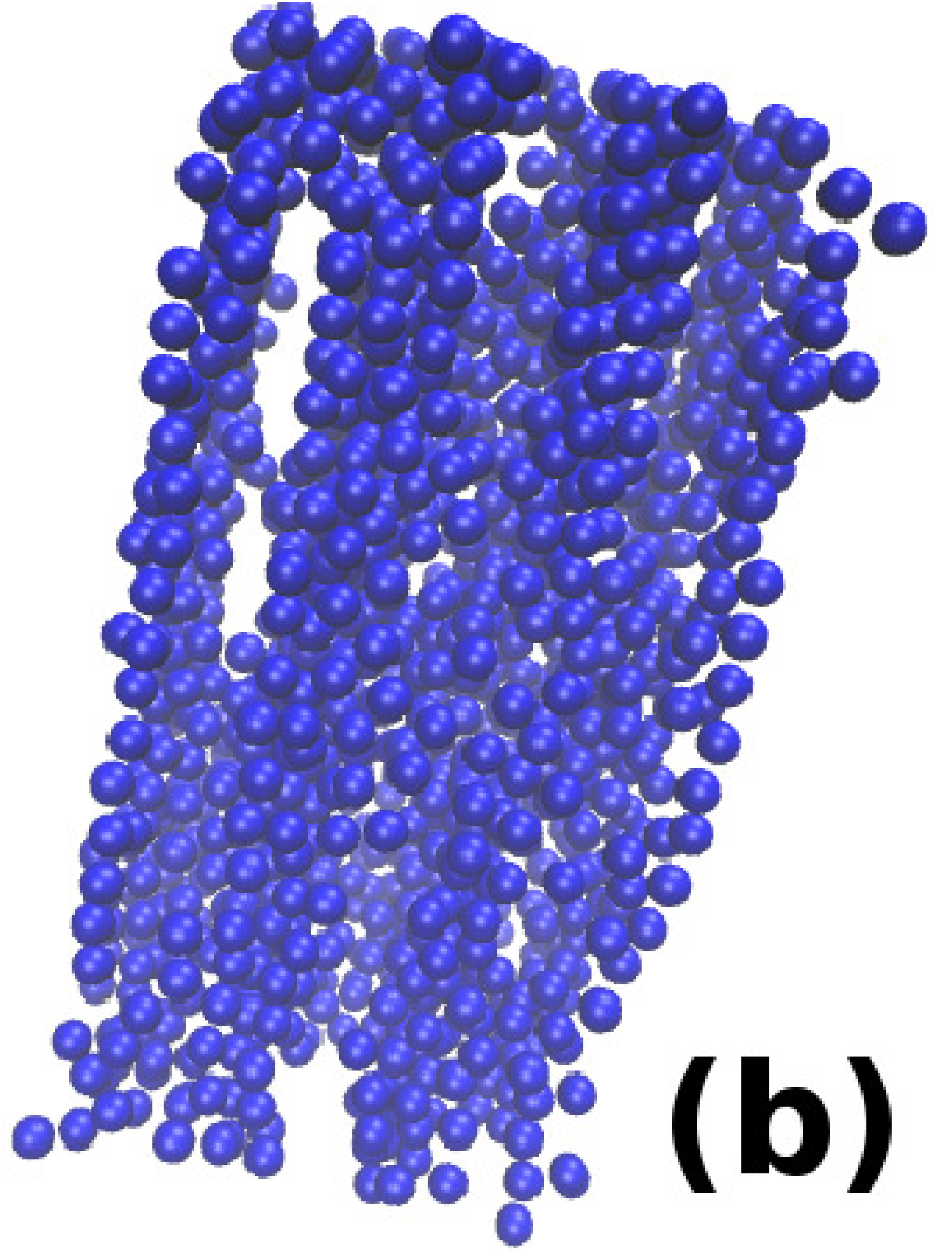}
\hspace{1cm}
\includegraphics[scale=0.2]{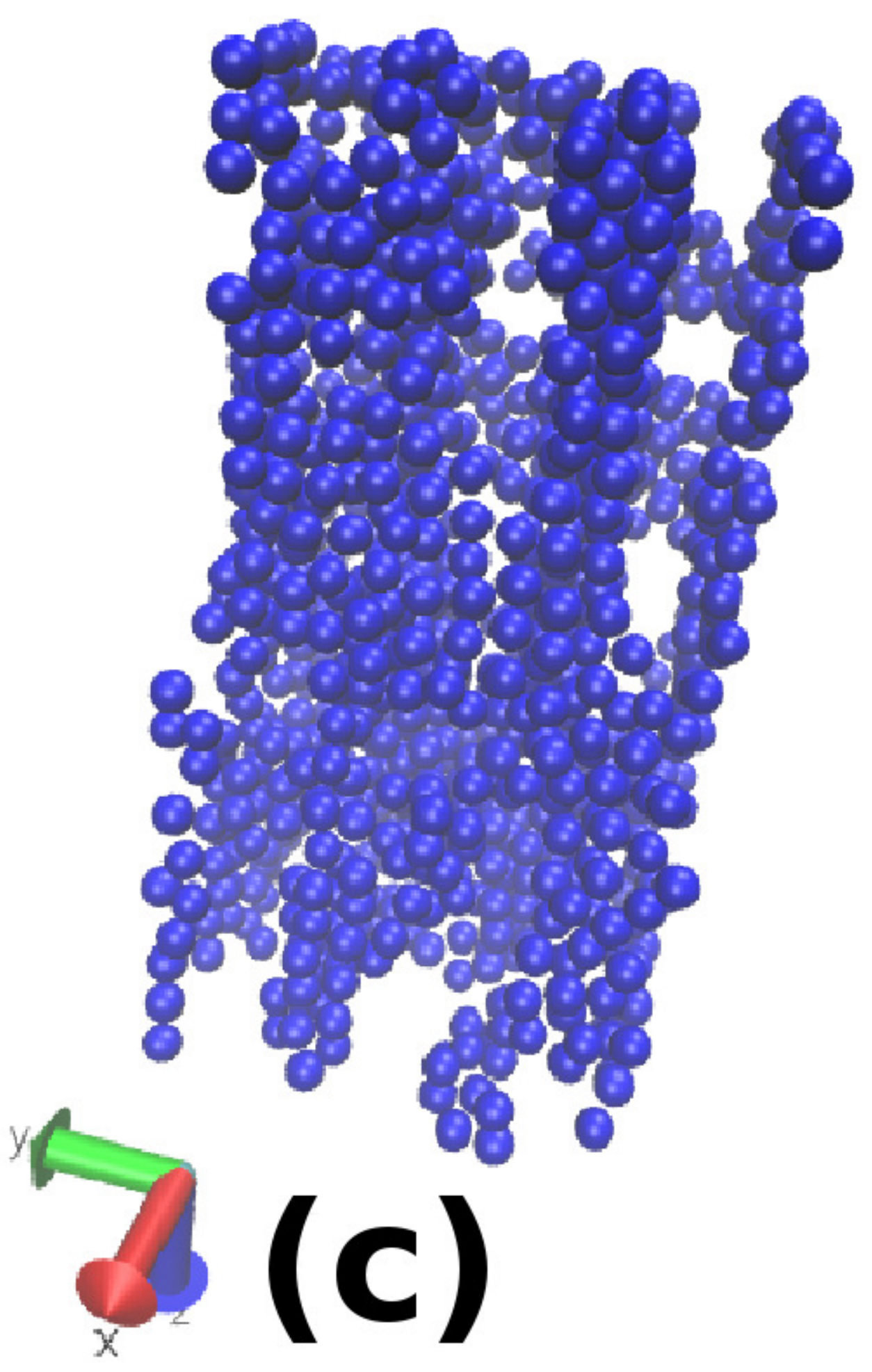}
\hspace{1cm}
\includegraphics[scale=0.2]{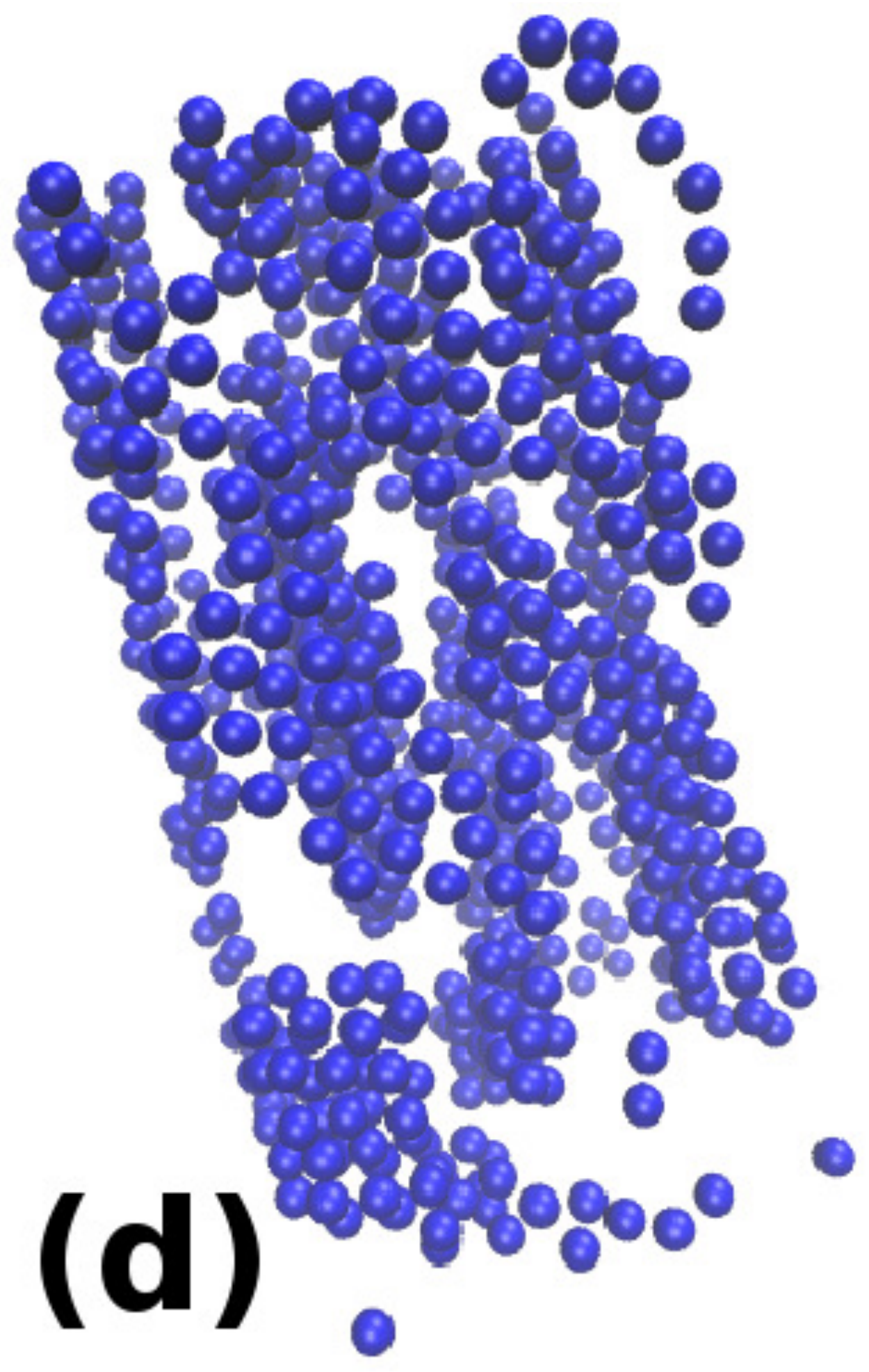}\\
\includegraphics[scale=0.3]{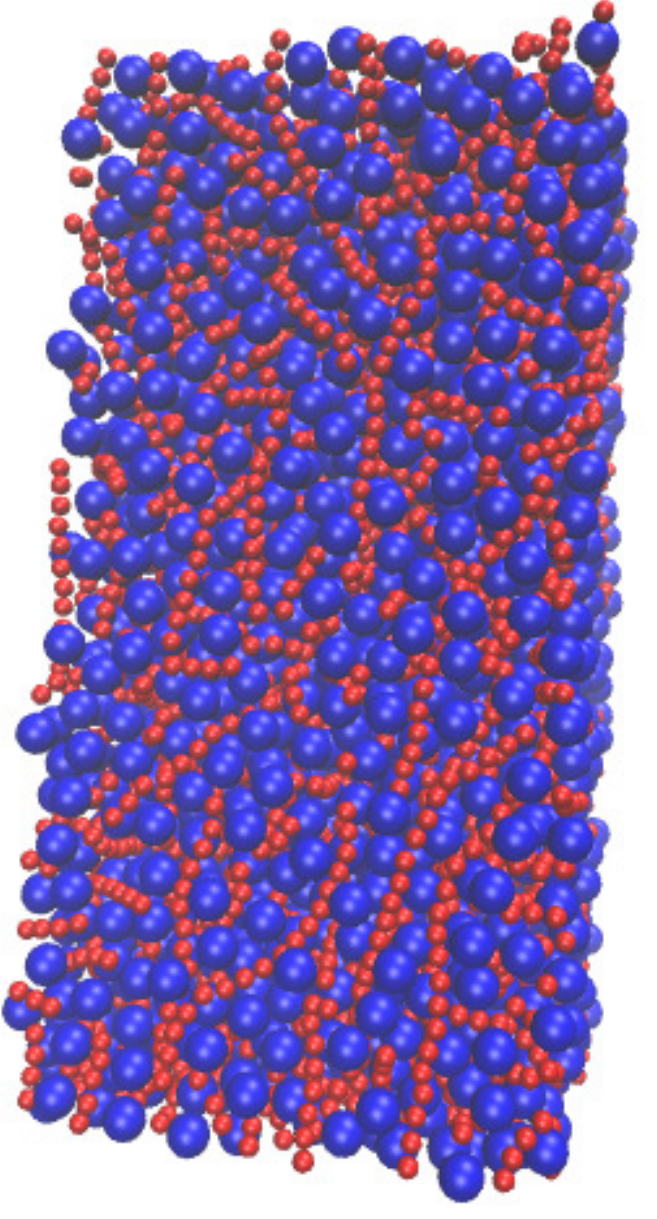}
\hspace{1cm}
\includegraphics[scale=0.3]{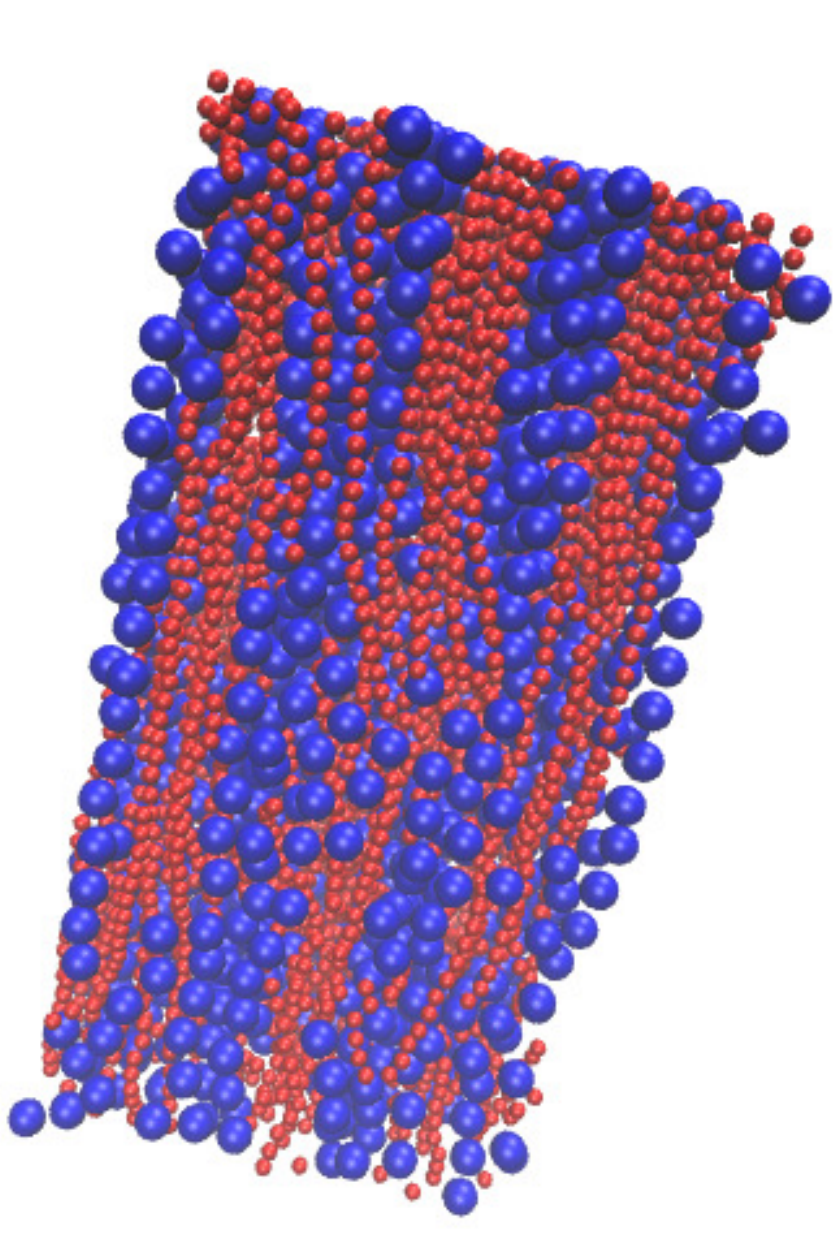}
\hspace{1cm}
\includegraphics[scale=0.3]{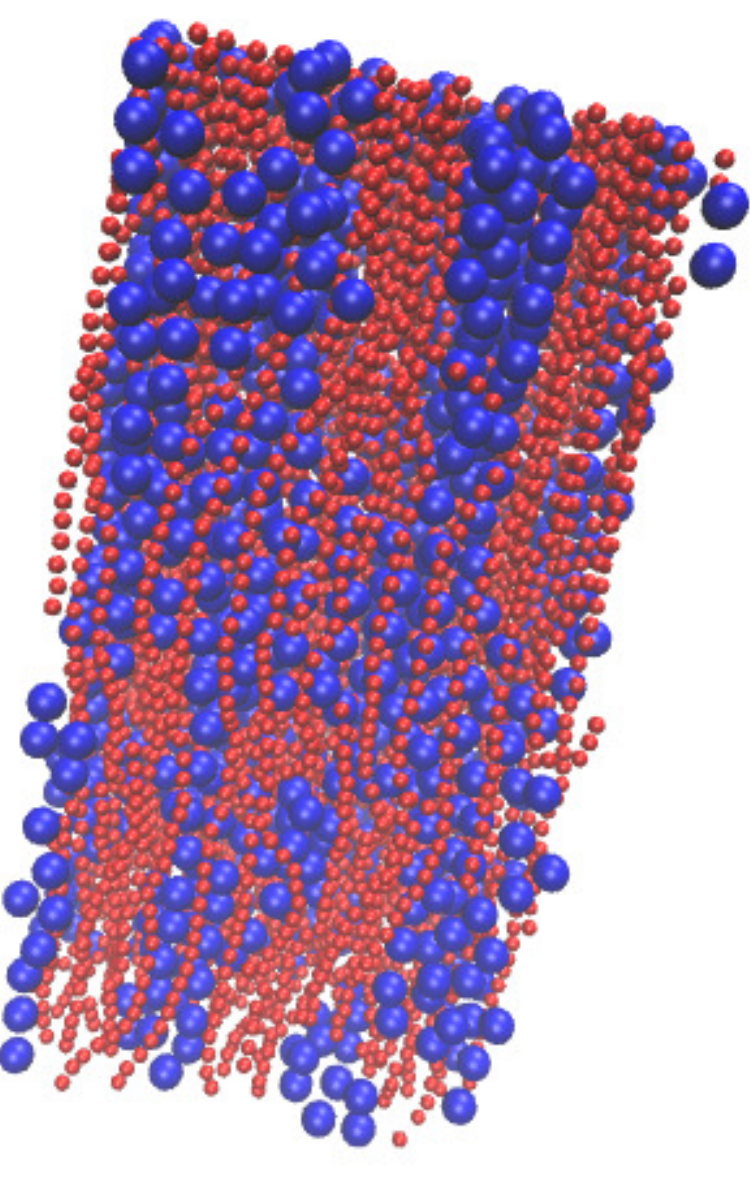}
\hspace{1cm}
\includegraphics[scale=0.3]{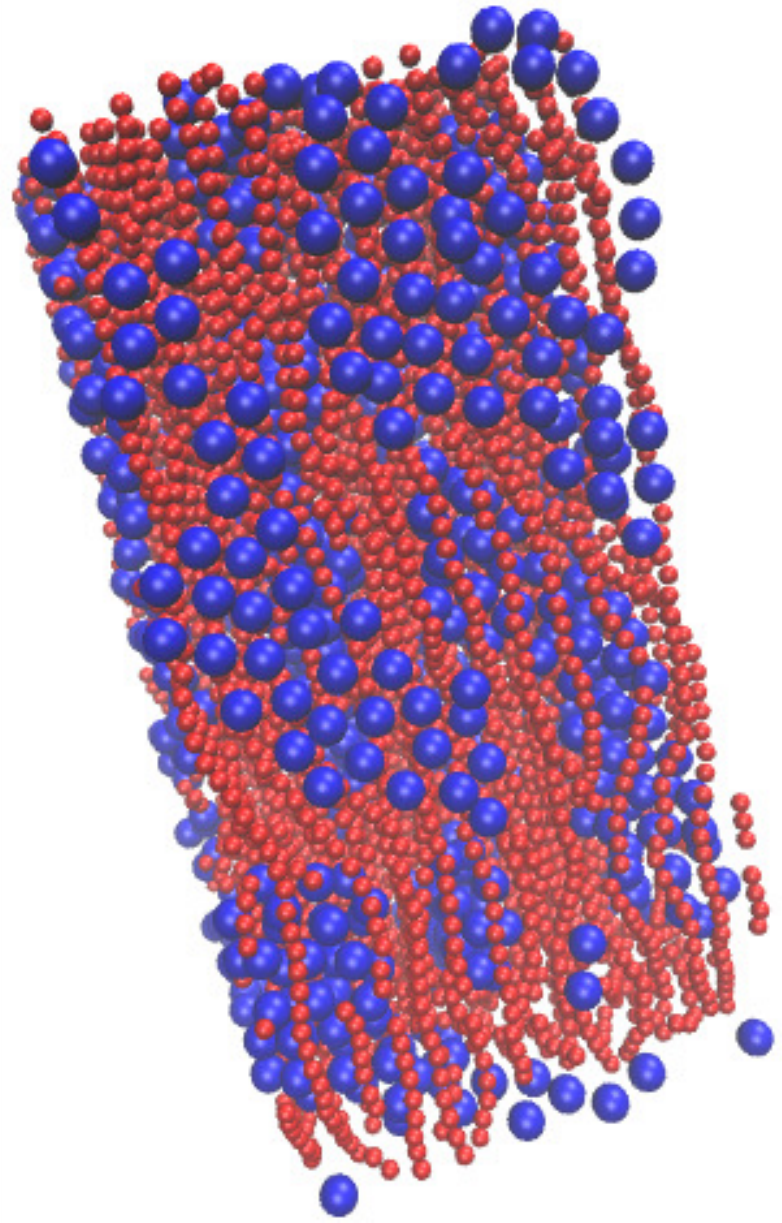} 
\caption{The figure shows four snapshots of NPs of size $\sigma_n=2.5\sigma$ for different values of $\sigma_{4n}$,  (a) $1.75\sigma, (b) 2\sigma, (c) 2.25\sigma$ and (d) $2.5\sigma$ respectively, in the upper row. The lower row shows the corresponding NP-monomer system with NPs and monomers indicated by red and blue colours, respectively. Figure (a) shows a uniformly dispersed mixed state of NPs and monomer chains for the lowest value of $\sigma_{4n}=1.75$. With the increase in $\sigma_{4n}$, the NPs form a percolating network as shown in (b). Further increase in $\sigma_{4n}$ leads to the NP network gradually breaking to form intermediate structures as in (c) and finally breaking into individual clusters represented by the figure (d).    }
\label{5000_250}
\end{figure*}

\begin{figure*}
\centering
\includegraphics[scale=0.2]{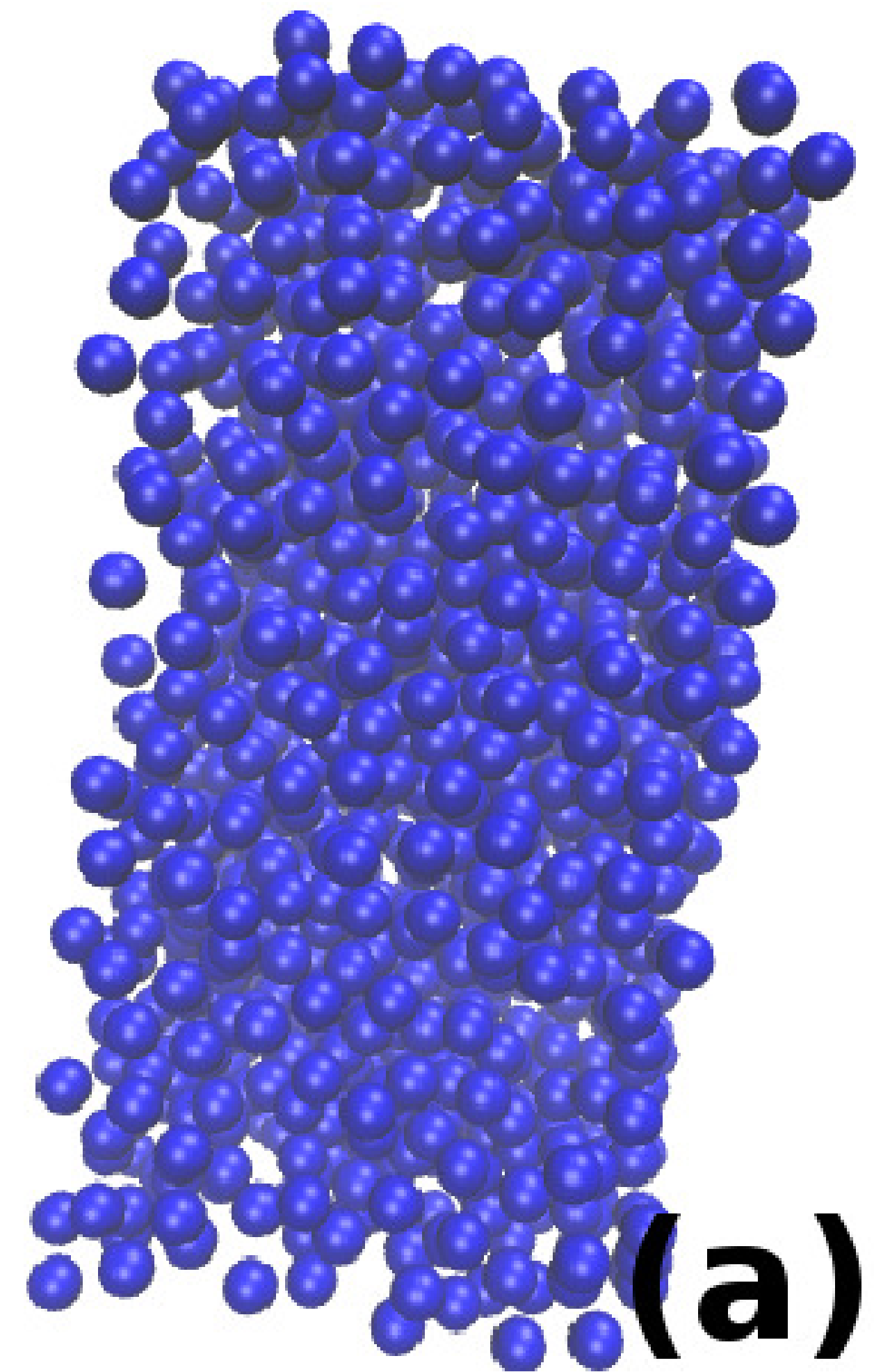}
\hspace{2cm}
\includegraphics[scale=0.2]{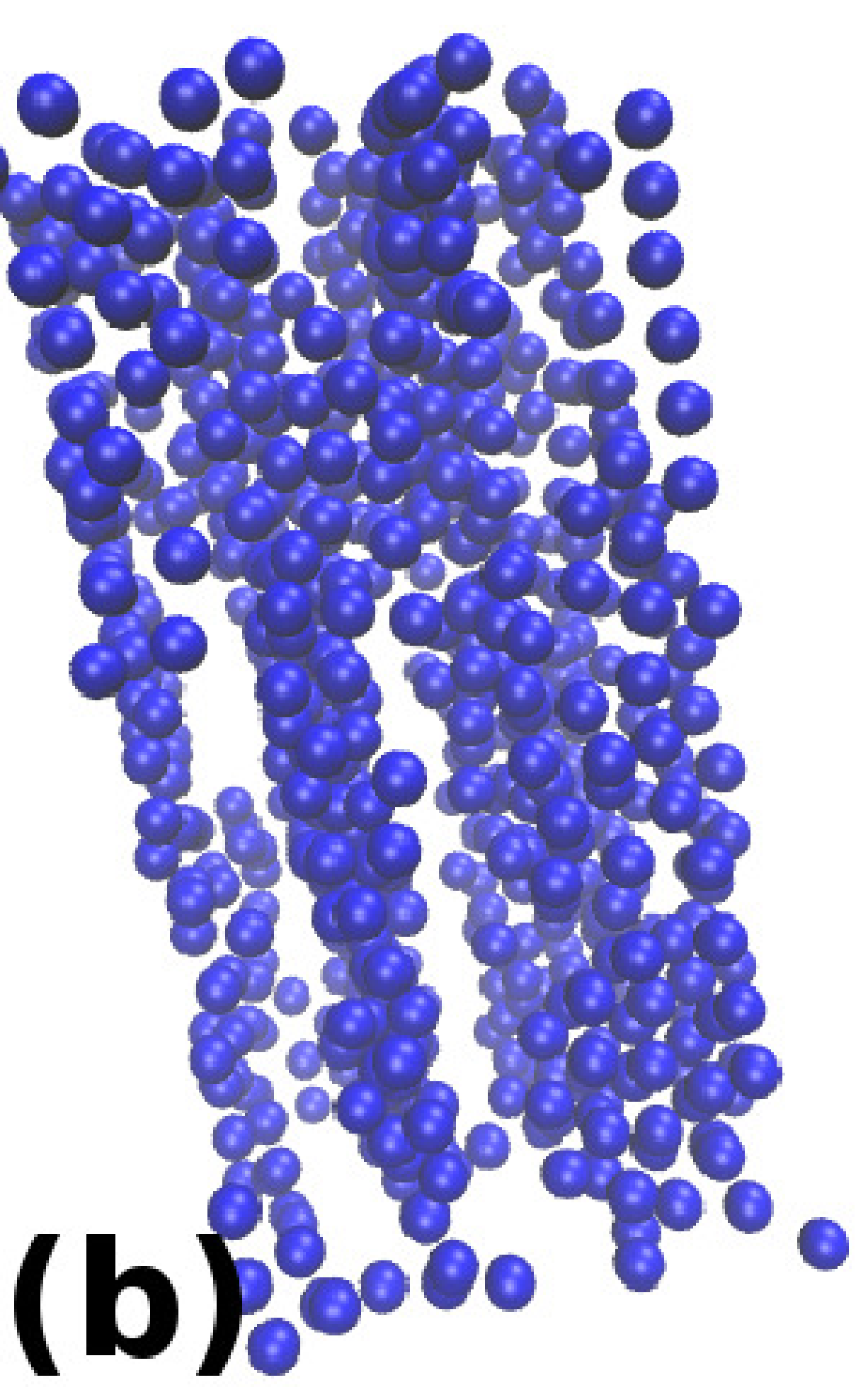}
\hspace{2cm}
\includegraphics[scale=0.2]{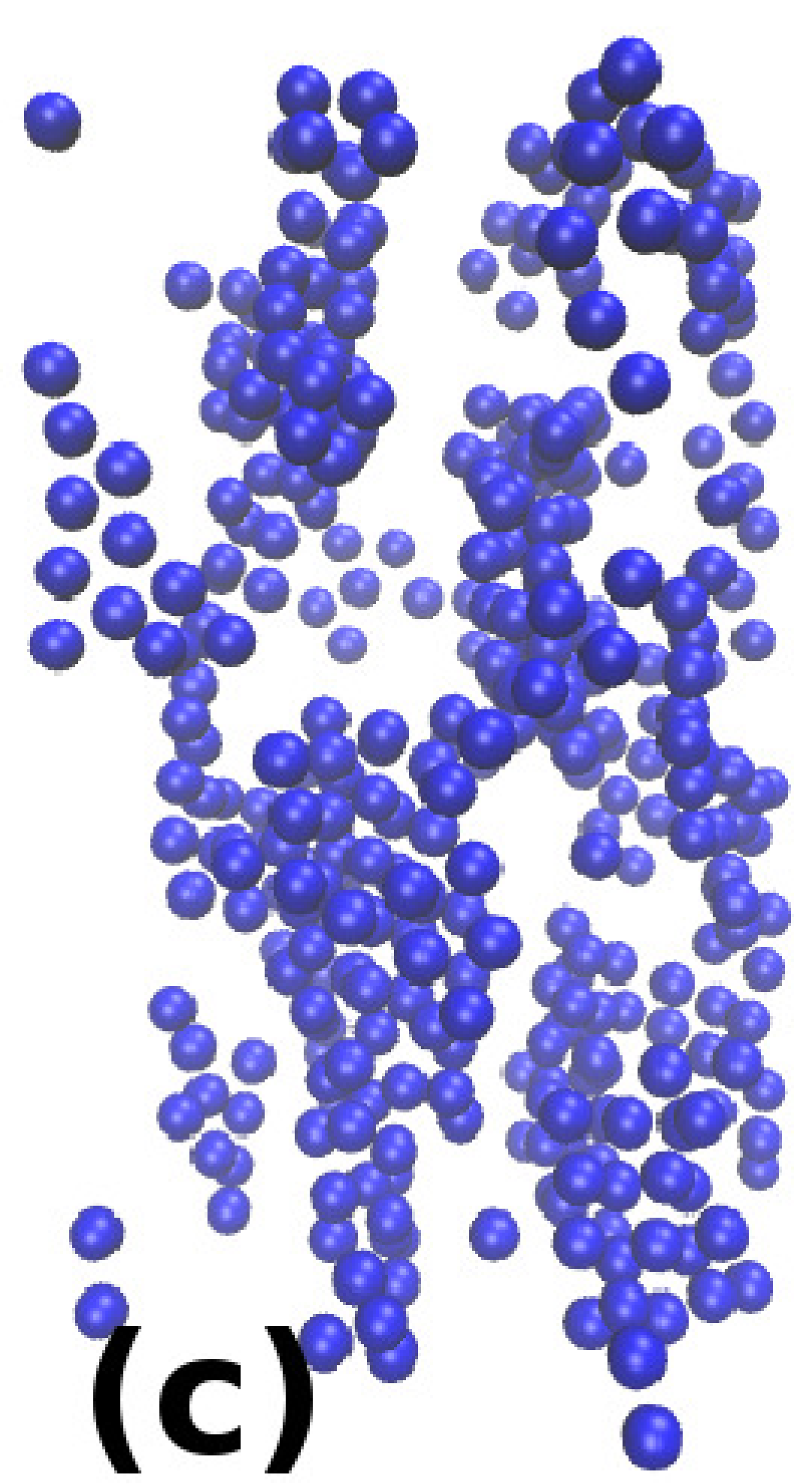} \\
\includegraphics[scale=0.3]{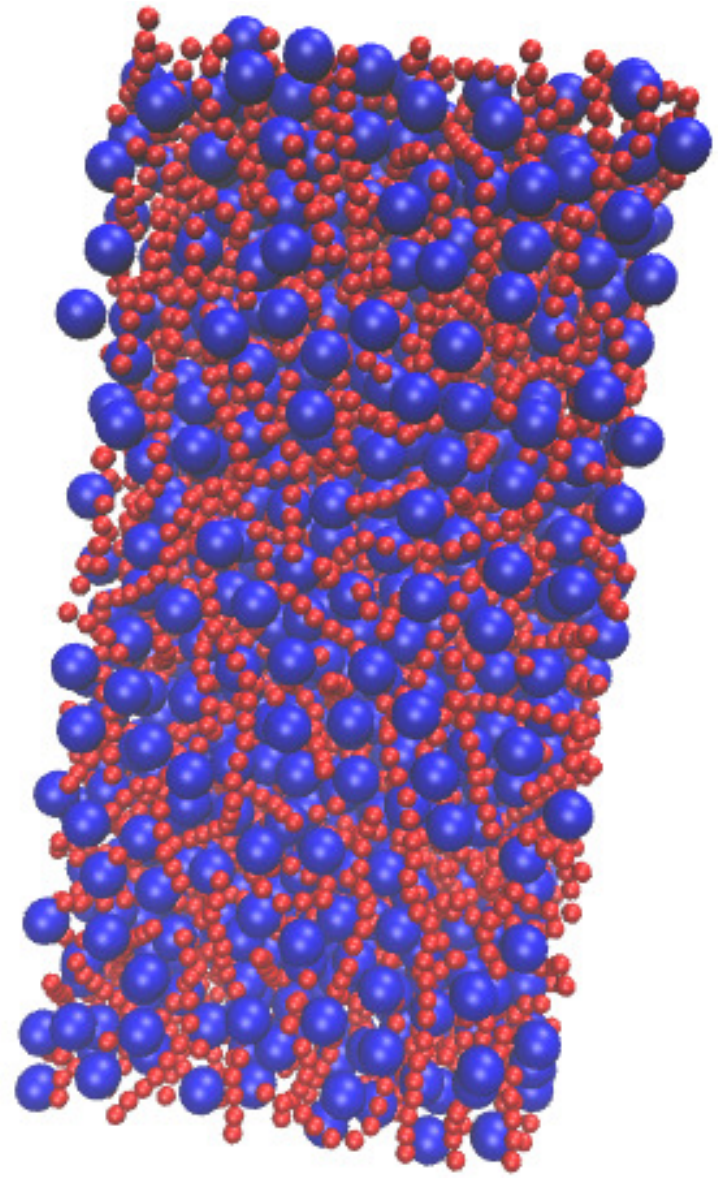}
\hspace{2cm}
\includegraphics[scale=0.3]{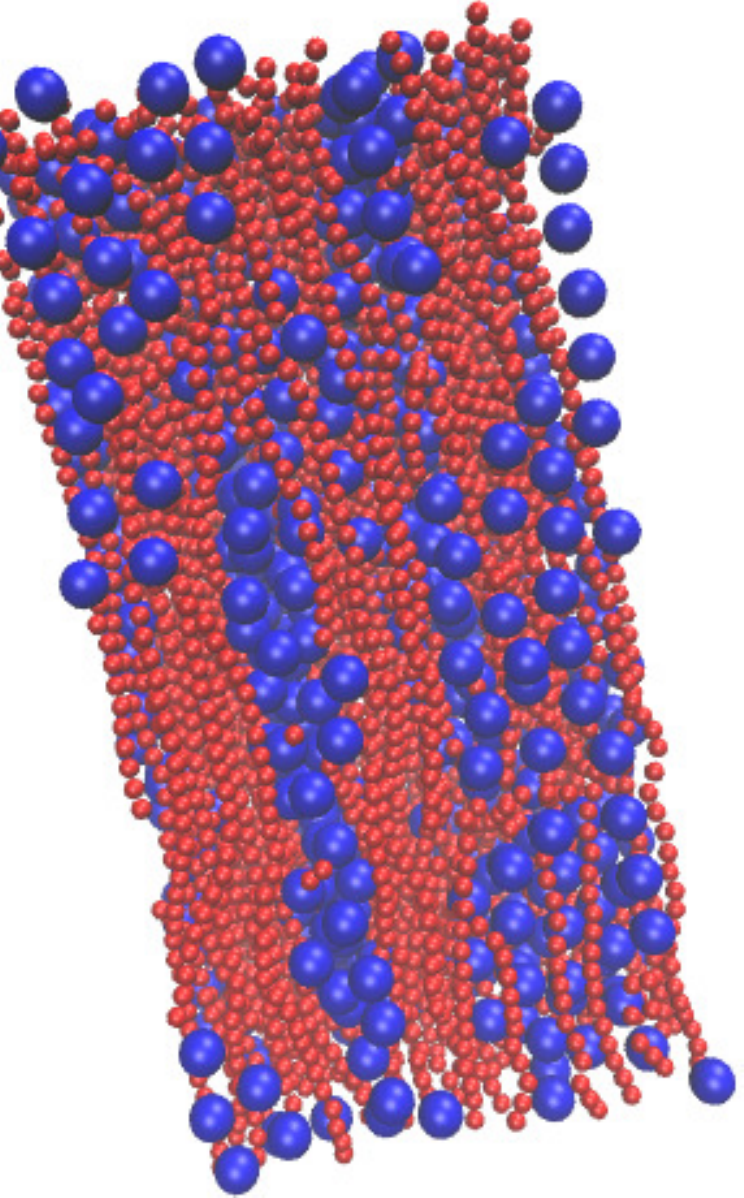}
\hspace{2cm}
\includegraphics[scale=0.3]{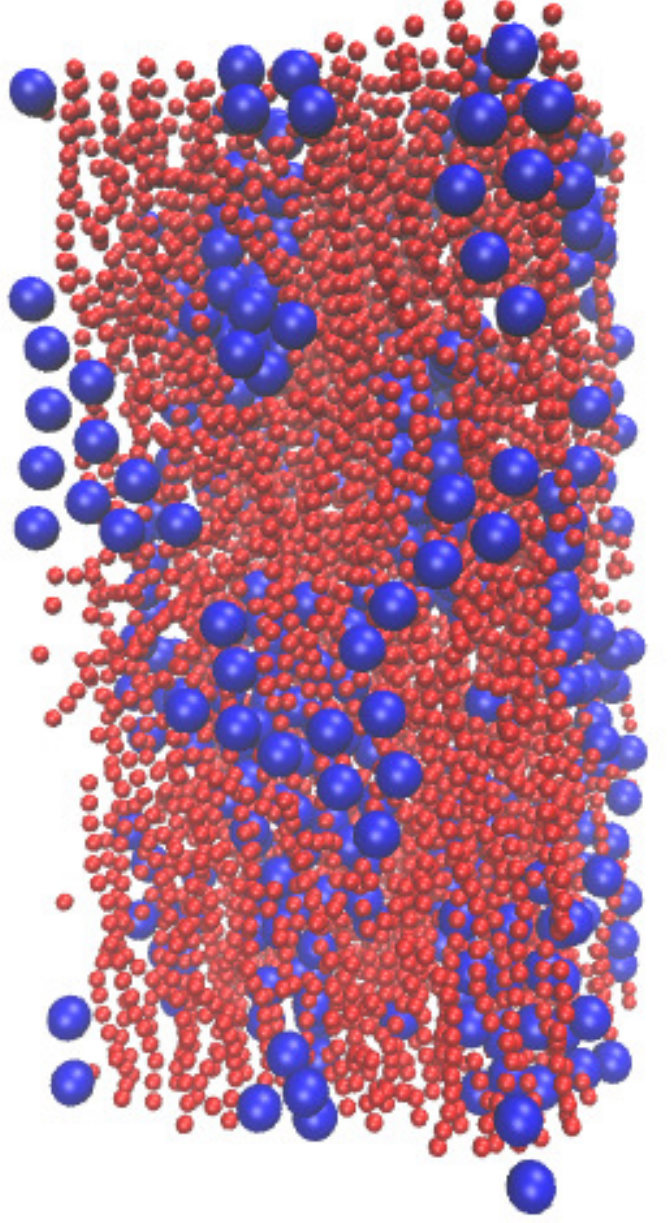}
\caption{The figure shows three snapshots of NPs of size $\sigma_n=3\sigma$ for different values of $\sigma_{4n}$, (a) $2\sigma, (b) 2.25\sigma$ and  (c) $2.5\sigma$, respectively, in the upper row. The corresponding systems of NP and monomers are shown in the lower row within the same columns. The increase in the value of $\sigma_{4n}$ first changes a uniformly mixed system of NPs and micellar chains (in (a)) to a percolating network of NPs (in (b)). With further increase in $\sigma_{4n}$, the NPs are not well packed and the structure is not well defined.}
\label{5000_300}
\end{figure*}

      For the size of NP $\sigma_n=1.5\sigma$, the structural changes in NPs from percolating network to non-percolating clusters as a result of an increase in $\sigma_{4n}$ is already reported in a previous communication for all the micellar densities considered here. The snapshots for the micellar number density $\rho_m=0.093\sigma^{-3}$ is reproduced here and shown in Fig.\ref{5000_150} for $\sigma_n=1.5\sigma$. The figure shows four snapshots for different values of $\sigma_{4n}=$  $(a)1.25\sigma$, $(b)2\sigma$, $(d)2.5\sigma$ and (d)$3\sigma$. The upper row shows only NPs from the WLM-NP system that is shown in the lower row, within the same column.  For a given value of NP size $\sigma_n$ and monomer size $\sigma$, there exists a lower bound to the value of $\sigma_{4n}$ due to the non-penetrable surfaces of the particles. For the value of $\sigma_n=1.5\sigma$, the lowest possible value of $\sigma_{4n}$ is calculated to be $\sigma_{4n}=\sigma/2+\sigma_n/2=1.25\sigma$. For the lowest value of $\sigma_{4n}=1.25\sigma$, the snapshot in Fig.\ref{5000_150}(a) shows a uniformly mixed state of micelles and NPs. Increasing the value of $\sigma_{4n}$ from $1.25\sigma$ to $1.5\sigma$ transforms the system structure from uniformly mixed state to a percolating network-like structure of NPs and micellar chains, as shown in Fig.\ref{5000_150}(b). With further increase in $\sigma_{4n}$, the NP network is observed to gradually break and produce non-percolating clusters for some high value of $\sigma_{4n}$. This can be observed in Fig.\ref{5000_150}(c) and (d), where NP network is shown to be breaking (in (c)) and finally producing non-percolating clusters (in (d)). It is shown that the shape anisotropy of the NP clusters is governed by the density of micellar matrix. For the density of micelles considered here $\rho_m=0.093\sigma^{-3}$, the NP clusters are found to have sheet-like structures as shown in Fig.\ref{5000_150}(d). 

       Now, to check whether the same kind of behaviour is shown by the system with larger NP size, the same set of runs are produced for different values of $\sigma_n$. The corresponding representative snapshots, after the systems show stable structures, are shown in Fig.\ref{5000_200}, \ref{5000_250} and \ref{5000_300} for values of $\sigma_n=2\sigma$, $2.5\sigma$ and $3\sigma$ respectively. The upper row shows only NPs from the corresponding WLM-NP systems which are shown in the same columns in lower row. In lower row, the snapshots show monomers in red and NPs in blue. In each figure, the value of $\sigma_{4n}$ increases from (a) to (d). The left most figures Fig.\ref{5000_200}(a), \ref{5000_250}(a) and \ref{5000_300}(a) are for the lowest possible value of $\sigma_{4n}$ calculated as $\sigma_{4n}=\sigma/2+\sigma_n/2$. As observed in Fig.\ref{5000_150}(a), (and also established for all the micellar densities considered \cite{arxive}) for all the values of $\sigma_n$, the snapshots for the lowest possible values of $\sigma_{4n}$  show a uniformly mixed state of micellar chains and NPs. From the figures \ref{5000_200}(a), \ref{5000_250}(a) and \ref{5000_300}(a), it can be seen that NPs do not form clusters. It is shown that micellar chains in this state are dispersed to form a uniformly mixed state \cite{arxive}. With an increase in the value of $\sigma_{4n}$, the NPs show formation of percolating network of clusters which breaks with further increase in the value of $\sigma_{4n}$. The righmost snapshots in all the figures (Fig.\ref{5000_150}, \ref{5000_200}(d) and \ref{5000_250}(d)) is for a higher value of $\sigma_{4n}$ for which the NP networks break into non-percolating clusters. All the snapshots in figures \ref{5000_150}, \ref{5000_200}, \ref{5000_250} and \ref{5000_300} that the WLM-NP system shows morphological changes from uniformly mixed state to percolating network-like structures with increase in $\sigma_{4n}$ which again transforms to for individual clusters of NPs at a higher value of $\sigma_{4n}$, for all the values of NP size $\sigma_n$ considered here. However, for the value of $\sigma_n=3\sigma$, it becomes difficult to introduce NPs inside the polymeric matrix due to the elastic cost. Therefore, for $\sigma_{n}=3\sigma$, the NPs form a network-like structure at $\sigma_{4n}=1.5\sigma$, but the breaking of this network into individual clusters is not observed due to difficulty in introducing NPs inside the matrix beyond $\sigma_n=2.5\sigma$.

\begin{figure}
\includegraphics[scale=0.3]{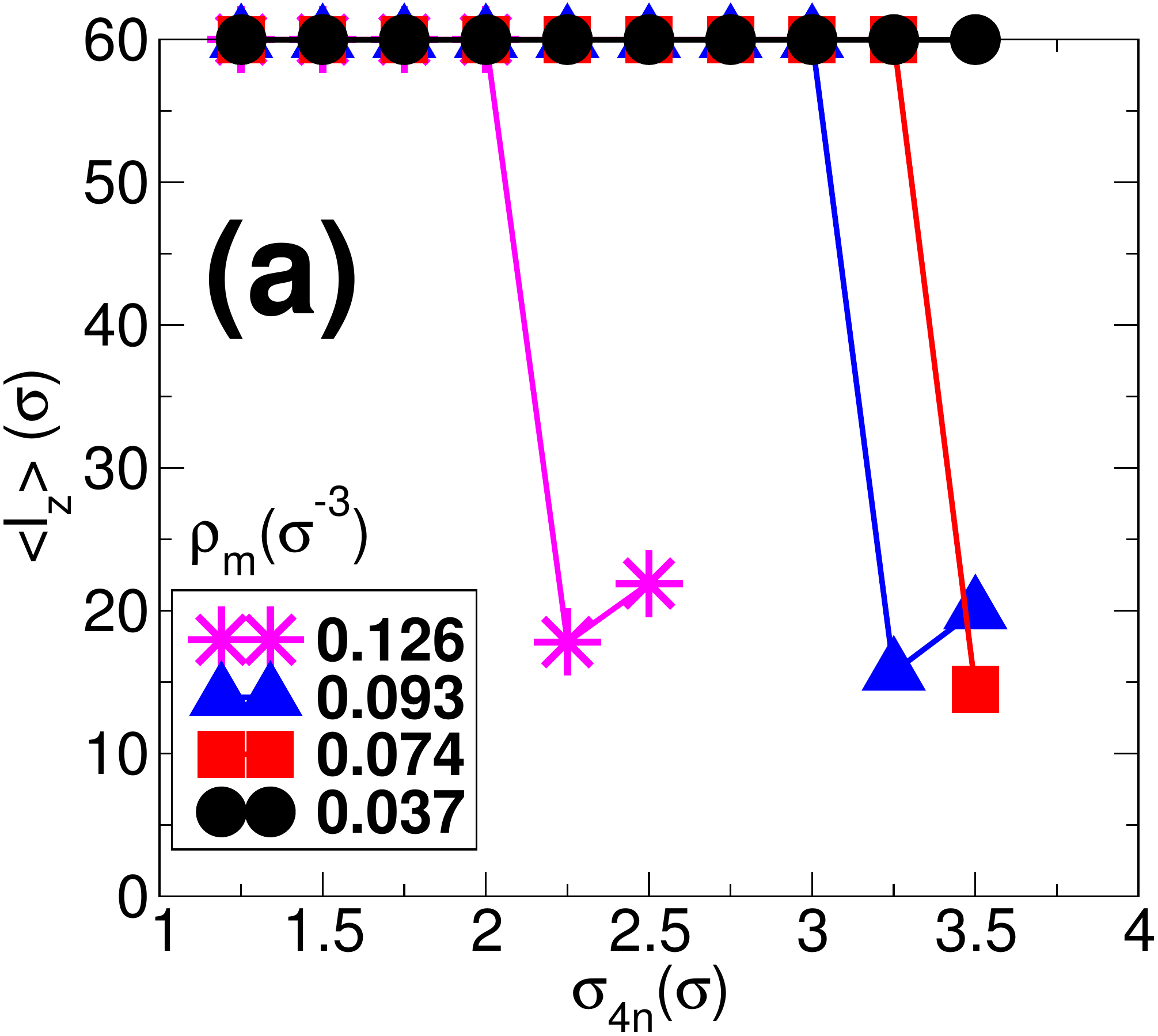}
\caption{The figure shows the average of the largest eigenvalue of the gyration tensor of NP clusters versus $\sigma_{4n}$, for $\sigma_n=1.5\sigma$. It shows graphs for different values of monomer number densities $\rho_m$ as indicated by the different symbols. The size of the clusters remains same as the system size ($60\sigma$) before showing a sudden decrease at a higher value of $\sigma_{4n}$. This sudden decrease in the graph at a higher value of $\sigma_{4n}$ indicates the transformation of NP networks to non-percolating clusters.}
\label{eigen}
\end{figure}

For $\sigma_n=1.5\sigma$, the transformation from a percolating network-like structure to non-percolating clusters is also confirmed by plotting the average size of the NP clusters in z-direction which is shown in Fig.\ref{eigen}. This is the average of the largest eigen-values of the NP clusters found by diagonalizing the gyration tensor of the NP clusters. 

      Thus, for all the values of NP size considered here, similar morphological transformations are recorded. However, the system behaviour is observed to show two major changes when the NP size is changed. One of the changes marked at once is the value of $\sigma_{4n}$ at which the transformation from a percolating network-like structure of NPs to non-percolating clusters occurs. This value seems to be decreasing with the increase in the size of NPs. This is evident from the snapshots in figures \ref{5000_150}, \ref{5000_200} and \ref{5000_250}. For snapshots shown in Fig.\ref{5000_150}, till the value of $\sigma_{4n}=2.5\sigma$ the snapshots show percolating network-like structures of NPs except for $\sigma_{4n}=1.25\sigma$. Non-percolating NP clusters are seen for $\sigma_{4n}=3\sigma$. In case of $\sigma_n=2.5\sigma$, the value of $\sigma_{4n}=2.5\sigma$ bears the non-percolating clusters of NPs. Thus, we can see that the value of $\sigma_{4n}$ at which the transformation from percolating to non-percolating NP clusters occurs, gets shifted to lower values of $\sigma_{4n}$ with the increase in the size of NPs. Finally, with further increase in NP size to $\sigma_{n}=3\sigma$, the introduction of NPs inside the simulation box becomes very difficult for $\sigma_{4n}=2.75\sigma$ and hence the non-percolating clusters of NPs are not observed. This behaviour is not only observed for $\rho_m=0.093\sigma^{-3}$, but also for the other values of micellar densities considered here ($\rho_m=0.074\sigma^{-3}$ and $0.126\sigma^{-3}$). For the lowest density of micelles, it has already been shown that for the range of $\sigma_{4n}$ considered here, the structural change from network to non-percolating clusters is not observed. This also comes out to be true for the larger size of particles. For all the size of NPs considered here, the NP-WLM system shows a system spanning percolating network-like structures of both NPs and micellar chains interpenetrating each other in case of $\rho_m=0.037\sigma^{-3}$. The snapshots corresponding to the effect of the size of NPs for densities $\rho_m=0.037\sigma^{-3}$ and $0.126\sigma^{-3}$ are provided in the supporting material \cite{supporting_material}.

\begin{figure}
\centering
\includegraphics[scale=0.25]{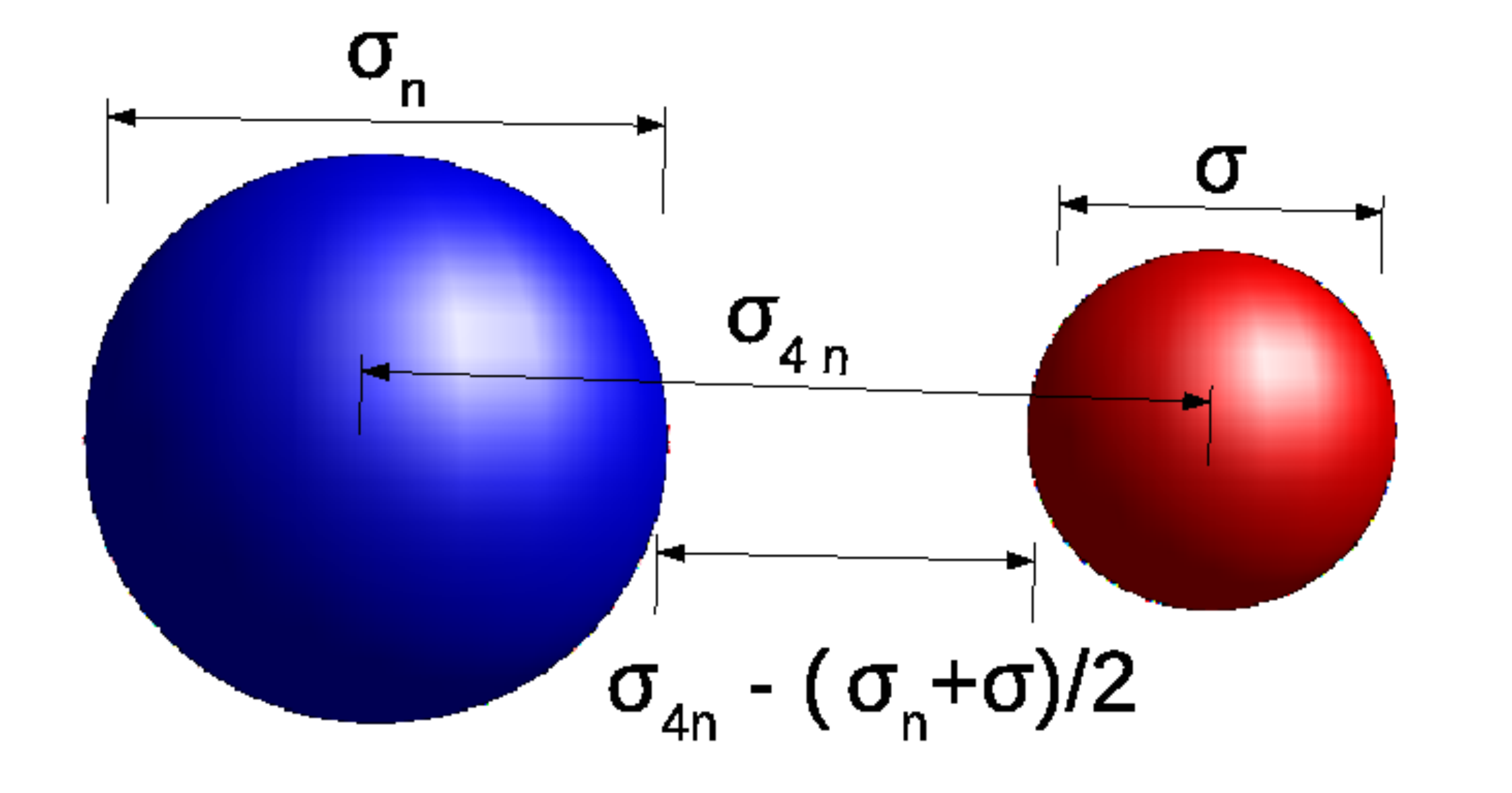}
\caption{ The figure explains the space parameter (SP) defined as SP = $ (\sigma_{4n} –(\sigma_n + \sigma/2))$. This quantity indicates the minimum approaching distance between the surfaces of a monomer (red) of size $\sigma$ and a NP (blue) with size $\sigma_n$, excluding the particles themselves. Thus, the value of $\sigma_{4n}$ is measured from centre-to-centre of the two particles, but the value of SP is measured from surface-to-surface of the particles. }
\label{space}
\end{figure}


\begin{figure*}
\centering
\includegraphics[scale=0.19]{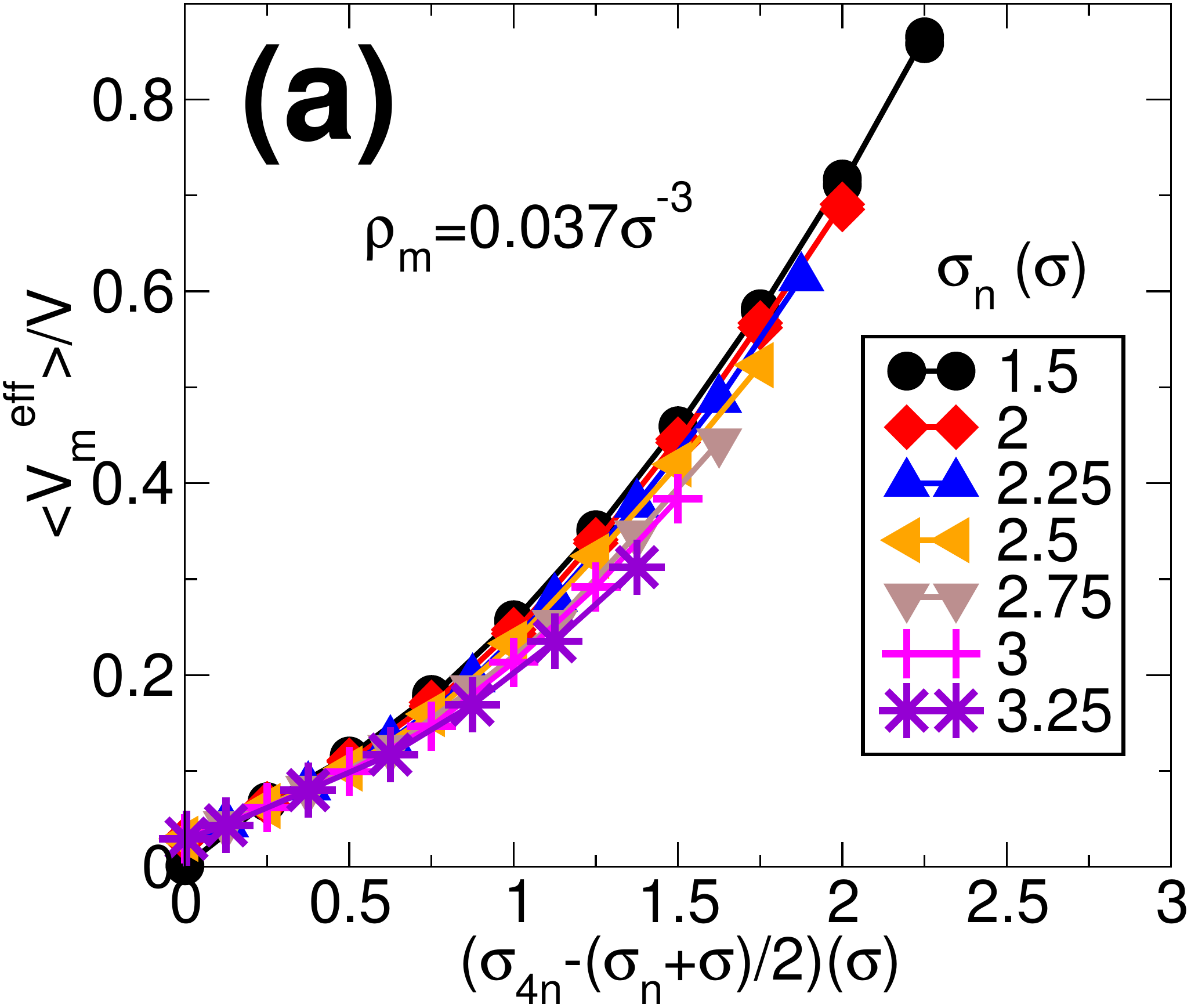}
\includegraphics[scale=0.19]{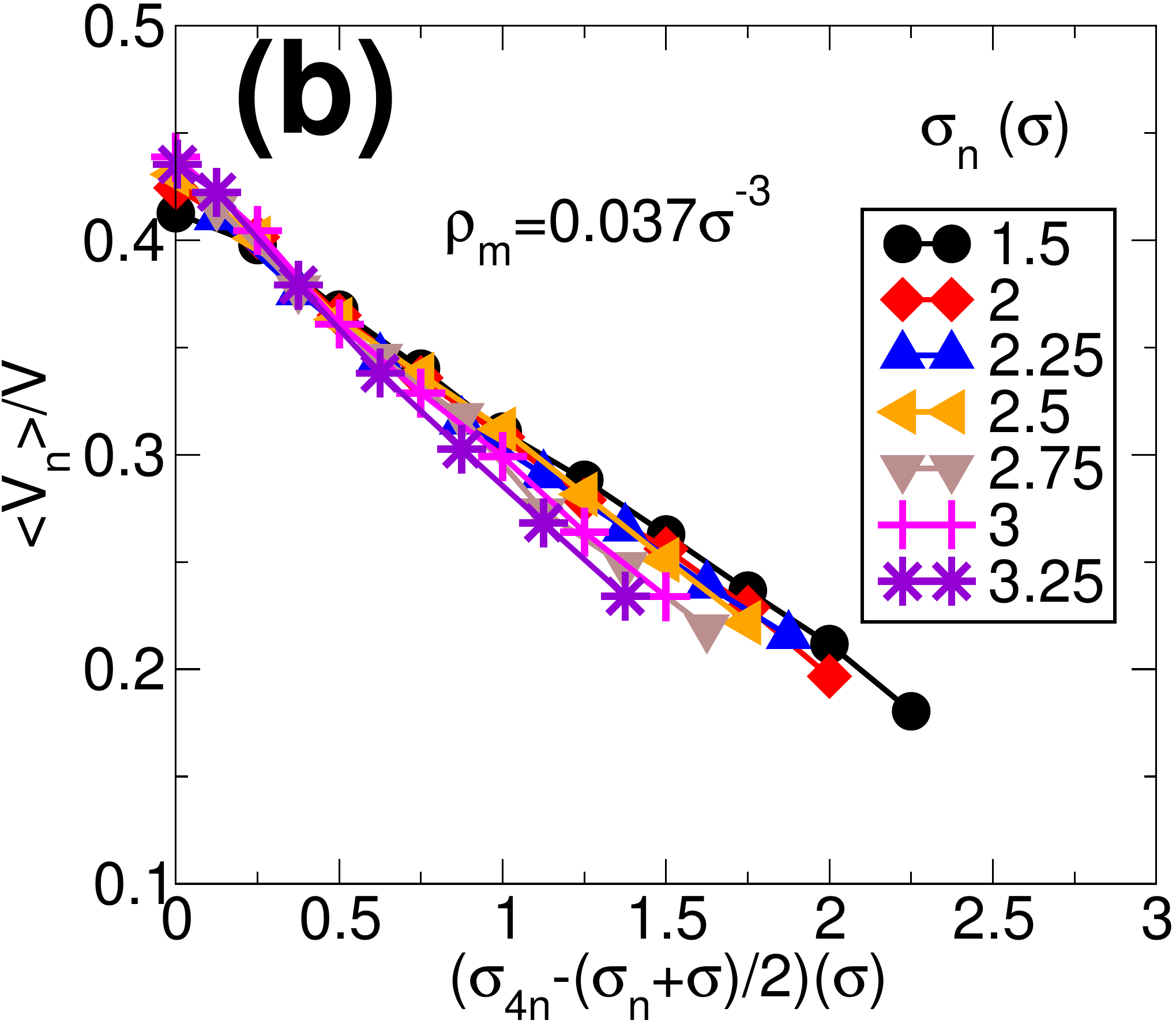}
\includegraphics[scale=0.19]{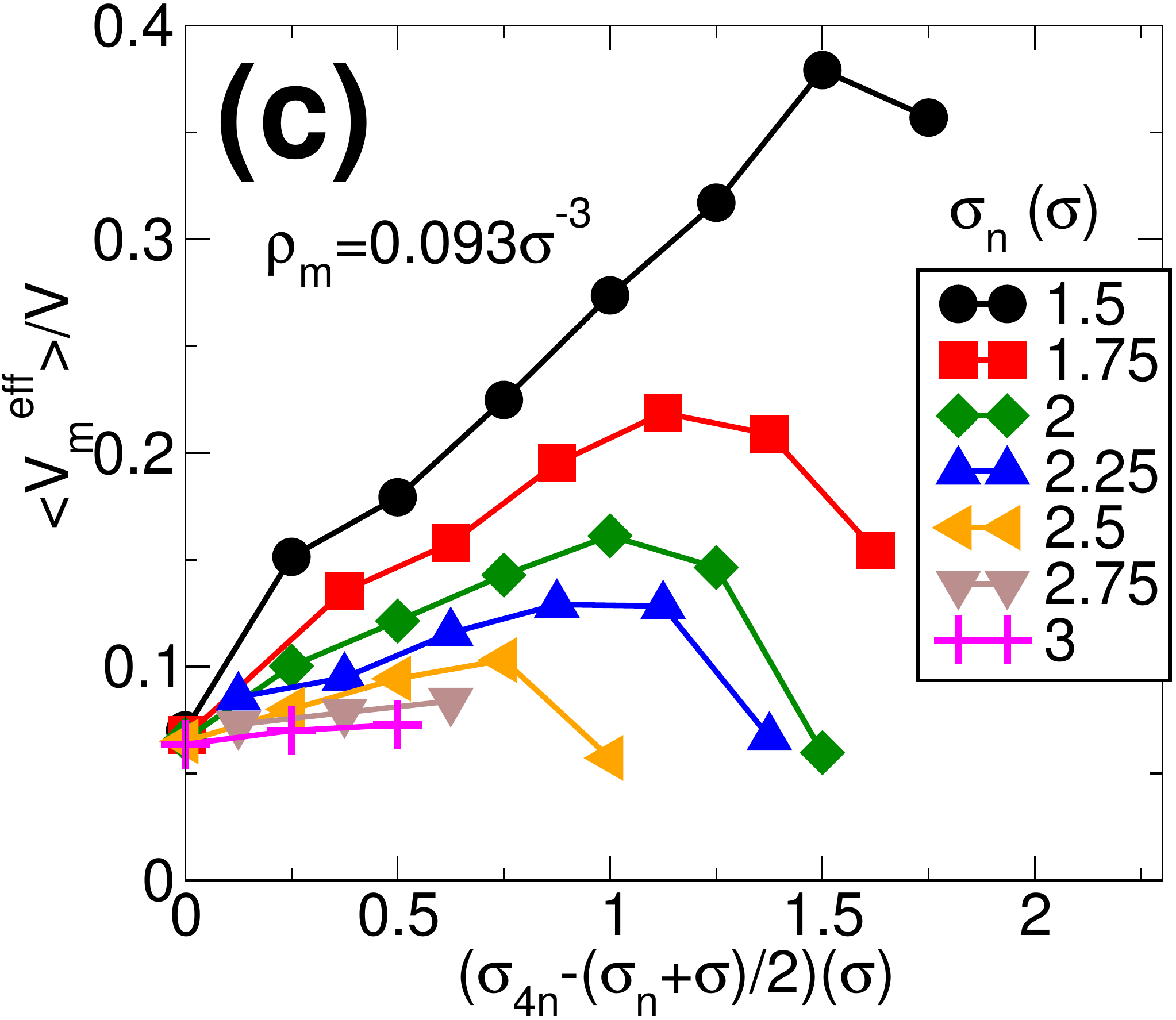}
\includegraphics[scale=0.19]{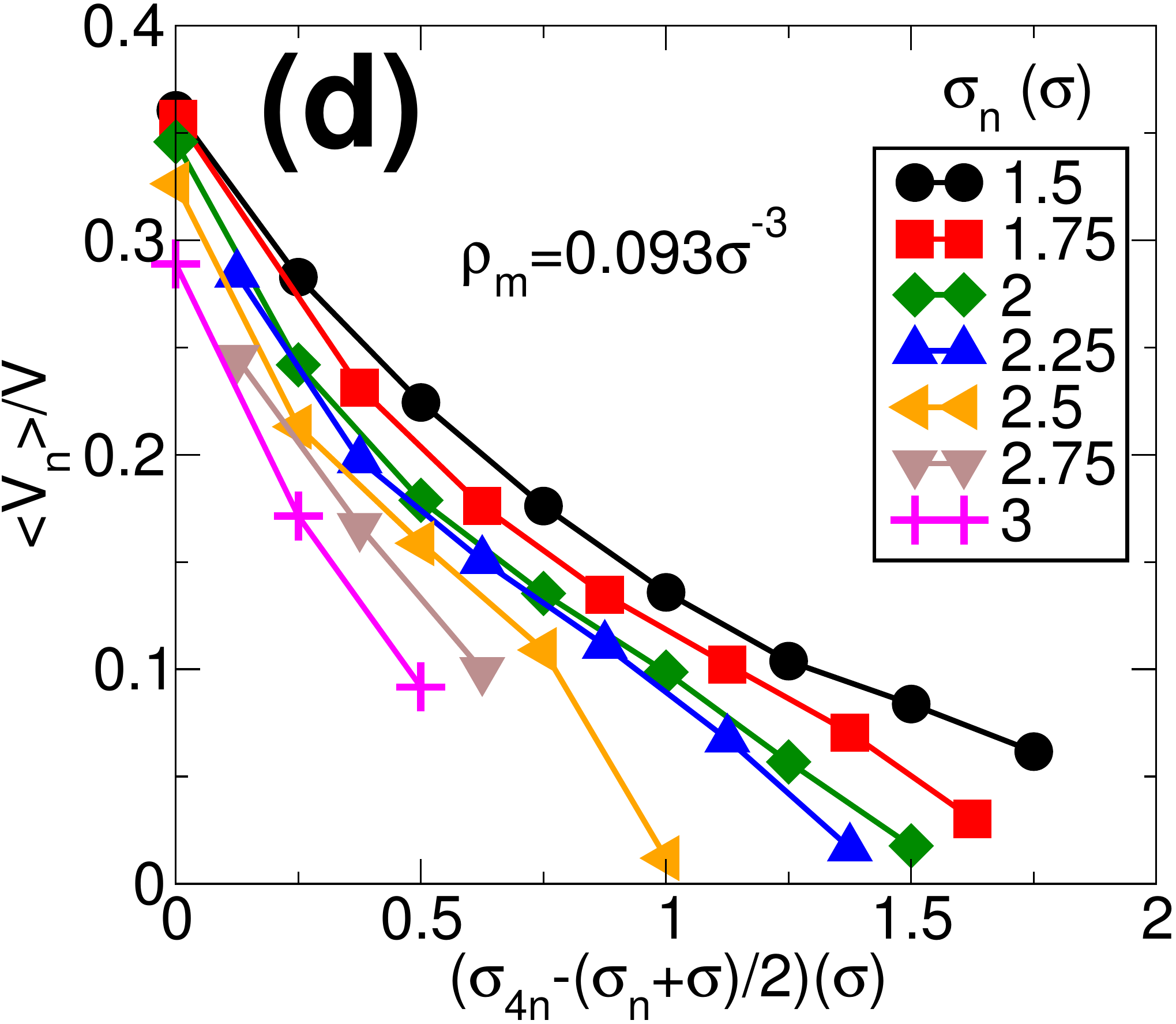}
\caption{ The figure shows the statistical average of the effective volume fraction of monomers $<{V_m}^{eff}>/V$ and the volume fraction of NPs $<V_n>/V$, where V is the volume of the simulation box, for two values of monomer number densities $\rho_m=0.037\sigma^{-3}$ ((a) and (b)) and $\rho_m=0.093\sigma^{-3}$ ((c) and (d)). For the case of the lower density of monomers, there is not much difference between the volume fraction for different values of NP size. The value of $<{V_m}^{eff}>$ increases and $<V_n>$ decreases with increase in $\sigma_{4n}$ for all the values of $\sigma_n$. However, there is a significant difference in the behaviour of $<{V_m}^{eff}>$ for the higher monomer density case (in (c)), as the size of NPs is changed. The effective volume fraction in this case also increases but show a decrease after a certain value of $\sigma_{4n}$. This is the point which marks the transformation of NP network to non-percolating clusters. Moreover, as the size of NP increases, this point of transformation (where effective monomer volume decreases), gets shifted to lower values of $\sigma_{4n}$. The volume fraction of NPs decreases with increase in $\sigma_{4n}$ for all the values of the size of NPs as shown in the figure (d).}
\label{mono_vol}
\end{figure*}

      The transformation from a percolating network-like structure of NPs to non-percolating clusters is marked by a decrease in the value of the effective volume fraction of micellar chains. It is due to the fact that inside a percolating network, micelles are surrounded by more NPs compared to polymers in between non-percolating clusters due to lower NP density. Therefore, inside a percolating network, a high value of repulsive interactions between monomers and NPs is expected due to $V_{4n}$. This increase in the repulsive interactions increases the effective volume of monomers (Fig.\ref{eff_vol}). Therefore, breaking of NP network to non-percolating clusters is marked by a decrease in the value of effective micellar volume fraction. Moreover, with the increase in the size of NPs, it becomes more difficult to introduce NPs inside the micellar matrix due to the elasticity of the matrix ($V_3$ potential). Therefore, with a larger size of NPs, a low value of NP volume fraction is observed. Hence, the systems with a larger size of NPs form non-percolating clusters at a lower value of $\sigma_{4n}$ compared to the systems with smaller NPs. To confirm these observations, we examine the behaviour of the effective volume fraction of micelles and volume fraction of NPs. The plots for these quantities are shown in Fig.\ref{mono_vol} for $\rho_m=0.037\sigma^{-3}$ (figures (a) and (b)) and $\rho_m=0.093\sigma^{-3}$ (figures (c) and (d)).


       The value of $\sigma_{4n}$ defines the minimum approaching distance between monomers and NPs. However, with the change in the size of NPs, the same value of $\sigma_{4n}$ gives out different values of the distance between the surfaces of the particles. For a given value of $\sigma_{4n}$, the minimum approaching distance between the surfaces of a monomer and a NP will be higher for a smaller NP compared to that of a bigger particle. Therefore, to study the effect of the change in $\sigma_{4n}$, a parameter which is more relevant is the minimum approaching distance between particle surfaces. We call this parameter as space parameter (SP) as it defines the space between the two surfaces and is calculated as $(\sigma_{4n}-(\sigma_n + \sigma)/2))$. The SP changes proportional to the value of $\sigma_{4n}$ and can be used synonymously with each other. Figure.\ref{space} explains the calculation of SP. The plots shown in Fig.\ref{mono_vol} are plotted against this parameter SP. Figure.\ref{mono_vol} shows the behaviour of effective volume fraction of monomers and NP volume fraction for two different values of micellar densities, $\rho_m=0.037\sigma^{-3}$ (\ref{mono_vol}(a) and (b)) and $0.093\sigma^{-3}$ (\ref{mono_vol}(c) and (d)). Each figure shows different graphs for different values of the NP size indicated by the different symbols. Figure. \ref{mono_vol}(a) shows the effective volume of monomers for the lowest density. As discussed earlier that no morphological transformation from network to non-percolating NP clusters is observed in this case, no decrease in the value of effective volume fraction of micelles is observed. For all the values of $\sigma_{4n}$ or SP, it only shows an increase in the effective micellar volume fraction with the increase in $\sigma_{4n}$ (or SP). The behaviour of NP volume fraction shown in Fig.\ref{mono_vol}(b) shows that with an increase in $\sigma_{4n}$ (SP), the NP volume fraction decreases. For all the values of NP size, the graphs of micellar effective volume fraction and NP volume fraction nearly overlap each other. This also indicates the similarity in the observed structures where all of them are forming system spanning percolating networks of NPs and micellar chains. Please refer the supporting materials for the snapshots for $\rho_m=0.037\sigma^{-3}$ with different $\sigma_{4n}$ and $\sigma_n$ \cite{supporting_material}.

      The behaviour of the effective volume fraction of micelles for $\rho_m=0.093\sigma^{-3}$ in Fig.\ref{mono_vol}(c) shows a very different behaviour compared to the behaviour shown by $\rho_m=0.037\sigma^{-3}$ in Fig.\ref{mono_vol}(a). As discussed previously, the breaking of NP network into non-percolating clusters is marked by a decrease in the value of effective volume fraction of micelles, the behaviour shown by \ref{mono_vol}(c) confirms the behaviour observed by their snapshots shown in  figure \ref{5000_150}, \ref{5000_200}, \ref{5000_250} and \ref{5000_300}. The effective monomer volume fraction shows an increase with the increase in the value of $\sigma_{4n}$ (or SP) but decreases for a higher value of $\sigma_{4n}$ indicating the formation of non-percolating clusters of NPs from a system spanning network-like structure. This behaviour is observed for all the values of NP size indicated by different symbols in the plot. It should be noted that the point where the effective monomer volume fraction shows a decrease in its value gets shifted to lower values of SP (or $\sigma_{4n}$)  with the increase in the size of NPs. This confirms the observed behaviour in Figs.\ref{5000_150}, \ref{5000_200} and \ref{5000_250} as discussed in earlier paragraphs. For bigger size particles $\sigma_n=2.75\sigma$ and $3\sigma$, the decrease in ${V_m}^{eff}$ is not observed which can be seen in Fig.\ref{5000_300}. This is due to the difficulty in introducing bigger particles at higher values of $\sigma_{4n}$ because of a huge elastic cost of the polymeric matrix. Moreover, with an increase in the SP parameter (or $\sigma_{4n}$), the NP volume fraction shows a decrease in its value as expected. The NP volume fraction not only decreases with increase in $\sigma_{4n}$ but also with an increase in NP size $\sigma_n$. Hence verifying the conclusion that with the increase in NP size, it becomes difficult to introduce NPs in the system.

\begin{figure}
\centering
\includegraphics[scale=0.185]{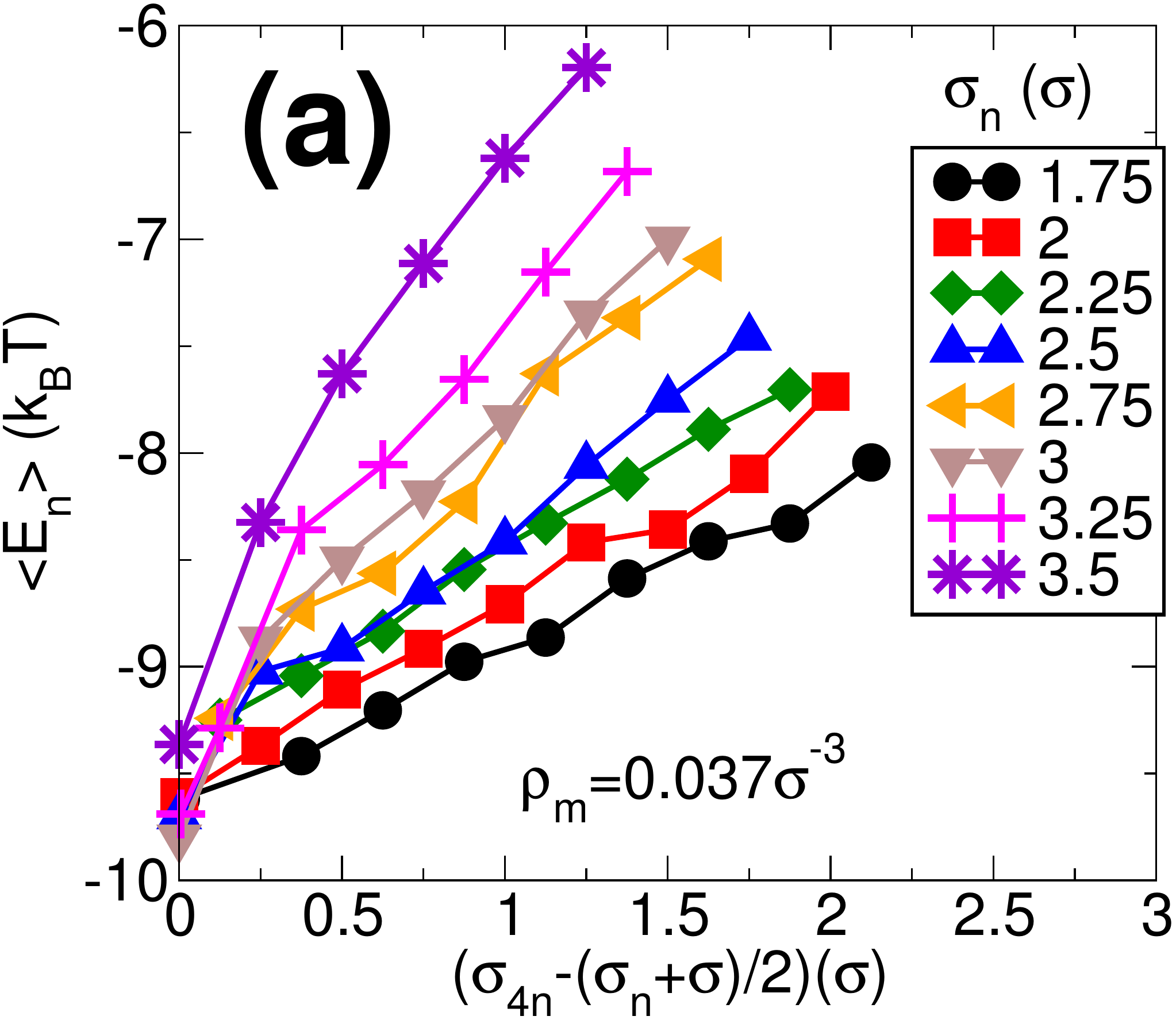} 
\includegraphics[scale=0.185]{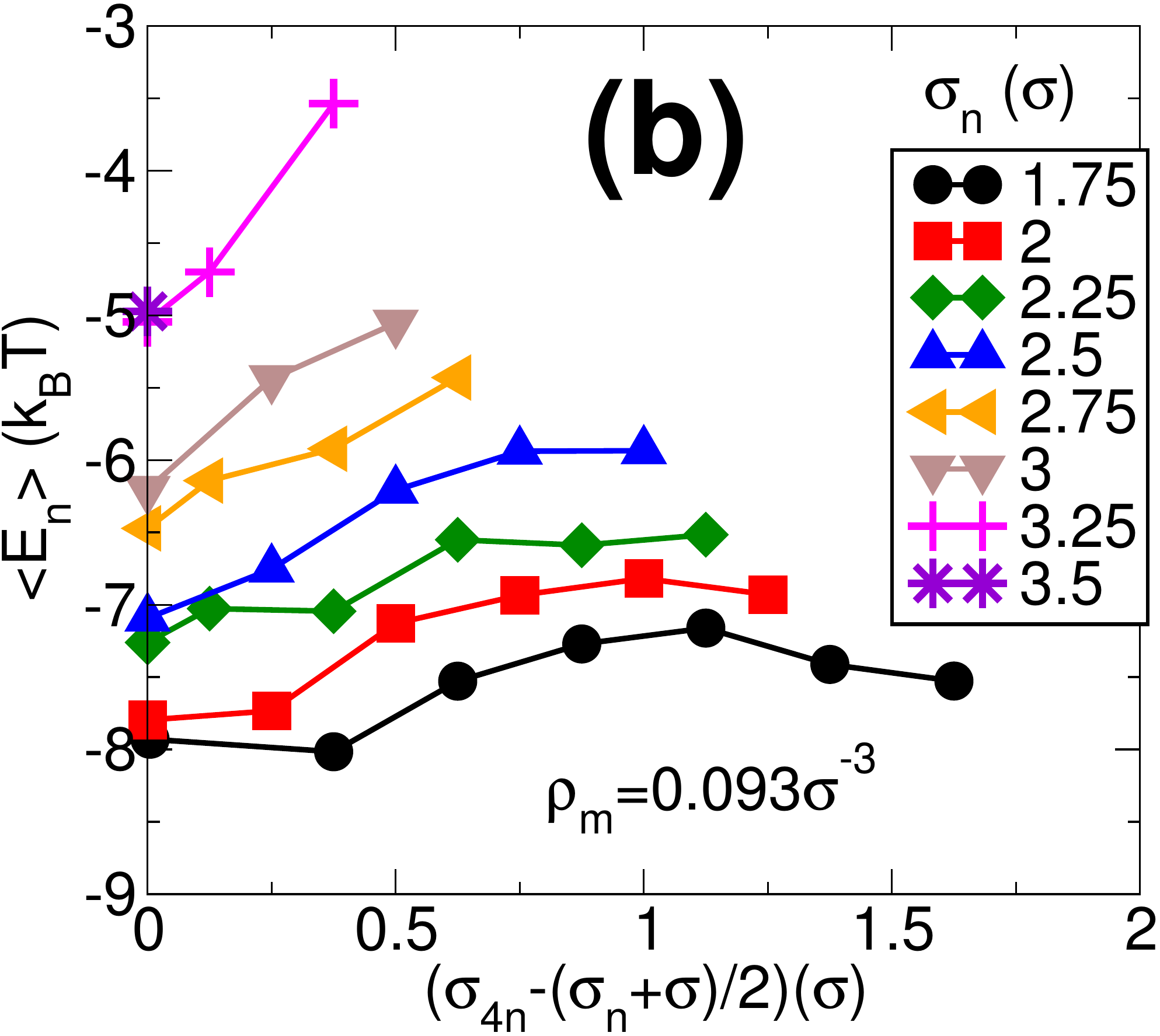}
\caption{The figure shows the average energy of NPs for two different values of monomer number densities (a) $\rho_m=0.037\sigma^{-3}$ and (b) $\rho_m=0.093\sigma^{-3}$. In each figure, the plots are shown for different values of NP size $\sigma_n$ as indicated by the different symbols. In both the cases, the NP energy is shown to be decreasing with increase in either $\sigma_{4n}$ (or the space parameter (SP)) or the size of NP $\sigma_{n}$. This decrease in the energy is due to the decrease in the surface to volume ratio leading to a decreased surface interactions with the increase in the NP size. Due to lower energy of NPs for larger particles, they are not able to form well-packed clusters.}
\label{en_nano}
\end{figure}

\begin{figure}
\centering
\includegraphics[scale=0.3]{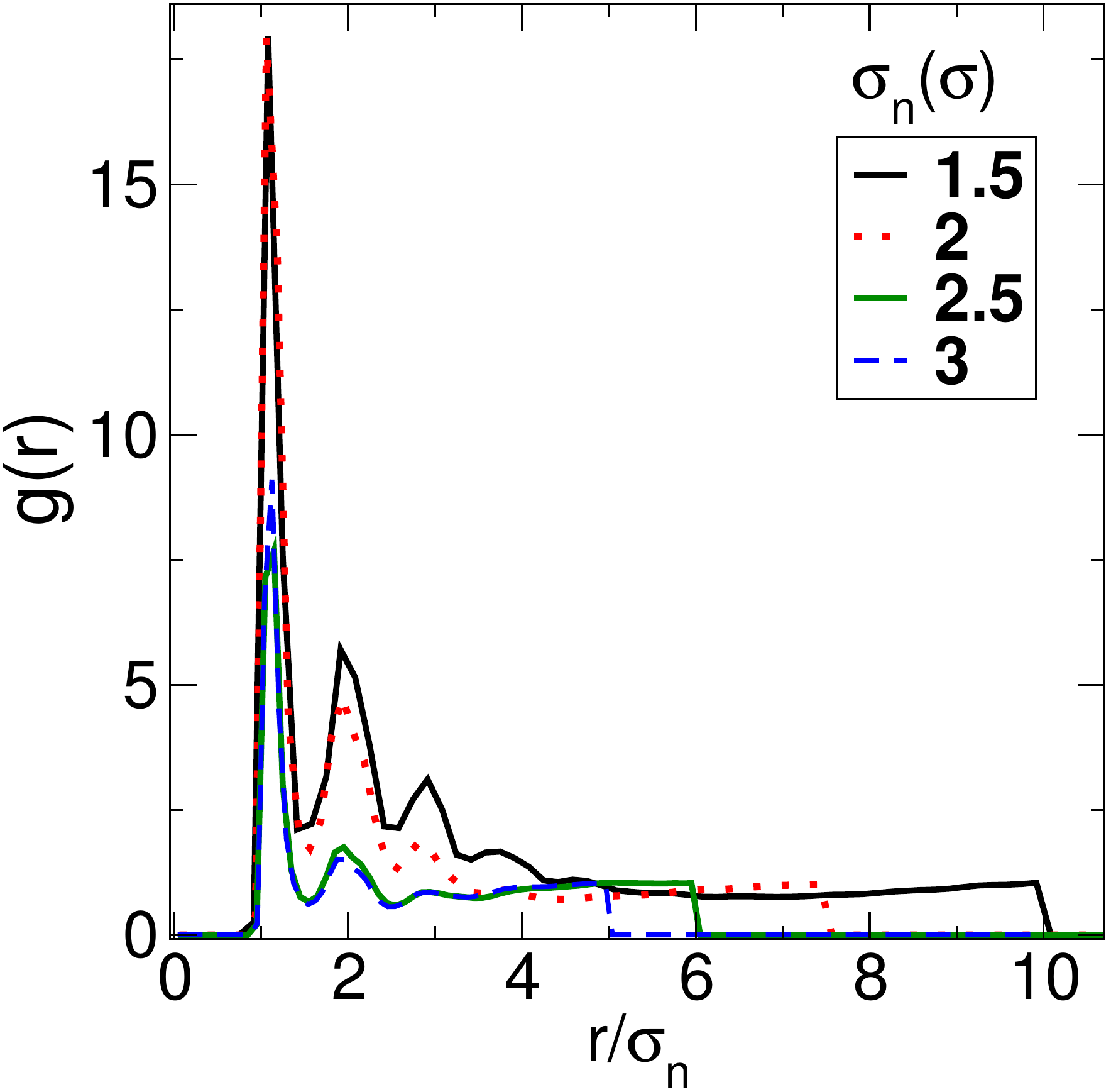} 
\caption{ The figure shows the pair correlation function for NPs for $\rho_m=0.093\sigma^{-3}$ and $\sigma_{4n}=2\sigma$. It shows different plots with different size of NPs indicated by the corresponding symbols. The x-axis has a parameter of distance between NPs divided by the NP size. This parameter suitably shifts the graphs along the x-axis in order to be able to compare the plots for different values of $\sigma_{n}$. The plots for $\sigma_n=1.5\sigma$ and $2\sigma$ shows relatively sharp peaks and a longer range of correlation compared to the plots for $\sigma_n=2.5\sigma$ and $3\sigma$. This shows that as the particle size increases, particles are unable to form a well-packed and well-defined structure of NPs.}
\label{gofr_nano}
\end{figure}

       Not only the bending cost of the elastic micellar matrix is responsible for the lower NP density for bigger NPs, but a higher average energy of NPs in a NP cluster for larger particles is also responsible. An increase in the size of NPs reduces the surface to volume ratio of NPs. This decreases the net surface interaction between them and hence increases the energy of the NPs in a cluster. Due to this increase of NP energy, it becomes difficult to form well-packed big clusters inside a micellar matrix. Therefore, a high elastic cost of micellar matrix compared to the energy of NPs is unable to bind NPs in a well-packed cluster leading to a reduced number density and low packing. The average energy plots for NPs shown in Fig.\ref{en_nano} confirms this. The figure shows the average energy of NPs for two different values of micellar number densities, (a) $\rho_m=0.037\sigma^{-3}$ and (b) $\rho_m=0.093\sigma^{-3}$. Each figure shows graphs for different values of NP size indicated by different symbols in the figure. The behaviour is again suitably shown against the parameter SP. Both the figures clearly show that the energy of NPs increases with increase in the size of NPs. For $\rho_m=0.037\sigma^{-3}$, we have seen that the NP volume fraction decreases with increase in $\sigma_{4n}$ (Fig.\ref{mono_vol}). Therefore, the increase in the value of NP energy is due to the decrease in the cluster size. Since the NPs always form percolating network-like structures for $\rho_m=0.037\sigma^{-3}$ with same periodicity for a given $\sigma_n$ (\cite{arxive}), the approximately linear decrease in the NP volume fraction and the corresponding increase in the NP energy indicates the decrease in the wall thickness of the NP network (or increase in the pore size). However, the NP network gradually breaks and finally splits into individual clusters in case of $\rho_m=0.093\sigma^{-3}$. Therefore, the increase in energy of NPs is not linear with $\sigma_{4n}$.

\begin{figure*}
\centering
\includegraphics[scale=0.2]{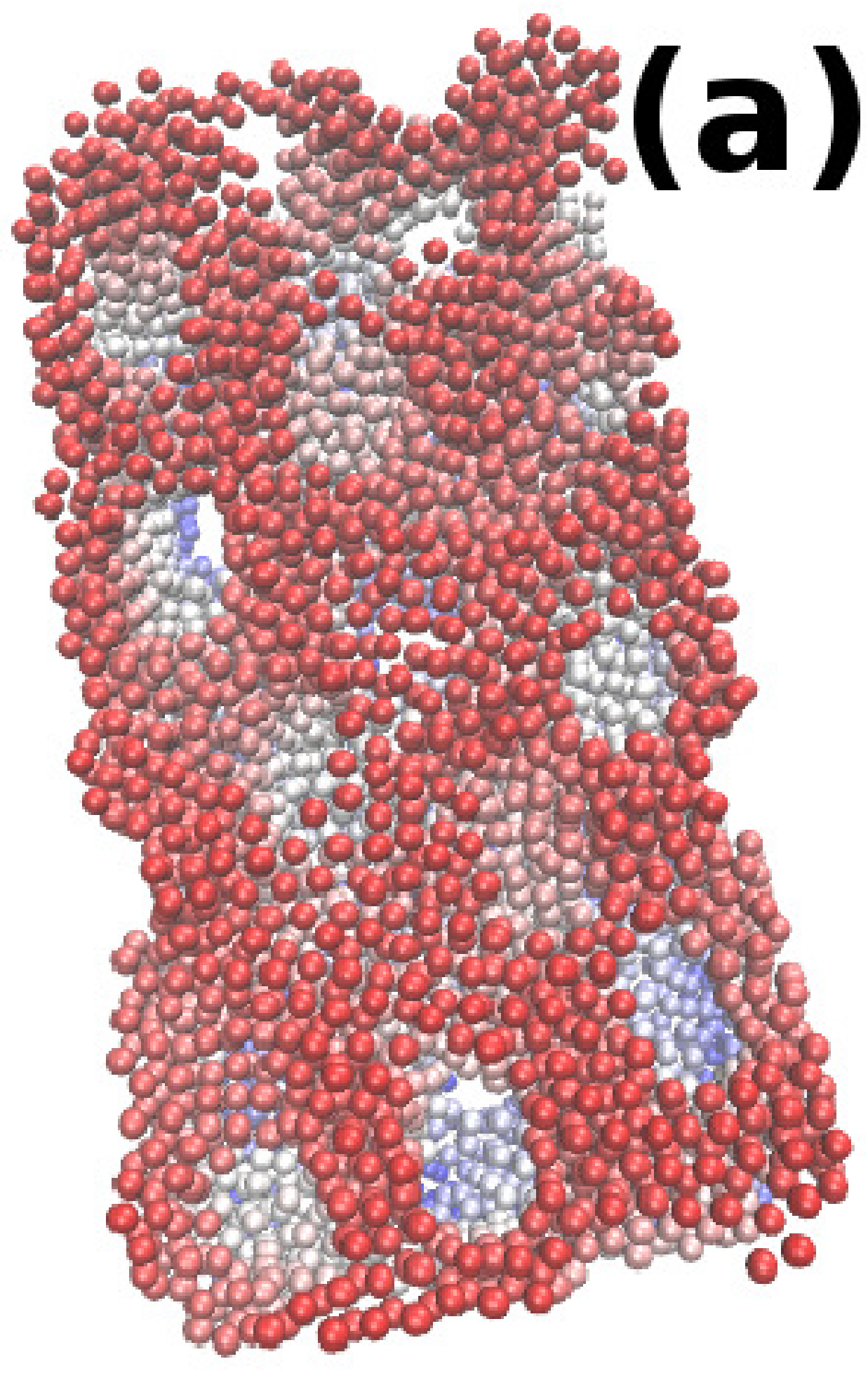}
\hspace{1cm}
\includegraphics[scale=0.2]{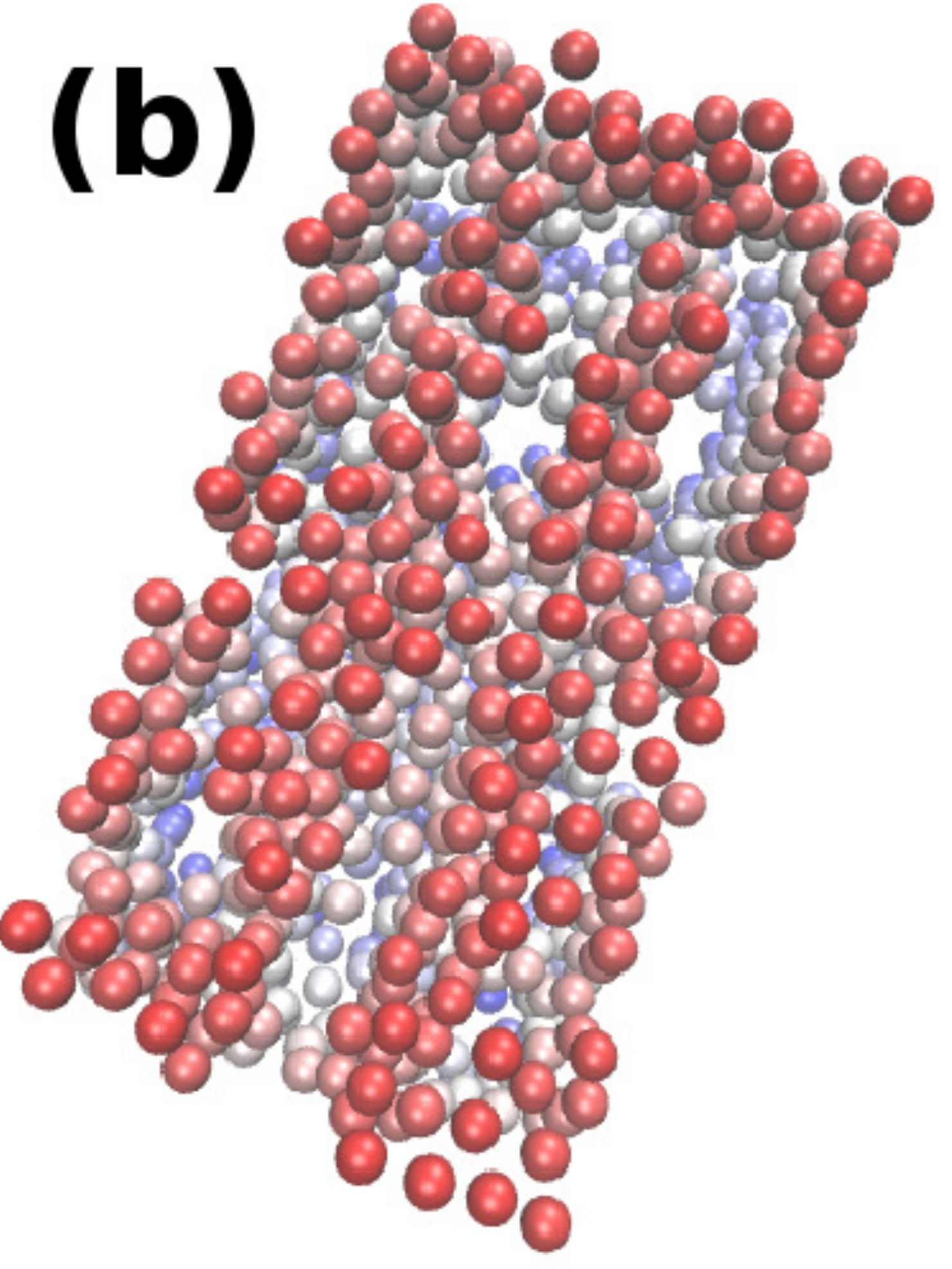}
\hspace{1cm}
\includegraphics[scale=0.2]{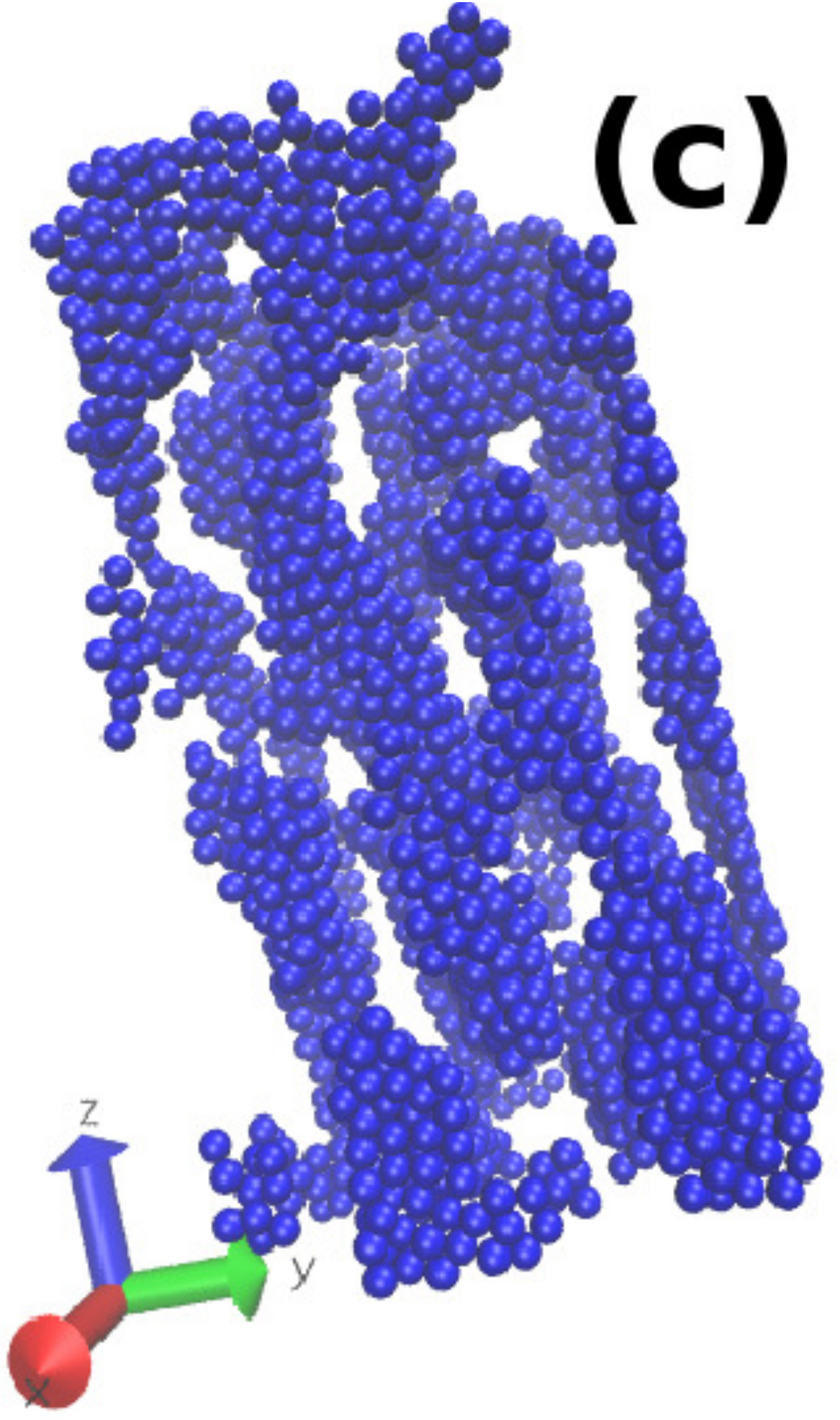}
\hspace{1cm}
\includegraphics[scale=0.2]{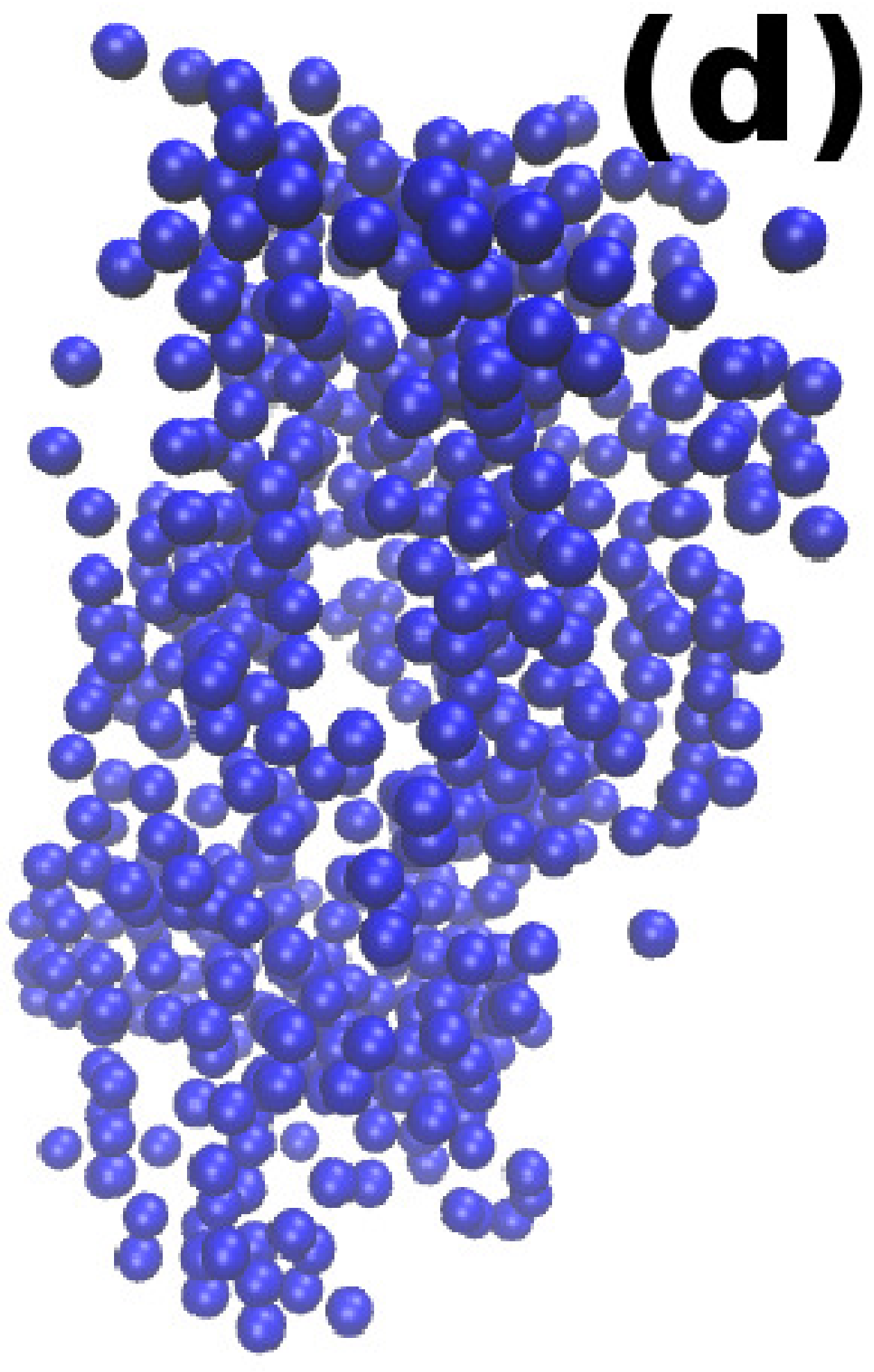}
\caption{ Figure shows the snapshots of NPs for $\rho_m=0.037\sigma^{-3}$ and $\sigma_{4n}=3\sigma$ (in (a) and (b)); and $\rho_m=0.126\sigma^{-3}$ and $\sigma_{4n}=2\sigma$ (in (c) and (d)). The size of NPs in Figs (a) and (c) is $\sigma_n=1.5\sigma$ while for Figs. (b) and (d) it is $\sigma_n=3\sigma$. The figures for the lower density case show a gradiant in color for better visualization. The effect of increase in the size of NP can be clearly seen from these figures with the NPs with a higher particle size not able to form a well packed clusters (in (b) and (d)) for both the cases of densities.}
\label{2000_6800}
\end{figure*}

 The effect of an increase in the NP energy with the increase in the size of NPs can be instantly seen to be affecting the arrangement of NP clusters. The low packing of the NP clusters for bigger NPs is very much evident if we compare the NP clusters in figures \ref{5000_150} with the clusters formed with larger size of NPs shown in figures \ref{5000_200}, \ref{5000_250} and \ref{5000_300}. As the value of NP size $\sigma_n$ increases, the NP packing decreases. This fact is best represented by comparing the graph of the pair correlation functions for NPs which is shown in Fig.\ref{gofr_nano}. The figure shows the pair correlation function for NPs for different values of NP size $\sigma_n$ as indicated by the different types of lines (or colours) in the plot. The value of $\sigma_{4n}$ is kept at $2\sigma$. The X-axis is normalized by dividing by the size of NP in order to compare the graphs with different particle sizes. Higher peaks and a long-range correlation in case of the lower size of particles indicate a well packed and denser system of NPs compared to larger sized particles. A sharp decrease in the peaks from $\sigma_n=1.5\sigma$ to $\sigma_n=3\sigma$ can be easily recognized supporting the observed snapshots for a well packed NP cluster in \ref{5000_150}(d) but a loosely packed structure of NPs in Fig.\ref{5000_300}(c) \cite{supporting_material}.

        As a result of the change in the size of NPs, we observed that the value of $\sigma_{4n}$ at which the transformation from NP network to non-percolating clusters occurs and the NP cluster packing gets affected. The transformation point shifts to the lower value of $\sigma_{4n}$ while, the packing of the NP clusters decrease with increase in the value of $\sigma_n$. Thus, it will be difficult to produce a well-defined NP structure with a larger sized NP. Summing up these observations, the figure \ref{2000_6800} represents the snapshots for two different values of micellar densities $\rho_m$ and NP size $\sigma_n$. The figure(a) and (b) shows snapshots for $\rho_m=0.037\sigma^{-3}$ with $\sigma_{4n}=3\sigma$. The two figures have different values of NP size, (a) $\sigma_n=1.5\sigma$ (b) $\sigma_n=3\sigma$. The snapshot with smaller particles (figure(a)) shows a well packed and a well defined percolating network-like structure of NPs compared to the snapshot shown in (b). Similar observations can be made for the snapshots shown in figures (c) and (d) which are for $\rho_m=0.126\sigma^{-3}$ and $\sigma_{4n}=2\sigma$. Figure (c) represents the system with $\sigma_n=1.5\sigma$ while, the value of $\sigma_n$ in figure (d) is $3\sigma$. Both the snapshots verify that the system with smaller sized particles has a better packing with a well-defined structure. Moreover, for both the densities, the value of $\sigma_{4n}$ is kept same for the two snapshots shown, but the snapshots show different states of their structures. The figure in (a) shows a system spanning percolating network-like structure for $\sigma_{4n}=1.5\sigma$, but the structure for $\sigma_n=3\sigma$ in (b) shows an intermediate structure with the network of NPs partially broken for the same value of $\sigma_{4n}$. Hence showing the shift of the value of $\sigma_{4n}$ for a network to non-percolating clusters transformation, with the change in the value of $\sigma_n$. These snapshots also show that the shape of the NP clusters is governed by the density of the polymeric matrix. We have seen sheet-like NP clusters in Figs.6, 7 and 8 for $\rho_m=0.093\sigma^{-3}$. However, the NP clusters in Fig.\ref{2000_6800}(a) are app. spherical for $\rho_m=0.037\sigma^{-3}$ while they are rod-like in Fig.\ref{2000_6800}(c) for $\rho_m=0.126\sigma^{-3}$ (please refer the supporting material for a detailed information \cite{supporting_material}).


\section{Conclusion}

        The paper examines in detail the effect of the NP size on the behaviour of the NP-Wormlike micellar system (OR NP- equilibrium polymer system). It shows that the system undergoes morphological changes for all the values of NP size considered here. However, for bigger NPs, it becomes difficult to introduce particles in the system and hence difficult to get the non-percolating clusters (e.g. for $\sigma_n=2.75\sigma$ and $3\sigma$). The effect of the size of NPs is observed to shift the point of transformation of NP network to  non-percolating clusters, to lower values of $\sigma_{4n}$ as a result of an increase in particle size. Moreover, the NP assembly displays well-packed clusters and well-defined structures for smaller particles compared to bigger particles. These two effects of the size of NPs are  presented by the snapshots of the systems and verified by plotting the effective volume fraction of monomers and pair correlation function of the NPs. The shift in the morphological transformation point is attributed to the increase in the average energy of NPs with the increase in NP size, because of a decrease in the surface to volume ratio. This is confirmed by plotting the average energy of NPs that show an increase with the increase in NP size. 

\section{Acknowledgement}

        I acknowledge the helpful discussion with Dr.Arijit Bhattacharyya, Dr.Guruswamy Kumaraswamy and Dr. Deepak Dhar. I am also thankful to the computational facility provided by National Param Super Computing (NPSF) CDAC, India for the use of Yuva cluster and the computer cluster in IISER-Pune, funded by DST, India by Project No. SR/NM/NS-42/2009 without which this work would not have been completed.

\nocite{author}
\bibliographystyle{spr-chicago}
\bibliography{epje}
\end{document}